\documentclass[prxquantum,preprint,superscriptaddress,longbibliography,amsmath,amssymb]{revtex4-2}
\usepackage{graphicx}
\graphicspath{{./figures/}}
\usepackage{bm,booktabs,array,tabularx}
\usepackage{physics}
\usepackage{siunitx}
\usepackage[colorlinks=true,citecolor=blue,linkcolor=blue,urlcolor=blue]{hyperref}
\usepackage[capitalize]{cleveref}
\usepackage{amsthm}
\newtheorem{proposition}{Proposition}
\AtBeginDocument{\RenewCommandCopy\qty\SI}

\providecommand{\dd}{\mathrm{d}}

\newcommand{\Ne}{\mathcal{N}_{W}}

\begin{document}

\title{The resource--invariant gap: a protected Floquet invariant need not protect Wigner negativity}

\author{Juste Deuyekbe}
\email{justedeuyekbe@gmail.com}
\affiliation{Department of Physics, Faculty of Science, University of Ngaoundere, P.O. Box 454, Ngaoundere, Cameroon}

\author{P. Djorw\'e}
\email{djorwepp@gmail.com}
\affiliation{Department of Physics, Faculty of Science, University of Ngaoundere, P.O. Box 454, Ngaoundere, Cameroon}
\affiliation{Stellenbosch Institute for Advanced Study (STIAS), Wallenberg Research Centre at Stellenbosch University, Stellenbosch 7600, South Africa}

\author{Nsangou Mama}
\email{mnsangou@yahoo.com}
\affiliation{Department of Physics, Faculty of Science, University of Maroua, P.O. Box 46, Maroua, Cameroon}

\author{S. G. Nana Engo}
\email{serge.nana-engo@facsciences-uy1.cm}
\thanks{Corresponding author.}
\affiliation{Department of Physics, Faculty of Science, University of Yaounde I, P.O. Box 812, Yaounde, Cameroon}

\begin{abstract}
Topological invariants constrain a driven Floquet generator, yet whether they protect a transient nonlinear bosonic resource remains open. We track boundary Wigner negativity of an even cat state in a chiral Floquet Su--Schrieffer--Heeger chain with boundary Kerr nonlinearity and structured loss. The quasienergy windings remain quantized at every chain length, yet the matched sector contrast flips sign when the chain grows under index-end shielding and is null within $\num{3e-5}$ under uniform loss at $t=T$ for $d\le\num{0.5}$. Prepared and shielded on the physical dangling end, the contrast remains positive, modulated by local dimer return rather than set by the invariant. A three-site causal light cone and a boundary-Laplacian erosion law trace the flip to strong-bond bulk exposure: the sign is set by where the bulk-rate loss sits relative to the prepared site, not by chain length. We call this decoupling the resource--invariant gap.
\end{abstract}


\maketitle\clearpage

\section{Introduction}
\label{sec:introduction}

Non-Gaussian states are essential resources for continuous-variable quantum advantage and bosonic error correction, underpinning cat-qubit encodings and fault-tolerant magic-state protocols~\cite{Weedbrook2012,Vlastakis2013,Touzard2018}. Wigner-function negativity provides a direct diagnostic of their nonclassicality~\cite{Hudson1974,Mari2012}, a formal resource monotone for the full state, evaluated here on a local marginal~\cite{Albarelli2018,Zhuang2018}, and operational lower bounds accessible through finite-sample witnesses~\cite{Chabaud2021witnessingwigner,Zaw2024}. A central architectural question is therefore whether a passive Floquet topological lattice can serve as a protected host for non-Gaussian cat states without converting nodal Wigner erosion into local phase decoherence, a prerequisite for any transfer between bosonic modules. In a dissipative lattice, however, the observed negativity depends on state preparation, bath geometry, observation window, and truncation, so a topological statement about the single-particle Floquet generator must be kept separate from a protection claim for a nonlinear mixed-state resource.

Where does the boundary of topological protection for a non-Gaussian resource lie? We test this as a falsifiable question: does the matched sector contrast between chiral Floquet sectors of a driven dissipative chain track the bulk invariant, which changes discontinuously at the phase boundary, the edge-mode localization, which varies continuously with size, or neither? And under which loss profile, system size, and drive phase does the answer change? Neither candidate explains the contrast: local light-cone branching and bath geometry, rather than bulk topology, govern the sign and survival of the resource.

The Su--Schrieffer--Heeger (SSH) model established that dimerization can support boundary modes protected by chiral symmetry~\cite{Su1979}. Floquet extensions add distinct zero- and $\pi$-quasienergy sectors. In one dimension, chiral symmetry makes these sectors accessible through winding numbers evaluated in symmetric time frames~\cite{Asboth2014}. Because the micromotion parameter is part of the driven evolution, a fixed-frame effective Hamiltonian is not generally sufficient to represent observables at all times~\cite{Xu2022micromotion}. Topological photonics and its optomechanical and phononic extensions provide platforms in which these structures can influence bosonic transport~\cite{Ozawa2019,Rudner2013,Xuereb2014,Peano2015,Peano2016}.

Once dissipation is present, the appropriate description is a periodically driven open-system master equation rather than a closed-system effective Hamiltonian alone~\cite{Mori2023,Chen2024,Sieberer2025}. Engineered loss can support protected or metastable structures under additional symmetry conditions~\cite{Diehl2011,Bardyn2013,Paszko2024}, but those conditions do not imply that arbitrary local observables inherit a protection law. Certain topological features of a mixed state survive broad classes of local noise even when the edge modes themselves do not~\cite{Paszko2024,Chirame2025}. Protection has nevertheless been demonstrated mostly for linear, Gaussian, or symmetry-resolved steady-state observables: photon blockade at SSH edge states~\cite{Li2024photonblockade}, transport of nonclassical light along topological domain walls~\cite{Perez2024domainwalls}, and two-photon correlations at a photonic-chip boundary state~\cite{Wang2019Optica}. A further route is chiral reservoir engineering of bosonic lattices~\cite{YanayClerk2018}.

Protection also has sharp limits. It weakens as two-photon entanglement grows~\cite{Tschernig2021}. Strong nonlinearity destabilizes chiral edge states~\cite{Chaunsali2021} and can drive a topological-to-chaotic transition in SSH chains~\cite{Sone2025chaotic}. Nonlinear interference can destroy Floquet edge modes~\cite{Michen2023}, local periodic driving depopulates edge states~\cite{Fedorova2019}, and helical interface states can survive without topological protection~\cite{Xu2023}; a review of the nonlinear-photonics interplay is given in Ref.~\cite{Smirnova2020}.

The complementary class is nonlinear functionals of a driven dissipative trajectory, evaluated in a finite window, and far less is known about it. To our knowledge, no systematic study has established whether the conditions that let a bulk invariant constrain a linear or steady-state observable also hold for such nonlinear, finite-time resources. Together with the linear-regime successes above (blockade~\cite{Li2024photonblockade}, domain-wall transport~\cite{Perez2024domainwalls}) and the graded entanglement failure~\cite{Tschernig2021}, this delimits the present claim to nonlinear trajectory functionals. On the coherent side, Floquet resilience has been demonstrated for timing-noise robustness in the kinematic regime~\cite{Bhargava2025} and for critical edge states~\cite{Zhou2025Floquet}.

Two companion analyses, of cavity-loss-limited metrology~\cite{Deuyekbe2026CavityLoss} and of non-Hermitian sensing~\cite{Deuyekbe2026MeanResponse}, reach the same conclusion for neighboring resources. A nominal gain, the loss-free Fisher information in one case and the amplification of a non-Hermitian lattice in the other, is not what the state delivers once the budget that pays for it is counted: the retained fraction and the mean-response share fall as the gain grows. Here we apply the same resource-accounting discipline at fixed preparation energy ($\mathcal{R}_{\rm enc}=\num{0.8866}$ quanta at $N_{\max}=\num{2}$; \cref{SI-sec:si_resource_convention}), reporting local negativity in raw Wigner units for the fixed preparation; the truncation residue is sector-asymmetric (\cref{SI-sec:artifact-sectorial}), so the contrast is not assumed to cancel it.

We establish the coherent two-step model with its integer frame windings, quasienergy gaps, and open-chain edge diagnostic. We then compare its micromotion-covariant edge observable and local Wigner negativity across the two sectors under matched dissipation (\cref{sec:results}). We define the \emph{resource--invariant gap} as the decoupling between the stroboscopic generator and the driven state: the bulk winding $(\nu_0,\nu_\pi)$ remains exact at every chain length, whereas the local negativity decays and changes sign under local loss at fixed invariant. Operationally, topological protection of the non-Gaussian resource is assessed against three criteria. First, the matched sector contrast must be positive with a two-sided \qty{95}{\percent} confidence interval excluding zero at every tested size. Second, for the state carrying that contrast, the negativity must either transfer into the covariant edge mode or its occupation must agree with conditional passive survival within $\num{1e-3}$. Third, the contrast must be null at $t=T$ for $d\le\num{0.5}$ when uniform loss removes edge-selective shielding---the signature of a shielding-borne, rather than protected, resource (\cref{SI-sec:statistics,SI-sec:exact-subspace}).

Within the parameter range studied, none of the linear topological guarantees transfers to the local non-Gaussian resource. Exact-$Q$ simulations show passive linear edge-mode survival below $\num{1e-3}$ when the state is initialized directly in the Floquet edge mode, yet the collective-mode Wigner negativity vanishes there, while sub-boundary cat preparation exceeds the passive residual and flips contrast sign beyond two unit cells. Even preparing the physical dangling end ($j=\num{1}$) yields a size-invariant positive contrast that nonetheless is consistent with zero under uniform loss at stroboscopic $t=T$ for $d\le\num{0.5}$ and is modulated by local dimer return rather than $\nu_0$, a modulation that is not one-to-one, so both boundary sites ($j=\num{0}$ and $j=\num{1}$) independently establish the resource--invariant gap. Through a three-site causal light cone, a boundary-Laplacian erosion law, exact excitation-subspace benchmarks through $N_{\max}=\num{6}$, a dangling-end control, and a grid-free displaced-Fock witness, these diagnostics delimit the conditions under which an open-system Floquet resource claim holds.

\section{Coherent chiral Floquet model}
\label{sec:model}

We consider a two-step drive of period $T=2\pi$ (normalized units). In step A (duration $T/2$) the intracell hopping $J_A$ is active; in step B the intercell hopping $J_B$ is active. The associated Bloch Hamiltonian is
\begin{equation}\label{eq:bulk}
  H(k)=\bigl(J_A+J_B\cos k\bigr)\sigma_x+J_B\sin k\,\sigma_y,
  \qquad\Gamma=\sigma_z,
\end{equation}
which anticommutes with the chiral operator, $\{\Gamma,H(k)\}=0$. The two symmetric time frames yield off-diagonal winding numbers $w_A$ and $w_B$, from which the integer Floquet invariants follow~\cite{Asboth2014}:
\begin{equation}\label{eq:invariants}
  \nu_0=\frac{w_A+w_B}{2},\qquad
  \nu_\pi=\frac{w_A-w_B}{2}.
\end{equation}

The same classification is recovered from the one-dimensional spectral localizer constructed on the symmetric-frame effective Hamiltonians (half-signature of the symmetry-reduced localizer); for details, step Hamiltonians, and the bulk-gap bound under on-site potential disorder see the Supplemental Material (\cref{SI-sec:model,SI-sec:localizer}). On the open $N$-cell lattice ($j=0,\dots,2N-1$), the step Hamiltonians take the explicit bipartite form $H_A = J_A\sum_{m=0}^{N-1}(a_{2m}^\dagger a_{2m+1}+\mathrm{H.c.})$ and $H_B = J_B\sum_{m=0}^{N-2}(a_{2m}^\dagger a_{2m+3}+\mathrm{H.c.})$. Because step~$A$ pairs $a_{2m}$ with $a_{2m+1}$ inside cell~$m$ while step~$B$ pairs $a_{2m}$ with $a_{2m+3}$ of cell~$m+1$, the physical one-dimensional backbone is ordered $1\text{--}0\text{--}3\text{--}2\text{--}\cdots\text{--}(2N-1)\text{--}(2N-2)$: sites $j=\num{1}$ and $j=2N-2$ are the dangling physical ends, whereas $j=\num{0}$ and $j=2N-1$ lie one bond inward (sub-boundary sites). These statements refer exclusively to the coherent single-particle model.

We searched for a mixed-state counterpart and did not find one: neither the eigenmode-resolved chiral marker nor the flattened-correlation-matrix signature separates the sectors once the state is driven and lossy, so no topological index is assigned to the driven state and the invariant quoted throughout is that of the generator. The working points are selected solely from the coherent invariant and gap criteria: $(J_A,J_B)=(\num{0.1},\num{1.2})$ for the chiral sector $\bigl(\nu_0,\nu_\pi\bigr)=(-1,0)$ and $(J_A,J_B)=(\num{1.2},\num{0.1})$ for the trivial control $\bigl(\nu_0,\nu_\pi\bigr)=(0,0)$. At both points the minimum zero-quasienergy gap is $\num{2.1991}$ and the minimum $\pi$-quasienergy gap is $\num{0.3142}$, and the maximum chiral-frame residual remains below $\num{1e-15}$ on the resolution ladder. Open-chain propagation produces a pair of near-zero edge modes for the chiral point and none for the control point, verifying bulk--boundary correspondence for the coherent single-particle generator (\cref{SI-fig:phase} of the Supplemental Material).

\section{Open-system protocol}
\label{sec:protocol}

\subsection{Lindblad dynamics and local observable}
\label{sec:lindblad_protocol}

The open-chain dynamics are generated by the Lindblad master equation
\begin{equation}\label{eq:lindblad}
  \dot\rho=-i[H(t)+H_{\rm K},\rho] +\sum_j\gamma_j\mathcal{D}[a_j]\rho,
  \qquad
  \mathcal{D}[a]\rho=a\rho a^\dagger-\tfrac12\{a^\dagger a,\rho\},
\end{equation}
where the local Kerr term is $H_{\rm K}=\chi a_0^{\dagger2}a_0^2$ with $\chi=\num{0.5}$. The Kerr strength is chosen comparable to the drive frequency ($\chi/\omega_d=\num{0.5}$ in normalized units). A Kerr-free control ($\chi=\num{0}$) shows that the nonlinearity amplifies rather than creates the two-cell contrast ($+\num{0.010467}$ against $+\num{0.022208}$): without it the three-cell contrast is null rather than negative, so the negative branch requires the Kerr term at the primary grid and solver (\cref{SI-sec:kerrfree}). The resource under test is prepared, not generated: the initial even cat of amplitude $\alpha=\num{1}$ already carries the local negativity $\delta_-(\num{0})=\num{0.1962}$, and what is tracked within one drive period is its decay and sector-dependent survival. The ideal cat at this amplitude is vacuum-dominated (Fock weights \qty{64.8}{\percent} at $n=\num{0}$, \qty{32.4}{\percent} at $n=\num{2}$, and \qty{2.79}{\percent} at $n\ge\num{4}$), while the primary state is the $N_{\max}=\num{2}$ truncated-displacement cat ($c_0=\num{0.7462}$, $c_2=\num{0.6658}$; \cref{SI-sec:fp-check}), so the $n\ge\num{4}$ tail is excluded by construction and the preparation ties directly to the cutoff sensitivity quantified in \cref{SI-sec:nmaxpanel}. All rates and couplings are expressed in normalized units of the drive frequency.

Three loss profiles are compared. The primary index-end edge-selective profile sets $\gamma_{\rm edge}=\num{1e-4}$ on the index boundaries ($j=\num{0}$ and $j=2N-1$) and $\gamma_{\rm bulk}=\num{0.2}$ elsewhere; the physical-end shielded profile places $\gamma_{\rm edge}$ on the physical dangling ends ($j=\num{1}$ and $j=2N-2$); and the uniform profile applies $\gamma_{\rm bulk}$ to every site. In drive units, the shielded site loses only $\gamma_{\rm edge}T=\num{6.3e-4}$ per period while a bulk site decays by $\gamma_{\rm bulk}T=\num{1.26}$ e-foldings, so within one drive period the two-cell cat evolves on an effectively isolated site (see \cref{sec:discussion}). The ratio scan in the Supplemental Material brackets where the index-end profile stops working (\cref{SI-sec:lossratio,SI-tab:lossratio}). Here and throughout, the stochastic perturbation $d$ is multiplicative loss-rate noise ($\gamma_j\to\gamma_j(1+\xi)$ with $\xi\sim\mathcal{U}(-d,d)$ and the resulting rate clamped at zero); the coherent-model localizer in the Supplemental Material instead uses an on-site perturbation (\cref{SI-sec:localizer}). A single noise factor is drawn per matched realization and applied to every site, preserving the loss-profile geometry exactly; noise is sampled independently across realizations at four amplitudes (\numlist{0;0.5;1.0;1.5}) in the primary scan, with $d=\num{0.35}$ added in the dimerization and truncation scans. The initial state is a normalized even cat $\ket{\alpha}+\ket{-\alpha}$ ($\alpha=\num{1}$) on the first site ($j=\num{0}$), or on the physical dangling end ($j=\num{1}$) in \cref{sec:sizelaw}. Unless labeled as a \qty{95}{\percent} bootstrap interval~\cite{Efron1979} or half-width, $\pm$ denotes one standard error.

\begin{figure}[htpb]
  \centering
  \includegraphics[width=0.98\linewidth]{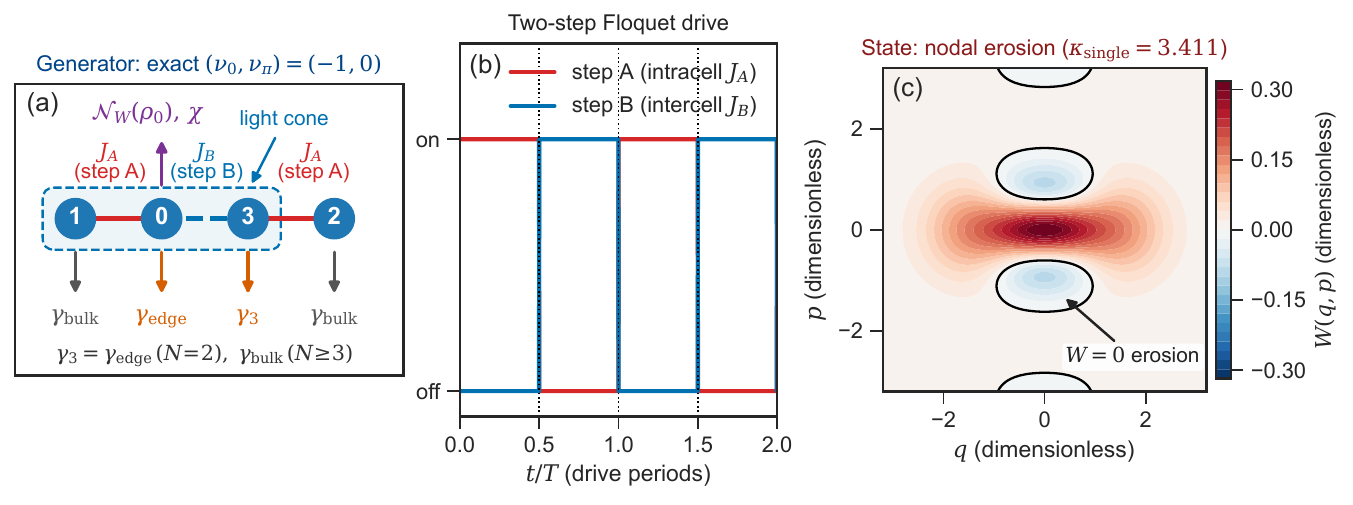}
  \caption{Conceptual overview of the resource--invariant gap and model protocol. (a)~Two-cell backbone $1\text{--}0\text{--}3\text{--}2$ with intracell ($J_A$, solid) and intercell ($J_B$, dashed) hoppings, bulk winding $(\nu_0,\nu_\pi)=(-1,0)$, and three-site causal light cone $\{0,1,3\}$ (shaded). Sites $j=\num{1}$ and $j=\num{2}$ are the dangling ends; $j=\num{0}$ hosts the even-cat preparation, $\chi$, and $\gamma_{\rm edge}$, while its strong-bond partner $j=\num{3}$ switches to $\gamma_{\rm bulk}$ from three cells onward. (b)~Two-step square drive of period $T=2\pi$. (c)~Even-cat Wigner distribution $W(q,p)$, positive and negative shading separated by the $W=0$ nodal contour under first-order diffusive erosion ($\kappa_{\rm single}=\num{3.411}$).}
  \label{fig:schematic}
\end{figure}

The primary observable is the integrated negative volume of the single-site Wigner function,
\begin{equation}\label{eq:negativity}
  \Ne(\rho)\equiv\delta_-(\rho)=\frac12\int\!\int\bigl(|W_\rho(q,p)|-W_\rho(q,p)\bigr)\,\dd q\,\dd p,
\end{equation}
evaluated on a $\num{31}\times\num{31}$ grid after one drive period, where we use the shorthand $\delta_-(t)\equiv\Ne(\rho_0(t))$ for the boundary-site marginal and $\Delta\Ne\equiv\delta_-^{\rm topo}-\delta_-^{\rm triv}$ for the matched sector contrast. Equation~\eqref{eq:lindblad} is integrated with the Monte Carlo wave-function method~\cite{Dalibard1992} and its QuTiP master-equation and trajectory implementations~\cite{Johansson2013}, while deterministic exact master-equation benchmarks use piecewise-constant half-period integration over the exact $U(1)$ excitation-conserving subspace $Q\le Q_{\max}$ (\cref{SI-sec:exact-subspace}). Quantum trajectories are grouped within each noise realization; sector contrasts are formed inside matched pairs before aggregation. Solver settings, parameter tables, and uncertainty estimation are given in the Supplemental Material (\cref{SI-sec:dynamics,SI-tab:parameters,SI-sec:statistics}). The model, drive protocol, loss channels and observable are summarized in \cref{fig:schematic}.

\subsection{Collective edge-mode diagnostic}
\label{sec:edgeobservable}

To test whether any observed contrast is carried by a collective edge mode, we define the Floquet edge vector $u_e(0)$ as the most edge-localized eigenvector of the coherent one-period propagator. Its micromotion-covariant continuation is $u_e(t)=U_{\rm coh}(t)u_e(0)$, and the corresponding mode operator and occupation are
\begin{equation}\label{eq:edge_mode}
  b_e(t)=\sum_j u_{e,j}(t)^*a_j,
  \qquad
  n_e(t)=\braket{b_e^\dagger(t)b_e(t)}.
\end{equation}
In the isolated-mode approximation the occupation would decay as
\begin{equation}\label{eq:passive_reference}
  n_e^{\rm ref}(t)=n_e(0)\exp[-\Gamma_e(t)],
  \qquad
  \Gamma_e(t)=\int_0^t\sum_j\gamma_j|u_{e,j}(s)|^2\,\dd s.
\end{equation}
Equation~\eqref{eq:passive_reference} is exact only for a decoupled passive mode; the shifts from edge--bulk transfer, Kerr nonlinearity and finite Fock truncation are bounded in \cref{SI-sec:covariantmethod}. We quantify deviations from this passive reference by the residual $R_e(t)\equiv n_e(t)-n_e^{\rm ref}(t)$ and compare sectors via the matched occupation contrast $\Delta n_e(t)\equiv n_e^{\rm topo}(t)-n_e^{\rm triv}(t)$. The core symbols and conventions used throughout this study are collected in \cref{SI-tab:simodelnotation} of the Supplemental Material.

\section{Results}
\label{sec:results}

This section establishes the resource--invariant gap through the size dependence of the matched contrast, the causal light cone and loss geometry that fix its sign, its dependence on the bulk invariant, and deterministic controls that isolate the mechanism (\cref{SI-sec:exact-subspace}).

\subsection{Size dependence of the local observable (at \texorpdfstring{$t=T$}{t=T})}
\label{sec:size}

\Cref{tab:contrast-summary} presents the primary zero-noise boundary-site Wigner negativity per sector and the resulting chiral-minus-trivial contrast from the exact piecewise-constant subspace master equation ($N=\num{2}\text{--}\num{5}$, cutoffs up to $N_{\max}=\num{6}$) and the \num{200}-trajectory Monte Carlo estimator. The \num{10}-trajectory noise-sweep ranges, solver bias ladder, and full Fock-cutoff panel are detailed in \cref{SI-tab:paired-comparison,SI-tab:trajladder,SI-tab:sivalidity,SI-tab:nmaxpanel}.

\begin{table}[htpb]
\centering
\caption{Zero-noise boundary-site Wigner negativity $\delta_-$ per sector and matched contrast (chiral $-$ trivial) at $t=T$ under index-end shielding ($\num{31}\times\num{31}$ grid; prepared state on $j=\num{0}$ carries $\delta_-(\num{0})=\num{0.1962}$ at $N_{\max}=\num{2}$). Exact piecewise-constant master-equation (ME) values in the conserved $Q\le Q_{\max}$ excitation subspace are strictly degenerate at $-\num{0.000113}$ across $N=\num{3},\num{4},\num{5}$ at $N_{\max}=\num{2}$ because every appended cell lies outside the three-site light cone (\cref{prop:bondselective}), and resolve a deeper negative contrast ($-\num{0.000955}$) at $N_{\max}=\num{4}$. Validated \num{200}-trajectory Monte Carlo (MC) rows at $N_{\max}=\num{2}$ bracket finite-trajectory variance; each MC row is labeled \num{200}-traj MC ($N_{\max}=\num{2}$), the \num{200} trajectories being the solver ensemble size per loss-realization draw; draw counts are $\num{100}$, $\num{50}$, $\num{200}$, and $\num{100}$ from two to five cells.}
\label{tab:contrast-summary}
\begin{tabular}{lcccc}
\toprule
Cells (modes) & Solver ($N_{\max}$) & Topological & Trivial & Contrast \\
\midrule
\num{2} (\num{4})  & Exact ME ($N_{\max}=\num{2}$)            & \num{0.02273} & \num{0.00053} & $+\num{0.02221}$ \\
\num{2} (\num{4})  & \num{200}-traj MC ($N_{\max}=\num{2}$)   & \num{0.02278} & \num{0.00101} & $+\num{0.02177}$ [\num{0.02136},\,\num{0.02218}] \\
\num{2} (\num{4})  & Exact ME ($N_{\max}=\num{4}$)            & \num{0.03835} & \num{0.00960} & $+\num{0.02875}$ \\
\num{2} (\num{4})  & Exact ME ($N_{\max}=\num{6}$)            & \num{0.03128} & \num{0.00523} & $+\num{0.02606}$ \\
\num{3}--\num{5} (\num{6}--\num{10}) & Exact ME ($N_{\max}=\num{2}$) & \num{0.00039} & \num{0.00050} & $-\num{0.000113}$ \\
\num{3} (\num{6})  & Exact ME ($N_{\max}=\num{4}$)            & \num{0.00859} & \num{0.00954} & $-\num{0.000955}$ \\
\num{3} (\num{6})  & \num{200}-traj MC ($N_{\max}=\num{2}$)    & \num{0.00068} & \num{0.00087} & $-\num{0.00019}$ [\num{-0.00058},\,\num{0.00020}] \\
\num{4} (\num{8})  & \num{200}-traj MC ($N_{\max}=\num{2}$)   & \num{0.00074} & \num{0.00106} & $-\num{0.00032}$ [\num{-0.00054},\,\num{-0.00010}] \\
\num{5} (\num{10}) & \num{200}-traj MC ($N_{\max}=\num{2}$)   & \num{0.00072} & \num{0.00100} & $-\num{0.00029}$ [\num{-0.00058},\,\num{0.00001}] \\
\bottomrule
\end{tabular}
\end{table}

The sign change between two and three unit cells is resolved in the exact master equation ($+\num{0.02221}$ to $-\num{0.000113}$ at $N_{\max}=\num{2}$, and $+\num{0.02875}$ to $-\num{0.000955}$ at $N_{\max}=\num{4}$) and persists at four unit cells in the \num{200}-trajectory estimator ($-\num{0.00032}\pm\num{0.00011}$, \qty{95}{\percent} interval excluding zero), with five cells corroborating the same negative plateau ($-\num{0.00029}\pm\num{0.00015}$; the interval crosses zero at this size, so the plateau is corroborating rather than confirming). In the \num{10}-trajectory noise ensemble (\cref{SI-tab:paired-comparison,SI-tab:sivalidity}), the finite-trajectory estimator shifts both zero- and small-negativity sectors upward, producing larger-magnitude operational contrasts ($+\num{0.01673}$ at two cells and $-\num{0.00210}$, $-\num{0.00201}$, $-\num{0.00218}$ at three through five cells at $d=\num{0}$) whose sign agrees with the exact master equation. An out-of-sample five-cell check ($-\num{0.00218}$) lands inside its predicted interval (\cref{SI-sec:size5,SI-tab:size5}), and at four cells ($N_{\rm MC}=\num{2000}$) the operational magnitude exceeds the minimum detectable effect by $\approx\num{2.6}\times$ ($\num{0.00201}$ versus $\Delta_{\min}=\num{0.00077}$; \cref{SI-sec:power}).

Physically, the sign reversal is carried by the rapid collapse of boundary-site negativity in the topological sector rather than by any change in the trivial control (\cref{tab:contrast-summary}). Within one drive period under the exact master equation ($N_{\max}=\num{2}$), the topological sector retains $\delta_-=\num{0.02273}$ at two cells (about \qty{12}{\percent} of the prepared $\delta_-(\num{0})=\num{0.1962}$, a cutoff-truncated value; the untruncated cat carries $\delta_-(\num{0})\approx\num{0.105}$, against which the retention is about \qty{22}{\percent}; \cref{SI-sec:obsvalidation}) and drops by a factor of $\approx\num{59}$ to $\num{0.00039}$ from three cells onward, whereas the trivial control remains nearly flat at \num{0.00050}--\num{0.00053}.

\begin{figure}[htpb]
  \centering
  \includegraphics[width=.98\linewidth]{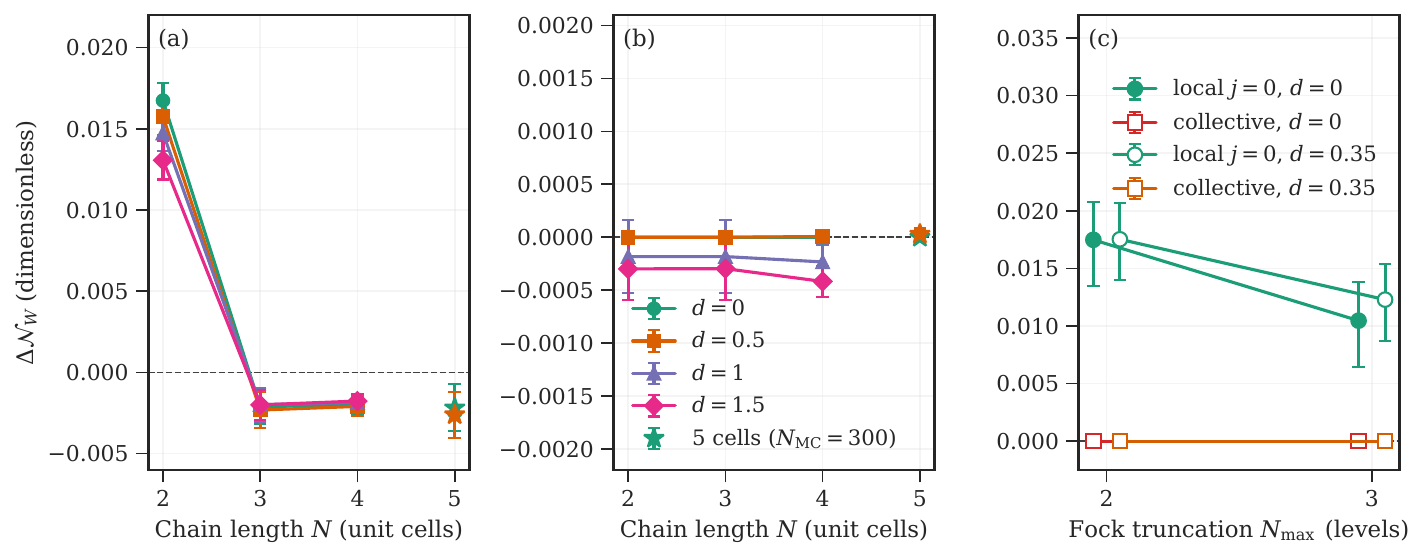}
  \caption{Size and loss-profile dependence of the local Wigner-negativity contrast in the \num{10}-trajectory noise ensemble.
  (a)~Edge-selective loss at $d\in\{\num{0},\num{0.5},\num{1.0},\num{1.5}\}$ against the zero reference: positive at two cells, negative thereafter; stars at five cells mark the out-of-sample check ($N_{\rm MC}=\num{300}$, $d\le\num{0.5}$).
  (b)~Uniform loss on the same scale: consistent with zero at all sizes for $d\le\num{0.5}$, small negative offsets at $d\ge\num{1.0}$. Error bars are \qty{95}{\percent} bootstrap intervals (normal-approximation $\pm\num{1.96}$ SE at five cells).
  (c)~Local site-$j=\num{0}$ versus collective edge-mode contrast at two cells, filled for $d=\num{0}$ and open for $d=\num{0.35}$, $\pm\num{1}$ SE; the collective-mode negativity vanishes at both Fock truncations.}
  \label{fig:contrast-summary}
\end{figure}

\begin{figure}[htpb]
  \centering
  \includegraphics[width=0.8\linewidth]{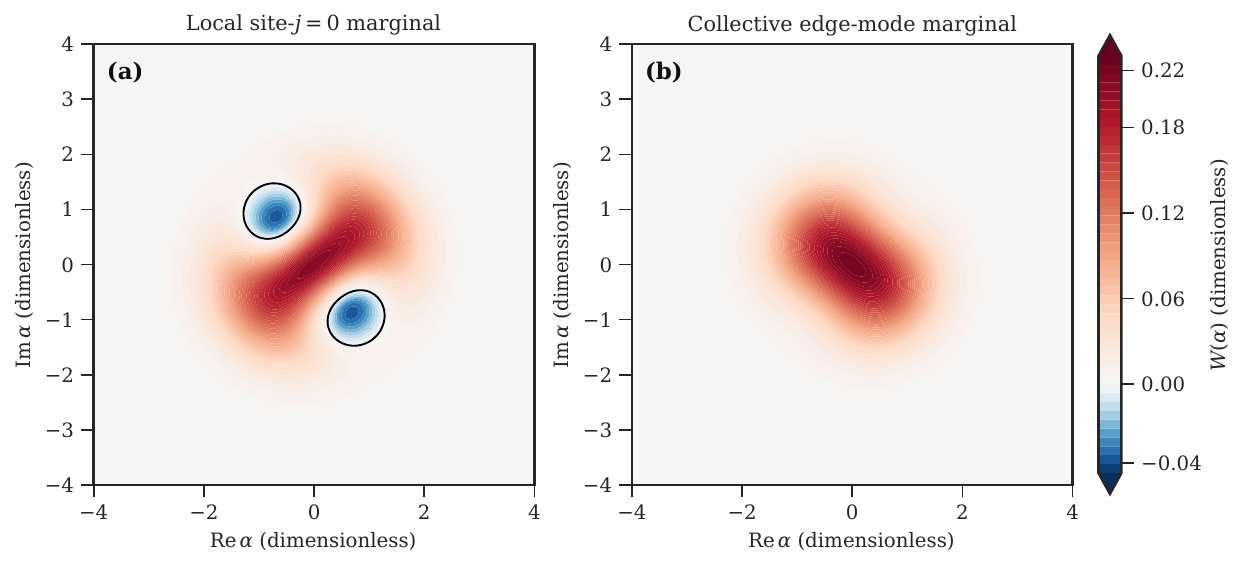}
  \caption{Local versus collective-mode Wigner function at two cells ($N_{\max}=\num{2}$, $d=\num{0}$, $t=T$, topological sector) under edge-selective loss in phase-space coordinates $\alpha=(q+ip)/\sqrt{2}$.
  (a)~Single-site Wigner function on boundary site~$j=\num{0}$, showing clear negative lobes enclosed by the black $W=0$ nodal contour.
  (b)~Wigner function of the same final state after rotation into the collective edge-mode basis; the distribution is non-negative everywhere.}
  \label{fig:collective-wigner}
\end{figure}

The same final states, rotated into the collective edge-mode basis, show no collective Wigner negativity at either Fock truncation ($N_{\max}=\num{2},\num{3}$) and either system size, nor under the mode-resolved preparation. Run counts and floors are given in \cref{SI-sec:collective-wigner}. The local site-$j=\num{0}$ contrast instead stays positive at two cells under index-end shielding (\cref{fig:contrast-summary}(c)). What is basis-dependent is the marginal Wigner negativity, not the non-Gaussianity itself, which passive rotations preserve; in the tested preparations the negativity appears in the local marginal and not in the collective-mode one. The mode-resolved preparation removes the basis for that expectation of collective negativity: the cat is prepared directly in the delocalized Floquet edge mode ($\alpha_j=\alpha u_{e,j}(0)$), so the initial collective marginal is non-Gaussian by construction. After one period the collective Wigner negativity stays below the $\num{1.2e-5}$ truncation floor across all realizations and the site-$j=\num{0}$ marginal is null as well (\cref{SI-sec:collective-wigner,fig:collective-wigner}). Under uniform loss, meanwhile, contrasts are consistent with zero at $d\le\num{0.5}$ at every size, and across all four loss-rate noise amplitudes the \num{10}-trajectory sign pattern remains stable (\cref{SI-fig:si-robustness}). At $d\ge\num{1.0}$ under uniform loss, the eight-mode contrasts resolve small negative offsets below $\num{4.2e-4}$, at least four times smaller than the index-end signal at the same noise level (\cref{fig:contrast-summary}(b)).

\subsection{Size-scaling fit and the crossover}

In the \num{10}-trajectory ensemble, fitting the four chain lengths ($N=\num{2}\text{--}\num{5}$) to a phenomenological signed power-law step versus a smooth $p=\num{1}$ form brackets the crossover $2<N^*<3$ along a degenerate ridge (\cref{SI-sec:si_sizelaw,SI-fig:si_sizelaw}). Here we isolate the exact physical mechanism behind that crossover and establish the corresponding result when the cat is prepared directly on the physical dangling end.

Two mechanisms could place a step at that size. In the first, the resource dies when Kerr processing of the prepared cat outruns the escape of the edge mode into the lossy boundary region. Both the Kerr scale and the escape time $\xi/v_{\max}=\num{4.0}\,\omega_d^{-1}$, with $\xi$ the edge-mode localization length and $v_{\max}$ the maximum group velocity of the drive, are set by the preparation and the drive, up to the weak size drift of the localization length ($|u_{e,0}|^{2}$ changes from \num{0.01343} to \num{0.01220} between two and four cells)---a drift far too small to flip a contrast spanning $+\num{0.0222}$ to $-\num{0.0001}$; a Kerr--escape competition would therefore return the same verdict at every size and cannot give a positive contrast at two cells together with a negative one at three. The exclusion rests not on where such a competition would place a crossover in $N$ but on the fixed-size measurement below, where moving the bulk-rate sink reverses the sign while neither the preparation nor the drive changes.

In the second, what matters is where the loss sits relative to the edge-mode weight. A single lossy site is not by itself a sink: in the two-cell chain ordered along the backbone $1\text{--}0\text{--}3\text{--}2$, the four sites $j=0,1,2,3$ carry single-particle edge weights $\num{0.0134}$, $\num{0.4866}$, $\num{0.4866}$, and $\num{0.0134}$, respectively, so the two dangling ends $j=\num{1},\num{2}$ carry \qty{97.3}{\percent} of the mode while the prepared sub-boundary site $j=\num{0}$ and its strong-bond partner $j=\num{3}$ are minority components. Whether the resource survives depends on which of these the bath reaches, and not on the number of lossy sites alone. The failure of the collective edge-mode transform points the same way: it concerns a cavity mode that does not exist once the sites carrying the mode are lossy.

The two readings are separated by two measurements. At two cells, placing a $\num{0.2}\,\omega_d$ sink moves the exact contrast across zero on both Wigner grids and at both truncations. Loss on $j=\num{1}$ (weak-bond dangling end carrying \qty{48.7}{\percent} of the edge weight) gives $+\num{0.0223}$ and $+\num{0.0228}$ on the two grids; loss on $j=\num{3}$ (the index-end strong-bond partner) gives $-\num{0.0231}$ and $-\num{0.0236}$; loss on the prepared site $j=\num{0}$ annihilates both sectors exactly; loss on the remaining dangling end $j=\num{2}$ leaves the zero-loss value ($-\num{7.9e-4}$) unchanged; and splitting the $\num{0.2}\,\omega_d$ budget evenly across both cells ($j=\num{1},\num{2}$) returns $+\num{0.0149}$. The same reversal persists at the next cutoff ($N_{\max}=\num{3}$): the $j=\num{1}$ sink gives $+\num{0.0156}$ and $+\num{0.0154}$ on the two grids and the $j=\num{3}$ sink $-\num{0.0164}$ and $-\num{0.0162}$. Against the zero-loss reference ($-\num{7.9e-4}$), the \num{0.046} span of this single-sink sweep is comparable to the \num{0.0223} exact-ME shift between two and three cells (\num{0.0188} in the \num{10}-trajectory ensemble).

The second measurement separates chain length from loss geometry directly. Holding the per-site rates and the total budget fixed, appending a third cell that carries no loss of its own leaves the contrast unchanged below $\num{1e-9}$, because the new cell is reached only through a site that carries no population at $t=T$. Terminating the appended cell with the edge rate, which is what the index-end boundary-preserving profile does, moves the contrast by an amount more than two orders of magnitude smaller than the size effect it is invoked to explain. Promoting site~$j=\num{3}$ from the low-loss terminal rate to the bulk rate at fixed length, without adding any cell, moves the contrast by $-\num{0.0223}$, which is the entire two-to-three-cell difference, and the four-mode profile that carries the promotion reproduces the three-cell value exactly. The step between two and three cells is therefore a threshold in the loss geometry, resolved below into a bond-selective rule: the sign is set by whether the bulk-rate loss sits on the strong-bond partner of the prepared site or on its weak-bond neighbor, and the apparent size dependence is a relabeling of the same profile that accompanies chain extension.

\subsection{Causal light cone, physical-end preparation, and the mechanism of the crossover}
\label{sec:sizelaw}

\begin{proposition}[Causal stroboscopic light cone and the bond-selective sign rule]\label{prop:bondselective}
For the two-step propagator $U_F=e^{-iH_B T/2}e^{-iH_A T/2}$ evaluated at $t=T$ on any chain with $N\ge\num{2}$ cells, the disjoint dimer structure of $H_A$ followed by $H_B$ restricts single-period propagation from the prepared site~$j=\num{0}$ to the closed three-site light cone $\{0,1,3\}$,
\begin{equation}\label{eq:uf_column}
U_F\ket{0}=\cos(J_A\pi)\cos(J_B\pi)\ket{0}-i\sin(J_A\pi)\ket{1}-i\cos(J_A\pi)\sin(J_B\pi)\ket{3}.
\end{equation}
Consequently: (i)~the dangling end~$j=\num{2}$ and every appended cell $N\ge\num{3}$ lie outside the one-period light cone ($P_2(t)\equiv 0$ identically throughout the first period in the linear propagator and $P_2(T)<\num{5e-18}$ in the full Kerr master equation). (ii)~The single-particle return probability $P_0^{(1)}(T)=\cos^2(J_A\pi)\cos^2(J_B\pi)=\num{0.5920}$ is strictly invariant under the sector exchange $(J_A,J_B)\leftrightarrow(J_B,J_A)$. (iii)~The two sectors therefore differ exclusively in their escape branching, routing $\num{0.3125}$ into $j=\num{3}$ versus $\num{0.0955}$ into $j=\num{1}$ in the topological sector and $\num{0.0625}$ into $j=\num{3}$ versus $\num{0.3455}$ into $j=\num{1}$ in the trivial control. Under an index-end boundary-preserving loss profile, the branching thus selects which sector surrenders its two-photon coherence to the bath: $\gamma_{\rm bulk}$ on $j=\num{1}$ damps the trivial escape channel and yields a positive contrast, whereas $\gamma_{\rm bulk}$ on $j=\num{3}$ damps the topological channel and reverses it. The resulting sign additionally requires the Kerr-dressed dwell of \cref{prop:erosion}: at $\chi=\num{0.5}$ the same profile reverses the contrast, while at $\chi=\num{0}$ it leaves a null three-cell contrast rather than a negative one (\cref{SI-sec:kerrfree}). Appending a cell that carries no loss of its own leaves the one-period contrast invariant below $\num{1e-9}$ at every Fock cutoff $N_{\max}$, while terminating that cell with $\gamma_{\rm edge}$ shifts it by more than two orders of magnitude less than the observed step.
\end{proposition}

The four test profiles that establish \cref{prop:bondselective} are collected in \cref{tab:decomposition}. Rows three and four are indistinguishable, and so are rows one and two: the length changes nothing once the per-site rates are held, while promoting the terminal site changes everything once the length is held. The two contributions close the observed two-to-three-cell difference to a residual of $\num{1e-9}$. As a reservoir-engineering rule, protecting a two-component cat on site~$j=\num{0}$ in a Floquet chain requires isolating its strong-bond partner; extending the lattice is not a control knob for the resource.

\begin{table}[htpb]
  \centering
  \caption{Length and profile decomposition of the two-to-three-cell change at zero noise, $\chi=\num{0.5}$, $N_{\max}=\num{2}$, on both Wigner grids. Columns two to four give the per-site loss rates in units of $\omega_d$, the cell count, and the total Markovian loss budget; the last two columns give the matched contrast on the $\num{31}\times\num{31}$ and $\num{61}\times\num{61}$ grids. Increasing the length at fixed per-site rates leaves the contrast unchanged below $\num{1e-9}$, while promoting the terminal site at fixed length moves it by $-\num{0.02232}$; matching the promoted three-cell budget to the two-cell value within $\num{2.5e-4}$ relative preserves the sign.}
  \label{tab:decomposition}
  \small
  \begin{tabularx}{\linewidth}{X c c c c c}
  \toprule
  & \multicolumn{3}{c}{Configuration} & \multicolumn{2}{c}{Contrast} \\
  \cmidrule(lr){2-4}\cmidrule(lr){5-6}
  Profile & Rates $\gamma_j/\omega_d$ & Cells & Budget & $\num{31}^{2}$ & $\num{61}^{2}$ \\
  \midrule
  Terminal protected & $\num{1e-4},\num{0.2},\num{0.2},\num{1e-4}$ & \num{2} & \num{0.4002} & $+\num{0.02221}$ & $+\num{0.02271}$ \\
  Same rates, one cell more & $\num{1e-4},\num{0.2},\num{0.2},\num{1e-4},\num{0},\num{0}$ & \num{3} & \num{0.4002} & $+\num{0.02221}$ & $+\num{0.02271}$ \\
  Terminal promoted & $\num{1e-4},\num{0.2},\num{0.2},\num{0.2}$ & \num{2} & \num{0.6001} & $-\num{0.000113}$ & $-\num{0.000182}$ \\
  Boundary preserving & $\num{1e-4},\num{0.2},\num{0.2},\num{0.2},\num{0.2},\num{1e-4}$ & \num{3} & \num{0.8002} & $-\num{0.000113}$ & $-\num{0.000182}$ \\
  Boundary preserving, & $\num{5.0e-5},\num{0.1},\num{0.1},$ & \num{3} & \num{0.4001} & $-\num{0.000602}$ & $-\num{0.000576}$ \\
  budget matched & \multicolumn{1}{l}{$\num{0.1},\num{0.1},\num{5.0e-5}$} & & & & \\
  \bottomrule
  \end{tabularx}
\end{table}

\begin{figure}[htpb]
  \centering
  \includegraphics[width=0.98\linewidth]{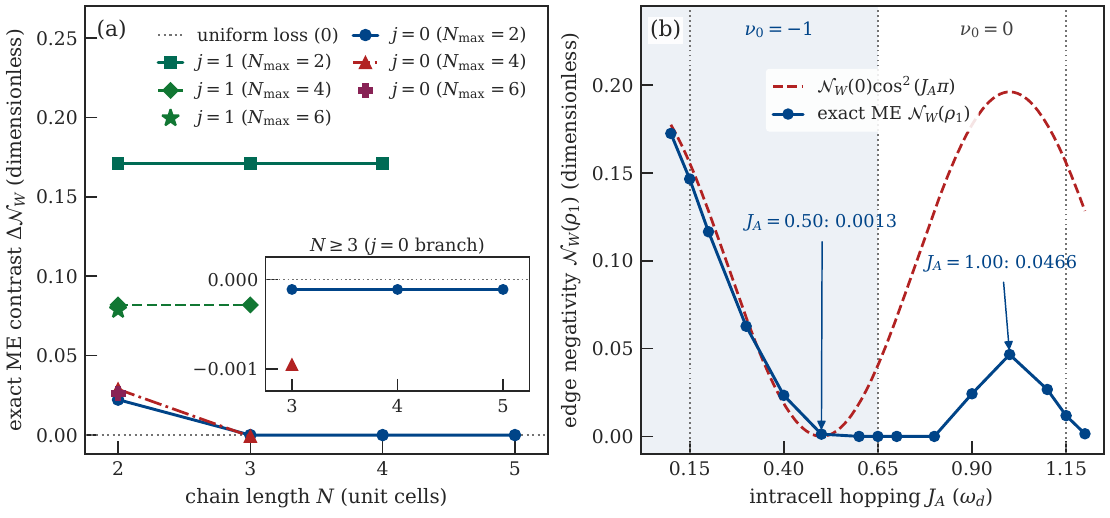}
  \caption{(a) Exact subspace master-equation contrast at the index-end ($j=\num{0}$, open circles) and physical-end ($j=\num{1}$, solid squares) profiles for $N_{\max}=\num{2},\num{3},\num{4}$ with shaded $\num{31}^2$ grid resolution. (b) Dangling-end one-period negativity versus $J_A$ along $J_A+J_B=\num{1.3}$ at $N=\num{3}$, $N_{\max}=\num{2}$, compared with the local return $\cos^2(J_A\pi)$; shading marks the winding regions and the undefined gap closings.}
  \label{fig:exact-mechanism}
\end{figure}

The complementary physical-end configuration shows the light-cone mechanism directly. Both the even-cat preparation and the Kerr nonlinearity ($\chi=\num{0.5}$) reside on the physical dangling end $j=\num{1}$, which carries \qty{48.7}{\percent} of the topological edge weight, and low loss shields the two physical dangling ends ($\gamma_1=\gamma_{2N-2}=\num{1e-4}$, $\gamma_{\rm bulk}=\num{0.2}$ elsewhere). In this configuration, step~$B$ acts on site~$j=\num{1}$ as an isolated monomer. Its single-period coherent evolution is $U_F\ket{1}=\cos(J_A\pi)\ket{1}-i\sin(J_A\pi)\cos(J_B\pi)\ket{0}-\sin(J_A\pi)\sin(J_B\pi)\ket{3}$, independently of $N\ge\num{2}$. Exact subspace master-equation integration accordingly yields a large, size-invariant positive contrast across every tested chain length: $+\num{0.17105}$ at $N_{\max}=\num{2}$ ($N=\num{2},\num{3},\num{4}$), $+\num{0.08187}$ at $N_{\max}=\num{4}$ ($N=\num{2},\num{3}$), and $+\num{0.07878}$ at $N_{\max}=\num{6}$ ($N=\num{2}$; \cref{fig:exact-mechanism}(a)). This survival persists over multiple drive periods ($+\num{0.17105}$ at $1T$ to $+\num{0.12181}$ at $8T$ for $N=\num{3}$, $N_{\max}=\num{2}$) and yields outright displaced-Fock certification at $N_{\max}=\num{6}$ (\cref{SI-tab:witness-exact}).

Even this positive physical-end regime, however, still exhibits the resource--invariant gap. Uniform loss annihilates the $j=\num{1}$ contrast identically ($\num{0.000000}$) at $t=T$ at every tested size at $d=\num{0}$ in the exact master equation, and leaves it consistent with zero for $0<d\le\num{0.5}$, and preparing on the primary sub-boundary site~$j=\num{0}$ while shielding only the physical ends likewise returns $\num{0.000000}$ (\cref{SI-sec:exact-subspace}). Scanning the dimerization along $J_A+J_B=\num{1.3}$ at $N=\num{3}$ further reveals that dangling-end survival is modulated by the local single-period return $\cos^2(J_A\pi)$ rather than set by the bulk invariant (\cref{fig:exact-mechanism}(b)). The negativity collapses to $\num{0.00128}$ inside the $(\nu_0,\nu_\pi)=(-1,-1)$ phase at $J_A=\num{0.50}$ (where $\cos(J_A\pi)=0$), vanishes identically over $J_A=\num{0.60}$--$\num{0.80}$, a window spanning the $\nu_0=-1\to0$ transition and the adjacent segment of the anomalous $\pi$-phase $[(\nu_0,\nu_\pi)=(0,-1)]$, and rises again outside that window, reaching $\num{0.04663}$ at $J_A=\num{1.00}$ where $|\cos(J_A\pi)|=1$ but $\num{0.02428}$ already at $J_A=\num{0.90}$, so the return factor modulates rather than sets the contrast (\cref{SI-tab:j1-dimerization}).

\subsection{Dependence on the bulk invariant}
\label{sec:invariantsweep}

The discriminating test asks whether the local contrast follows the bulk invariant, which changes at a gap closing, or the edge-mode localization, which varies continuously. A dimerization scan along $J_A+J_B=\num{1.3}$ (\num{19} points with $J_A$ from \num{0.1} to \num{1.2}, two and three cells, $N_{\rm MC}=\num{500}$ realizations per point at $d=\num{0}$ and \num{0.35}, noise realizations and trajectory seeds shared across points) crosses all three gap closings of the model. The sweep observable is the edge-minus-center negativity at each dimerization point; because the center-site negativity vanishes identically on this grid it equals the edge negativity itself. It is not the matched sector contrast of \cref{tab:contrast-summary}, and only the difference between the two dimerization endpoints maps onto that sector contrast. Continuity across closings is therefore a property of the sweep observable; its implication for the sector contrast rests on the endpoint differences.

This observable is non-monotone: it peaks inside the strongly dimerized topological regime ($+\num{0.05928}$ at two cells, $J_A=\num{0.20}$), is zero to within \num{2e-5} over $\num{0.6}\le J_A\le\num{0.8}$, a window containing the $\nu_0=-1\to0$ transition and points gapped on both sides, and becomes finite again in the strongly dimerized trivial regime. On matched realizations the contrast is continuous with its bracketing points at two of the three closings; at the low-$J_A$ $\pi$ closing the residual is $+\num{0.00086}\pm\num{0.00017}$ (\qty{1.95}{\percent} of the local signal), a resolved sub-percent excursion rather than a weak one, and the difference between the two dimerization endpoints reproduces the primary sign pattern ($+\num{0.01673}$ and $-\num{0.00189}$ against $+\num{0.01673}$ and $-\num{0.00210}$). Along this ordered one-dimensional sweep, the two-cell contrast correlates negatively with the measured edge-mode localization length (Pearson $r=\num{-0.680}$, reported descriptively without a parametric $p$-value because adjacent sweep points are autocorrelated; \cref{SI-sec:dimerization,SI-tab:dimerization}), with the same unresolved trend at three cells. Of the two alternatives posed in the Introduction, the contrast follows neither the bulk invariant nor the edge-mode localization---the sink swap at fixed length reverses the sign without changing the localization---and is governed by local light-cone exposure instead.

\subsection{Micromotion-covariant edge mode}
\label{sec:covariant}

\begin{figure}[htpb]
  \centering
  \includegraphics[width=0.90\linewidth]{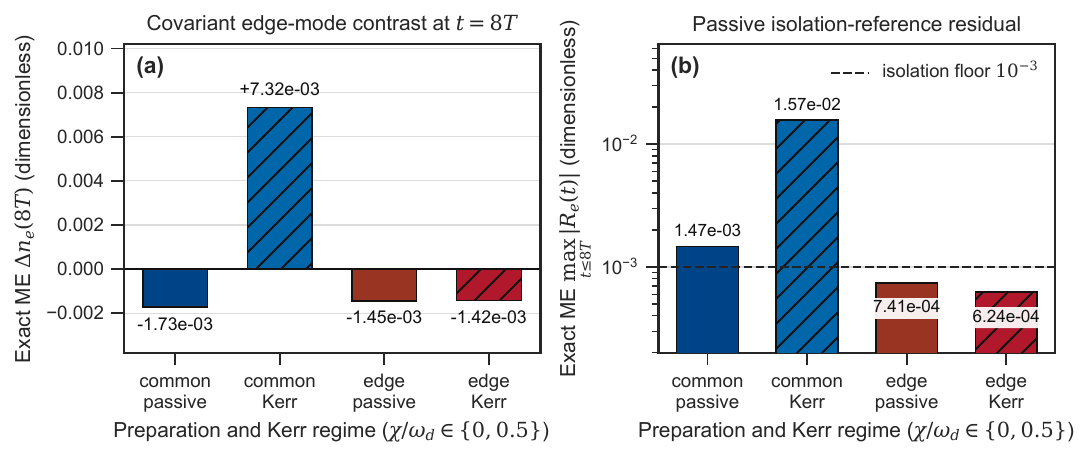}
  \caption{Micromotion-covariant edge-mode benchmark at $N=\num{2}$ under edge-selective loss and zero loss-rate noise. Evaluated in the exact total-excitation subspace $Q\le\num{4}$ ($\dim=\num{70}$). Bars distinguish common-site ($j=\num{0}$, blue) from Floquet-mode (red) initialization and passive ($\chi=\num{0}$, solid) from interacting ($\chi=\num{0.5}$, hatched) regimes. (a)~Matched edge-population contrast $\Delta n_e(8T)$. (b)~Maximum passive-isolation residual $\max_{t\le 8T}|R_e(t)|$ relative to \cref{eq:passive_reference} alongside the $\num{1e-3}$ isolation floor (dashed line).}
  \label{fig:covariant}
\end{figure}

Evaluating the Floquet edge-mode occupation \cref{eq:edge_mode} via an exact piecewise-constant master equation in the excitation-conserving subspace $Q\le\num{4}$ ($\dim=\num{70}$) cleanly separates linear mode isolation from non-Gaussian resource protection (\cref{fig:covariant}; see \cref{SI-sec:covariantmethod}). Under uniform passive loss ($\chi=\num{0}$), the exact simulation reproduces \cref{eq:passive_reference} to machine precision (\cref{SI-sec:covariantmethod}). Under index-end shielding, initializing directly in the Floquet edge mode preserves linear passive survival within the $\num{1e-3}$ isolation budget, with $\max_{t\le 8T}|R_e(t)|=\num{7.41e-4}$ ($\Delta n_e(8T)=\num{-1.45e-3}$) at $\chi=\num{0}$ and $\num{6.24e-4}$ ($\Delta n_e(8T)=\num{-1.42e-3}$) at $\chi=\num{0.5}$, even though the collective Wigner negativity remains null. By contrast, preparing the even cat locally on site~$j=\num{0}$ violates the passive-isolation bound ($\max_{t\le 8T}|R_e(t)|=\num{1.47e-3}$ at $\chi=\num{0}$ with $\Delta n_e(8T)=\num{-1.73e-3}$, rising to $\num{1.57e-2}$ with $\Delta n_e(8T)=+\num{7.32e-3}$ at $\chi=\num{0.5}$).

\subsection{Deterministic controls and finite-size mechanism}
\label{sec:controls}

Two exactly solvable benchmarks isolate kinematic and dissipative effects. In the absence of loss and Kerr nonlinearity the covariant edge population remains unity to machine precision ($1-p_e(t)<\num{1.33e-15}$), while the laboratory-frame population falls to $\approx\num{0.895}$. Under uniform loss the integrated-exposure reference is recovered to \num{1.94e-15}; under the edge-selective profile a residual of order \num{1e-3} appears, consistent with loss-generated edge--bulk mixing during micromotion (\cref{SI-sec:unitarybenchmark}).

\begin{figure}[thpb]
  \centering
  \includegraphics[width=.98\linewidth]{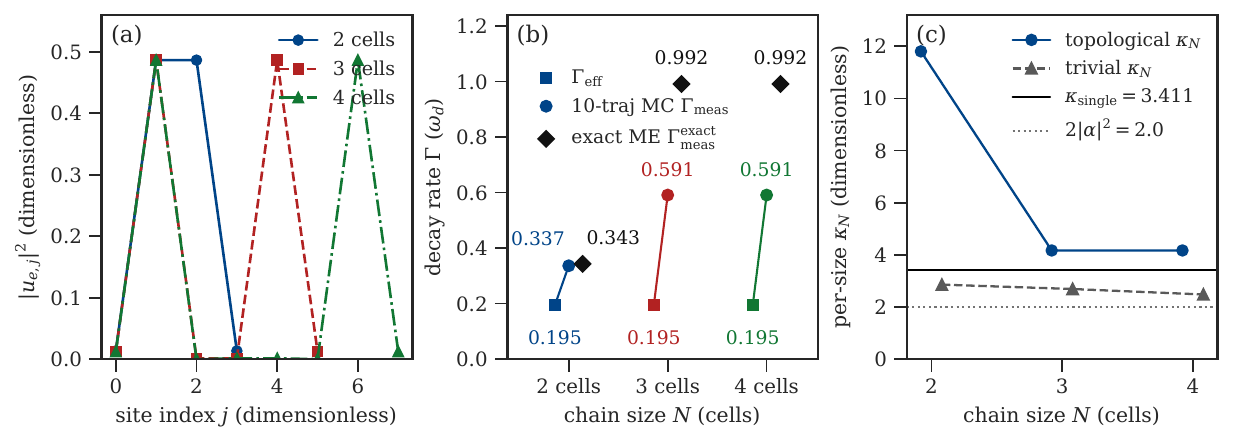}
  \caption{Finite-size mechanism diagnostics.
  (a)~Edge-mode weight $|u_{e,j}|^2$ for chains of two, three, and four unit cells, peaking equally at the two dangling ends ($j=\num{1}$ and $j=2N-2$) with minority sub-boundary weight on $j=\num{0}$.
  (b)~Ten-trajectory Monte Carlo decay rate $\Gamma_{\rm meas}$ and exact master-equation rate $\Gamma_{\rm meas}^{\rm exact}$ ($\num{0.343}$ at $N=\num{2}$, $\num{0.992}$ at $N=\num{3},\num{4}$) against the isolated-mode rate $\Gamma_{\rm eff}$.
  (c)~Size-dependent prefactor $\kappa_N$ from the two-photon dwell time for both sectors, against $\kappa_{\rm single}=\num{3.411}$ (solid) and the large-cat asymptote $2|\alpha|^2=\num{2.0}$ (dotted).}
  \label{fig:mechanism}
\end{figure}

The mechanism summarized in \cref{fig:mechanism} has two candidate sources: the edge mode delocalizes as cells are added, and the loss profile changes under the same extension. The measurements below assign the sign to the second. Mechanistically, single-site photon loss acts as a Gaussian amplitude-damping channel whose exact Wigner propagator in coherent-state coordinates $\boldsymbol\beta=(\Re\beta,\Im\beta)=(q+ip)/\sqrt{2}$ is
\begin{equation}\label{eq:fp_main}
  W_t(\boldsymbol\beta) = \frac{2}{\pi(1-e^{-\gamma t})} \int\!\dd^2\beta'\, \exp\!\left[-\frac{2|\boldsymbol\beta - e^{-\gamma t/2}\boldsymbol\beta'|^2}{1-e^{-\gamma t}}\right] W_0(\boldsymbol\beta').
\end{equation}

\begin{proposition}[Boundary-Laplacian erosion law]
\label{prop:erosion}
Let $W_0$ be a normalized Wigner function whose negative set $\Omega=\{\boldsymbol\beta:W_0(\boldsymbol\beta)<0\}$ has a smooth boundary $\partial\Omega$ with outward normal $\mathbf n$ and arc element $\dd\ell$, and let $\delta_-(0)$ be its integrated negative volume. Under single-site Gaussian amplitude damping of rate $\gamma$, the deterministic contraction $\boldsymbol\beta\mapsto e^{-\gamma t/2}\boldsymbol\beta$ maps $\Omega$ onto itself up to scale and leaves $\delta_-$ unchanged at all $t$, whereas expanding \eqref{eq:fp_main} to first order in $\gamma t$ and applying the divergence theorem on $\partial\Omega$ gives the exact short-time erosion prefactor
\begin{equation}\label{eq:kappa_single_main}
  \kappa_{\rm single} = -\frac{1}{\gamma}\frac{\dd\ln\delta_-}{\dd t}\bigg|_{t=0} = \frac{1}{8}\,\frac{\int_\Omega \Delta W_0\,\dd^2\beta}{\delta_-(0)} = \frac{1}{8\,\delta_-(0)}\oint_{\partial\Omega}\nabla W_0\cdot\mathbf n\,\dd\ell .
\end{equation}
The erosion rate of the resource is therefore a boundary integral over the nodal curve where the Wigner function changes sign, and depends on the state only through the normal gradient there (\cref{SI-sec:localizer-transfer,SI-sec:fp-check}).
\end{proposition}

For the $N_{\max}=\num{2}$, $\alpha=\num{1}$ truncated even cat ($c_0=\num{0.7462}$, $c_2=\num{0.6658}$) the contour integral evaluates to $\kappa_{\rm single}=\num{3.411}$. Across the chain, the residual edge--bulk decay rate $\gamma_{\rm eb}=\kappa\,\Lambda_2^{\rm coh}$ is governed by the rate-weighted two-photon coherent dwell $\Lambda_2^{\rm coh}$ (\cref{SI-tab:closedform-kappa}). Comparing this nonlinear dwell against its Kerr-free counterpart ($\Lambda_2^{\rm coh,lin}=\num{0.0256}$ at $N=\num{2}$; \cref{SI-tab:closedform-kappa}) uncovers a Kerr two-photon self-trapping mechanism: the boundary interaction $\chi=\num{0.5}$ detunes the two-photon manifold $\ket{2}_0$ by $2\chi=\num{1.0}\,\omega_d$ relative to single-particle tunneling, suppressing $\Lambda_2^{\rm coh}$ by a factor of $\num{2.13}$ at $N=\num{2}$ ($\num{0.0256}\to\num{0.0120}$). This suppression mirrors the roughly twofold enhancement of the exact two-cell contrast from the $\chi=\num{0}$ control ($+\num{0.01047}$) to the interacting protocol ($+\num{0.02221}$; \cref{SI-sec:kerrfree}). Because the inter-cell Floquet step transfers population to site~$j=\num{3}$ with period-averaged dwell $\langle P_3\rangle_T=\num{0.198}$, the index-end rule shields site~$j=\num{3}$ at two unit cells ($\gamma_3=\num{1e-4}$) and exposes it from three cells onward ($\gamma_3=\num{0.2}$), raising $\Lambda_2^{\rm coh}$ by a factor of $\num{7.9}$ from $\num{0.0120}$ to $\num{0.0950}$ and overwhelming this local anharmonic self-trapping. The fixed-length isolation in \cref{prop:bondselective,tab:decomposition} shows that this $\num{7.9}$-fold jump originates from promoting site~$j=\num{3}$ rather than from chain extension. In \cref{fig:mechanism}, $\kappa_N$ is compared with the \num{10}-trajectory diagnostic (\num{11.8} at two cells versus \num{4.2} from three onward) and with the exact master-equation decay rate $\Gamma_{\rm meas}^{\rm exact}$, which steepens from $\num{0.343}$ at $N=\num{2}$ to $\num{0.992}$ at $N=\num{3},\num{4}$ (\cref{SI-sec:localizer-transfer,SI-tab:bound-verification,SI-tab:closedform-kappa}).

\section{Experimental implementation and measurement protocol}
\label{sec:exp_protocol}

\subsection{Design guidelines for bosonic edge-resource preservation}
\label{sec:design_guidelines}

Together, \cref{prop:bondselective,prop:erosion} translate the resource--invariant gap into four design rules for driven-dissipative bosonic circuits:
\begin{enumerate}
  \item \emph{Do not rely on passive band invariants for non-Gaussian protection:} Quantized Floquet windings $(\nu_0,\nu_\pi)$ guarantee single-particle edge modes but do not suppress first-order nodal Wigner erosion ($\kappa_{\rm single}=\num{3.411}$; \cref{prop:erosion}) without active reservoir stabilization~\cite{Touzard2018,Zapletal2022multimode,Rojkov2026catmanifolds}.
  \item \emph{Isolate the strong-bond partner rather than extending the lattice:} Extending the chain beyond $N=\num{2}$ unit cells leaves the single-period resource invariant at fixed local rates (\cref{prop:bondselective}); preserving a cat state on site~$j=\num{0}$ instead requires shielding its intercell strong-bond partner ($j=\num{3}$) against bulk dissipation.
  \item \emph{Lock readout strictly to stroboscopic times:} Intra-period micromotion causes order-of-magnitude swings and sign reversals in the matched contrast (\cref{SI-sec:micromotion}), so parity and tomography pulses must be synchronized to $t=T$. The readable contrast depends on the drive-period count $n$: on the two-cell branch it vanishes at $n=\num{2}$ and $n=\num{6}$ and is resolved at $n=\num{1},\num{3},\num{4},\num{8}$ (\cref{SI-tab:longtime}). The $n=\num{2}$ zero is exact while \num{10}-trajectory estimates there fluctuate about it. Readout therefore locks to the $n=\num{1}$ period and its $n=\num{3},\num{4},\num{8}$ revivals.
  \item \emph{Certify contrast with the $n=\num{1}$ displaced-Fock witness:} Replacing full phase-space tomography with the one-photon witness certifies the two-cell asymmetry at Fock support $N_{\max}\ge\num{4}$ at modest shot budgets (\cref{sec:hardware_protocol}), whereas higher-order rungs ($n\ge\num{2}$) remain below threshold.
\end{enumerate}

\subsection{Hardware scales and falsifiable measurement protocol}
\label{sec:hardware_protocol}

To relate the normalized units ($\omega_d=1$, $T=2\pi$) to physical hardware scales, \cref{tab:hardware} lists idealized cyclic frequencies $X/(2\pi)$ for representative superconducting circuit-QED ($\omega_d/2\pi=\qty{100}{\mega\hertz}$, $T=\qty{10}{\nano\second}$) and optomechanical ($\omega_d/2\pi=\qty{10}{\mega\hertz}$, $T=\qty{100}{\nano\second}$) arrays (\cref{SI-sec:hardware}). Both rows preserve $\chi/J_B=\num{0.42}$, $\gamma_{\rm bulk}/\chi=\num{0.4}$, and $\gamma_{\rm edge}/\gamma_{\rm bulk}=\num{5e-4}$. In circuit QED, $\gamma_{\rm bulk}=\num{0.2}\,\omega_d$ corresponds to an engineered dump port with drive-scale quality factor $Q_{\rm drive}=\num{5.0}$~\cite{Blais2021Circuit} ($Q_{\rm res}=\num{250}$ for a \qty{5}{\giga\hertz} cavity alongside $Q_{\rm edge}=\num{5e5}$ on the shielded end); in optomechanics, the \qty{5.0}{\mega\hertz} Kerr scale requires qubit-mediated rather than bare radiation-pressure anharmonicity.

\begin{table}[htpb]
\centering
\caption{Idealized hardware cyclic-frequency lookup table mapping normalized parameters ($\omega_d=1$, $T=2\pi$) to representative superconducting circuit-QED and optomechanical resonator targets (\cref{SI-sec:hardware}). Columns three and four list cyclic frequencies $X/(2\pi)$ for circuit-QED ($\omega_d/2\pi=\qty{100}{\mega\hertz}$) and optomechanical ($\omega_d/2\pi=\qty{10}{\mega\hertz}$) targets.}
\label{tab:hardware}
\small
\begin{tabular}{lccc}
\toprule
Parameter & Normalized ($\omega_d$) & Circuit QED & Optomechanical \\
\midrule
Period $T$ & $2\pi$ & \qty{10}{\nano\second} & \qty{100}{\nano\second} \\
Strong hopping $J_B/2\pi$ & \num{1.2} & \qty{120}{\mega\hertz} & \qty{12}{\mega\hertz} \\
Weak hopping $J_A/2\pi$ & \num{0.1} & \qty{10}{\mega\hertz} & \qty{1.0}{\mega\hertz} \\
Boundary Kerr $\chi/2\pi$ & \num{0.5} & \qty{50}{\mega\hertz} & \qty{5.0}{\mega\hertz} \\
Bulk loss $\gamma_{\rm bulk}/2\pi$ & \num{0.2} & \qty{20}{\mega\hertz} & \qty{2.0}{\mega\hertz} \\
Edge loss $\gamma_{\rm edge}/2\pi$ & \num{1e-4} & \qty{10}{\kilo\hertz} & \qty{1.0}{\kilo\hertz} \\
\bottomrule
\end{tabular}
\end{table}

A parity measurement at the phase-space origin returns $\mathrm{sign}\,W(0)$ without reconstructing the distribution, and a displaced-Fock witness converts displaced photon statistics into a binary certificate that the sampled state's Wigner function takes negative values, at a cost independent of grid resolution, so the two-cell signal is available without full tomography. The volume of negativity is a different object, requiring a reconstruction on a stated resolution: we report it on grids spanning $\num{21}^{2}$ to $\num{61}^{2}$ points and treat the parity witness and the reconstructed volume as complementary instruments with distinct error budgets. In an experiment the parity read-out would come from a qubit dispersively coupled to the cavity mode, and the reconstruction from heterodyne tomography of the same mode in a repeated run.

Displaced-Fock witnesses of Wigner negativity~\cite{Chabaud2021witnessingwigner,Chakrabarty2026} make this certificate quantitative: they certify nonclassicality whenever $\max_\alpha F_1(\alpha)=\max_\alpha\braket{1|D^\dagger(\alpha)\rho_0 D(\alpha)|1}>\omega_1=\num{0.5}$. At the primary cutoff $N_{\max}=\num{2}$, the exact-ME topological state sits just below threshold ($F_1^{\rm topo}=\num{0.4944}$, $F_1^{\rm triv}=\num{0.4482}$, $\Delta F_1=+\num{0.0461}$), so \num{10}-trajectory threshold crossings reflect finite-trajectory estimator variance rather than single-shot certification (\cref{SI-sec:witness}). Once higher Fock components enter at the higher cutoffs ($N_{\max}=\num{4}$ and $\num{6}$), the exact-ME topological state crosses the analytical threshold ($F_1^{\rm topo}=\num{0.5100}$ and $\num{0.5056}$, versus $F_1^{\rm triv}=\num{0.4591}$ and $\num{0.4565}$). Displaced photon counting then separates the two sectors via $\Delta F_1=+\num{0.0491}$ (\cref{SI-tab:witness-exact}), while reversing sign at $N\ge\num{3}$ ($\Delta F_1=-\num{0.0031}$ at $N_{\max}=\num{4}$, both below $\num{0.5}$). Over the \num{10}-trajectory ensemble, the paired relative-entropy non-Gaussianity contrast~\cite{Albarelli2018} is $\num{0.1316}\pm\num{0.0025}$.

The falsifiable experimental test proceeds in five stages (shot budgets and acceptance criteria in \cref{SI-sec:measurement-protocol}):
\begin{enumerate}
  \item \emph{Coherent calibration:} Tune couplings to $(J_A,J_B)=(\num{0.1},\num{1.2})$ (chiral) and $(\num{1.2},\num{0.1})$ (control), verifying the minimum zero and $\pi$ quasienergy gaps (\num{2.1991} and \num{0.3142}) and boundary localization at the chiral point only.
  \item \emph{Loss-profile engineering:} Calibrate $\gamma_{\rm bulk}=\num{0.2}$ and $\gamma_{\rm edge}=\num{1e-4}$ (or an intermediate ratio $\le\num{1e-2}$, over which the two-cell contrast remains on its positive plateau) via ringdown, alongside a uniform-loss control ($\gamma_j=\num{0.2}$).
  \item \emph{State preparation and stroboscopic drive:} Prepare the even cat $\alpha=\num{1}$ on site~$j=\num{0}$ (or on the dangling end~$j=\num{1}$ under physical-end shielding, with the Kerr term relocated to the prepared site) and evolve for one period $T$, locking readout strictly to $t=T$ to avoid intra-period micromotion sign flips.
  \item \emph{Grid-free witness and tomography:} For near-term two-cell certification without tomography at high-cutoff Fock support ($N_{\max}=\num{6}$), measure the displaced photon-counting witness $F_1(\alpha)$. On site~$j=\num{0}$, resolving $\Delta F_1=+\num{0.0491}$ requires \num{1622} shots per sector at \qty{80}{\percent} power (\num{2685} at \qty{95}{\percent}) and \num{49062} shots for outright certification above $\omega_1=\num{0.5}$; preparing on the dangling end~$j=\num{1}$ cuts these budgets to \num{577} shots per sector and \num{822} shots outright (\cref{SI-tab:witness-exact}). For the full sign pattern, reconstruct the edge-site Wigner function on a $\num{31}\times\num{31}$ grid.
  \item \emph{Falsification:} The boundary claim is refuted if the measured contrast tracks the bulk invariant across $J_A+J_B=\num{1.3}$ (failing to collapse near $J_A=\num{0.50}$ on $j=\num{1}$), remains nonzero at $t=T$ under uniform loss for $d\le\num{0.5}$, or fails to invert under the bond-selective two-cell sink swap (from $+\num{0.0223}$ to $-\num{0.0231}$).
\end{enumerate}

All hardware rows and shot budgets in this section assume the zero-temperature regime ($n_{\rm th}\lesssim\num{0.1}$) and unit-efficiency detection; the $j=\num{0}$ witness certification budgets further require $\eta\gtrsim\num{0.96}$, as stated in the Supplemental Material (\cref{SI-sec:hardware,SI-sec:measurement-protocol}); a finite-thermal Lindblad evolution with thermal excitation and dephasing channels was not simulated.

\section{Discussion}
\label{sec:discussion}

\subsection{Generator topology versus phase-space resource survival}
\label{sec:interpretation_A}

The $\mathbb{Z}\times\mathbb{Z}$ pair $(\nu_0,\nu_\pi)=(-1,0)$ classifies the stroboscopic \emph{generator}: it is an integer invariant of the unitary Floquet operator and stays an exact integer across all chain lengths, with chiral-frame residuals below $\num{1e-15}$. Local Wigner negativity, by contrast, is a basis-dependent functional of the driven dissipative \emph{state} evaluated after a finite window. Under local amplitude damping, \cref{eq:fp_main,eq:kappa_single_main} show that phase-space contraction leaves $\Ne$ invariant while Gaussian diffusion smooths negative interference troughs along the nodal boundary at rate $\kappa_{\rm single}=\num{3.411}$. Whether topology shields $\Ne$ therefore reduces to whether the dynamics keeps two-photon cat coherence away from bulk decay channels during the period. The intercell Floquet step does the opposite: it transfers weight onto site~$j=\num{3}$ with period-averaged dwell $\langle P_3\rangle_T=\num{0.198}$: shielding $j=\num{3}$ at two unit cells ($\gamma_3=\num{1e-4}$) preserves a positive contrast, whereas exposing $j=\num{3}$ to bulk loss ($\gamma_3=\num{0.2}$) from three cells onward overwhelms Kerr self-trapping and reverses the sign. The same local exposure mechanism accounts for the continuity across all three bulk gap closings and for the absence of collective edge-mode negativity despite $|u_{e0}|^2=\num{0.013}$.

Recent work on mixed-state Floquet topology clarifies the general principle: for noninteracting fermions, the $\mathbb{Z}\times\mathbb{Z}$ classification of a closed chiral chain collapses in the steady state to at most a weaker symmetry-protected invariant~\cite{Dinc2026}. We observe the bosonic, finite-time counterpart for a nonlinear phase-space resource: the invariant of the generator remains exact while the state-level resource vanishes or flips sign with the spatial loss profile. Shielding and preparing the physical dangling end ($j=\num{1}$) keeps the contrast positive and size-invariant, yet uniform loss removes it at stroboscopic $t=T$ for $d\le\num{0.5}$. Along the dimerization scan this survival is modulated by the local single-period return probability $P_{\rm return}=\cos^2(J_A\pi)$ rather than by the nonlocal bulk invariant: it collapses at the dynamical decoupling point, vanishes across the segment of the anomalous $\pi$-phase adjacent to the transition, and rises again outside that window, peaking where $|\cos(J_A\pi)|=1$ (\cref{SI-tab:j1-dimerization}). Equal $\cos^2(J_A\pi)$ in different bulk phases does not give equal contrast, so the modulation is not a one-to-one correspondence.

\subsection{What would change this conclusion}
\label{sec:portee}

Three developments would require the conclusion to be revised. The first is a mechanism that suppresses edge-region loss while retaining bulk loss. A bath whose spectral density vanishes on the boundary sites while remaining finite in the bulk would remove the direct boundary decay, though leakage onto bulk-damped sites still limits survival, as the decay from $\num{0.17105}$ to $\num{0.11997}$ over eight periods at $\gamma_{\rm edge}=\num{1e-4}$ shows. Suppressing that leakage as well would preserve the dangling-end state at fixed dimerization, and the two-cell sink-placement benchmark established in \cref{sec:sizelaw} is the cheapest way to test it. The second is a resource functional that is insensitive to the loss channel by construction. Our observable is a convex, piecewise-linear functional of the density matrix whose negative volume erodes at first order under single-photon loss (\cref{prop:erosion}); a functional protected by an exact symmetry of the dissipator, as in symmetry-resolved mixed-state orders~\cite{Paszko2024,Chirame2025}, would not. The third is a regime in which the edge mode is pumped or stabilized macroscopically rather than prepared as a finite-energy two-component cat ($\alpha=\num{1}$), shifting the Kerr self-trapping scale relative to single-particle tunneling. None of the three is available within the finite-energy cat protocol studied here at the single preparation amplitude $|\alpha|=\num{1}$ and Kerr strength $\chi=\num{0.5}$; all three are experimentally addressable.

For neighboring resources the accounting is the same: boundary dissipation governs the local non-Gaussian coherence here just as port dissipation bounds metrological extraction there~\cite{Deuyekbe2026CavityLoss,Deuyekbe2026MeanResponse}. The bond-selective sink rule of \cref{prop:bondselective} is the discrete-lattice counterpart of spatial channel separation in non-Hermitian ladders.

\subsection{Boundary conditions and literature placement}
\label{sec:lit_placement_A}

Unlike coherent Floquet classification~\cite{Asboth2014,Xu2022micromotion}, which concerns the driven generator alone, driven-dissipative quantum resources require an engineered edge-to-bulk loss ratio to sustain a positive two-cell signal, robust up to $\gamma_{\rm edge}/\gamma_{\rm bulk}=\num{1e-2}$ ($+\num{0.0158}$, positive in \qty{90}{\percent} of realizations; \cref{SI-tab:lossratio,SI-sec:lossratio,tab:hardware}). Tunable-coupler circuit-QED and optomechanical lattices already provide the experimental primitives---cat preparation~\cite{Vlastakis2013}, parity readout~\cite{Lutterbach1997}, engineered reservoir stabilization~\cite{Touzard2018}, and homodyne tomography~\cite{Aspelmeyer2014}---needed to test this crossover. For modular bosonic error-correction architectures, the implication is direct: passive Floquet band topology cannot replace active reservoir stabilization~\cite{Touzard2018,Zapletal2022multimode,Rojkov2026catmanifolds} or selective isolation of strong-bond dissipation channels in the local preservation step that routing cat-code states between cavities would require. Together, the displaced-Fock witness budgets and the exact-$Q$ covariant benchmark (\cref{sec:covariant,sec:hardware_protocol,SI-tab:witness-exact}) reinforce the generator--state split: passive linear mode isolation does not transfer non-Gaussian negativity across bases~\cite{Albarelli2018}.

The negative result complements mixed-state symmetry-protected topological (SPT) theory~\cite{Paszko2024,Chirame2025,Hannukainen2026,Sun2025holographic}, whose steady-state order is itself fragile to symmetric perturbations via strong-to-weak symmetry breaking~\cite{Shah2025instability}. Likewise, driven-dissipative bosonic platforms can engineer steady-state topological phases~\cite{Diehl2011,Bardyn2013,GomezLeon2023resonator,Peano2016}, stabilize Floquet quasienergy states~\cite{Ritter2025autonomous,Roy2025} or long-range correlations~\cite{ClaveroRubio2026}, and preserve parity- or Kerr-stabilized Wigner negativity independently of band topology~\cite{Liu2026,Raza2026}. None of these steady-state mechanisms protects the finite-time trajectory functional tracked here. Similarly, experiments showing topological protection of linear correlations, photon blockade, or parametric enhancement~\cite{Wang2019Optica,Perez2024domainwalls,Li2024photonblockade} do not conflict with our findings. Even for two-photon states, topological protection degrades as entanglement grows~\cite{Tschernig2021}, and an undriven Kitaev Majorana plateau~\cite{Plodzien2026} rests on quadratic parity conservation, not on a finite-time non-Gaussian functional evaluated under a strongly edge-shielded local loss profile.

Our findings isolate three necessary conditions for a bulk Floquet invariant to protect a local non-Gaussian resource: (1)~a dissipator respecting the protecting symmetry, (2)~an observable supported on the protected mode basis, and (3)~size-stable mode localization. Spatial dissipation engineering in monitored SSH chains highlights the sensitivity of edge dynamics to bath geometry~\cite{Salatino2025}. Yet testing condition~(1) explicitly via single-sublattice loss commuting with the chiral operator yields a small negative matched contrast ($-\num{0.00081}\pm\num{0.00005}$ at two cells, unchanged at $-\num{0.00079}\pm\num{0.00009}$ at matched total loss budget; \cref{SI-sec:chiral-loss,SI-tab:chiralloss}) because both sectors retain nearly identical negativity ($\num{0.02285}$ versus $\num{0.02366}$), erasing sector distinction rather than protecting an asymmetry. While weak symmetries can protect string-order diagnostics~\cite{Huang2025weaksymmetries}, they do not protect this local non-Gaussian monotone, and conditions~(2) and~(3) fail by construction. Any topological transfer theorem would therefore need open-system generalizations beyond the coherent spectral localizer~\cite{CerjanLoring2025Tutorial,WongCerjanCole2024Quantitative,Ghosh2024Local}, which bounds the single-particle gap (\cref{SI-sec:localizer}) but not many-body Lindblad trajectories.

Within the normalized parameter regime mapped here (\cref{tab:hardware}), the conclusions hold across the Fock cutoffs, the single-edge HEOM check performed at five times the operating bulk rate, the $\num{64}$-period evolution, and the intra-period drive-phase sweep (\cref{SI-sec:exact-subspace,SI-sec:crosssolver,SI-sec:longtime,SI-sec:micromotion}). The $\chi=\num{0}$ control preserves the two-cell sign at about half its amplitude and the three-cell null, but not the negative branch, which exists only with the Kerr term (\cref{SI-sec:kerrfree}).

\section{Conclusion}
\label{sec:conclusion}

We tested whether the quantized bulk topological invariants of a chiral Floquet SSH generator protect a transient non-Gaussian resource under dissipation, and found a clear \emph{resource--invariant gap}. While $(\nu_0,\nu_\pi)$ remain exact integers at every chain length, the matched Wigner-negativity contrast on sub-boundary site~$j=\num{0}$ under index-end shielding flips sign from $+\num{0.02221}$ at two unit cells to $-\num{0.000113}$ at three, four, and five cells at $N_{\max}=\num{2}$ ($+\num{0.02875}$ versus $-\num{0.000955}$ at $N_{\max}=\num{4}$; $+\num{0.02606}$ at $N_{\max}=\num{6}$), the appended cells carrying no additional population. The sweep observable is continuous across two of the three quasienergy gap closings and carries a resolved sub-percent excursion at the third; the matched sector contrast is constrained only by the endpoint differences. The contrast is consistent with zero under uniform loss at stroboscopic $t=T$ for $d\le\num{0.5}$ and is not restored by chiral sublattice loss.

Two analytical results explain why: a three-site causal Floquet light cone (\cref{prop:bondselective}) places the contrast sign under the control of strong-bond bulk exposure on site~$j=\num{3}$, while Gaussian phase-space diffusion along the nodal contour (\cref{prop:erosion}, $\kappa_{\rm single}=\num{3.411}$) erodes Wigner negativity whenever coherent two-photon dwell overwhelms local Kerr self-trapping. Preparing and shielding the dangling end~$j=\num{1}$ sustains a size-invariant contrast by the light cone of \cref{prop:bondselective} ($+\num{0.07878}$ at $N=\num{2}$, $N_{\max}=\num{6}$). Yet this survival is non-topological: it is modulated by the local single-period return $P_{\rm return}=\cos^2(J_A\pi)$ rather than set by the bulk winding, equal return values within one bulk phase still giving unequal contrast (\cref{fig:exact-mechanism}(b), \cref{SI-tab:j1-dimerization}). Complementary exact-$Q$ covariant benchmarks show passive linear mode survival ($\le\num{7.41e-4}$) for Floquet-mode initialization without transferring non-Gaussian negativity. Together with the bond-selective sink benchmark (\cref{tab:decomposition}) and the grid-free displaced-Fock witness (\cref{SI-tab:witness-exact}), these results show that a claim of topological protection for a nonlinear bosonic resource must rest on direct verification of the state functional, not on the coherent generator.

\section*{Supplementary material}

Supplementary material accompanying this article presents the model conventions and equations, the matched statistical analysis and uncertainty quantification, deterministic benchmarks, numerical convergence and control scans, and robustness and collective-mode analyses. It further documents the grid-free witness certification, the short-time Lindblad expansion, the hardware parameter mapping, the validity and parameter-scope tables, and the size-scaling fit underlying the main text~\cite{SupplementalMaterial}.

\section*{Data availability}

All data supporting the findings of this study are available within the article and its Supplementary material. Manuscript sources, simulation scripts, and exact-subspace benchmarks accompany the submission and will be deposited in a persistent public archive with a registered identifier upon acceptance.

\section*{Acknowledgments}

We thank the Stellenbosch Institute for Advanced Study for hospitality. Computational resources were provided by the penavoraserver cluster (ThinkStation P720, \qty{128}{\giga\byte} RAM, \num{48} cores) through Penabei Samafou; supporting data are provided with the manuscript. This work received no external funding.

\section*{Author contributions}

JD and PD performed the numerical simulations and analyzed the data. NM contributed to the theoretical framework and provided critical feedback on the results. SGNE conceived, supervised, and directed the project. PD and SGNE wrote the final version of the manuscript. All authors reviewed and approved the final manuscript.

\section*{Competing interests}

The authors declare no competing interests.

\section*{Use of AI}

AI-based language models were used to support literature searches, to refine and debug analysis scripts, and to verify bibliographic metadata. All authors reviewed and edited all AI-assisted content and take full responsibility for the accuracy, originality, interpretation, and final wording of this manuscript.

\bibliographystyle{apsrev4-2}
\bibliography{04_track_A_topo_references}

\end{document}


\title{Supplemental Material for ``The resource--invariant gap: a protected Floquet invariant need not protect Wigner negativity''}

\author{Juste Deuyekbe}
\email{justedeuyekbe@gmail.com}
\affiliation{Department of Physics, Faculty of Science, University of Ngaoundere, P.O. Box 454, Ngaoundere, Cameroon}

\author{P. Djorw\'e}
\email{djorwepp@gmail.com}
\affiliation{Department of Physics, Faculty of Science, University of Ngaoundere, P.O. Box 454, Ngaoundere, Cameroon}
\affiliation{Stellenbosch Institute for Advanced Study (STIAS), Wallenberg Research Centre at Stellenbosch University, Stellenbosch 7600, South Africa}

\author{Nsangou Mama}
\email{mnsangou@yahoo.com}
\affiliation{Department of Physics, Faculty of Science, University of Maroua, P.O. Box 46, Maroua, Cameroon}

\author{S. G. Nana Engo}
\email{serge.nana-engo@facsciences-uy1.cm}
\affiliation{Department of Physics, Faculty of Science, University of Yaounde I, P.O. Box 812, Yaounde, Cameroon}

\maketitle

\section{Conventions and notation}
\label{sec:scope}

\subsection{Resource convention}
\label{sec:si_resource_convention}

One budget is consumed to encode the state and a different one is degraded in reading it out; the two are never added. The encoded resource is
\begin{equation}
\mathcal{R}_{\rm enc}=\sum_{j\in\text{carrier}}\langle \hat a_j^\dagger \hat a_j\rangle+\langle \hat n_{\rm prep}\rangle ,
\label{eq:si_renc}
\end{equation}
summed over the modes that carry the signal, in quanta. Added noise $n_{\rm add}$ is deliberately absent from \cref{eq:si_renc}: it enters only as the denominator of a signal-to-noise ratio, because a quantum added by the detector characterizes the readout rather than paying for the encoding.

Contrast values are properties of a solver, loss profile, and preparation protocol and are not transferable across them: the noise-free ($d=\num{0}$) edge-selective contrast in the \num{10}-trajectory ensemble is $+\num{0.0167}$ ($+\num{0.02221}$ in the exact master equation), whereas removing dissipation altogether ($\gamma_j=\num{0}$) gives $-\num{7.9e-4}$, an offset comparable to the negative branch. Every contrast here is therefore paired within one protocol, never compared across protocols. The preparation is fixed and $\mathcal{R}_{\rm enc}$ follows from it: the primary even cat is truncated at $N_{\max}=\num{2}$ with $c_0=\num{0.7462}$ and $c_2=\num{0.6658}$, so $\mathcal{R}_{\rm enc}=2c_2^2=\num{0.8866}$ quanta on the boundary site, no preparation noise is counted, and both sectors share that preparation exactly. The paired contrast is thus already a per-preparation quantity: the zero-noise two-cell edge-selective contrast $+\num{0.016731}$ corresponds to $+\num{0.01887}$ per prepared photon. Ratios of contrasts are unaffected by the choice, which is why the main text quotes the contrast in raw Wigner units and reports $\mathcal{R}_{\rm enc}$ once, here; dividing a negativity by the ideal-state value of a different preparation would report that state's own truncation as a resource, a quantity not used anywhere in this work.

\subsection{Notation}
\label{sec:si_notation}

\Cref{tab:simodelnotation} fixes the symbols, units, and working points used throughout this study. Throughout, $\Delta_r=Y_{r,{\rm chiral}}-Y_{r,{\rm trivial}}$ denotes the matched sector contrast, $\Delta_{\min}$ is its detection floor, and the encoded resource $\mathcal{R}_{\rm enc}$ and added detection noise $n_{\rm add}$ follow \cref{sec:si_resource_convention}.

\begin{table}[htbp]
\centering
\caption{Notation and convention glossary for the chiral Floquet SSH model, open-system observables, and analytical decay bounds. All rates and couplings are expressed in normalized drive-frequency units ($\omega_d = 2\pi/T = 1$, $T = 2\pi$).}
\label{tab:simodelnotation}
\begin{tabularx}{\linewidth}{l l >{\raggedright\arraybackslash}X}
\toprule
Symbol & Value / Units & Physical meaning \\
\midrule
$J_A, J_B$ & $(\num{0.1},\num{1.2})$ / $(\num{1.2},\num{0.1})$ & Intracell and intercell hopping amplitudes (chiral / trivial working points) \\
$w_A, w_B$ & $\in \mathbb{Z}$ & Symmetric-frame chiral winding numbers \\
$\nu_0, \nu_\pi$ & $(-1,0)$ / $(0,0)$ & Floquet bulk topological invariants at quasienergy $0$ and $\pi$ \\
$\chi$ & \num{0.5} ($\chi/\omega_d$) & On-site Kerr nonlinearity on the prepared boundary site \\
$\gamma_{\rm edge}, \gamma_{\rm bulk}$ & \num{1e-4}, \num{0.2} & Boundary and bulk single-photon loss rates (edge-selective profile) \\
$d$ & \num{0}--\num{1.5} & Multiplicative loss-rate noise amplitude \\
$N_{\max}$ & \num{2} (exact subspace through \num{6}) & Local Fock-space truncation cutoff \\
$\mathcal{N}_W(\rho),\,\delta_-$ & Dimensionless & Integrated negative volume of the single-site Wigner function \\
$\Delta_r,\,\Delta\mathcal{N}_W,\,\Delta_{\min}$ & Dimensionless & Matched chiral-minus-trivial contrast ($Y_{r,{\rm chiral}}-Y_{r,{\rm trivial}}$) and detection floor \\
$\mathcal{R}_{\rm enc},\,n_{\rm add}$ & Quanta & Encoded probe resource and added detection-chain noise (\cref{sec:si_resource_convention}) \\
$b_e(t),\,u_e(t),\,\xi_{\rm part}$ & Dimensionless & Micromotion-covariant edge-mode operator, state vector, and participation length \\
$R_e(t),\,\Delta n_e(t)$ & Dimensionless & Covariant passive-isolation residual $n_e(t)-n_e^{\rm ref}(t)$ and matched occupation contrast $n_e^{\rm topo}(t)-n_e^{\rm triv}(t)$ \\
$F_n,\,\omega_n$ & Dimensionless & Displaced-Fock witness fidelity and analytical/upper-bound negativity threshold \\
$\Lambda_1,\,\Lambda_2^{\rm coh}$ & Dimensionless & Rate-weighted single- and two-photon coherent dwell integrals \\
$\kappa_{\rm single},\,\kappa_N$ & \num{3.411} (single-site) & Short-time Wigner-negativity decay prefactor and $N$-cell effective prefactor \\
\bottomrule
\end{tabularx}
\end{table}

\section{Two-step chiral model}
\label{sec:model}

For a unit cell containing sublattices $A$ and $B$, the bulk Hamiltonian is
\begin{equation}\label{eq:bulk}
H(k)=\left(J_A+J_B\cos k\right)\sigma_x+J_B\sin k\,\sigma_y,
\end{equation}
with chiral operator $\Gamma=\sigma_z$. The drive consists of
\begin{equation}\label{eq:drive}
H(t)=\begin{cases}
H_A,&0\leq t<T/2,\\
H_B,&T/2\leq t<T,
\end{cases}
\qquad T=2\pi,
\end{equation}
where $H_A$ contains only the intracell hopping and $H_B$ only the intercell hopping. On the open $N$-cell chain ($j=0,\dots,2N-1$), these step Hamiltonians read
\begin{equation}\label{eq:ha_hb_realspace}
H_A = J_A\sum_{m=0}^{N-1}\bigl(a_{2m}^\dagger a_{2m+1}+\mathrm{H.c.}\bigr),\qquad
H_B = J_B\sum_{m=0}^{N-2}\bigl(a_{2m}^\dagger a_{2m+3}+\mathrm{H.c.}\bigr),
\end{equation}
which wire the modes into the real-space nearest-neighbor backbone $1\text{--}0\text{--}3\text{--}2\text{--}\cdots\text{--}(2N-1)\text{--}(2N-2)$, placing sites $j=\num{1}$ and $j=2N-2$ at the dangling physical ends while sites $j=\num{0}$ and $j=2N-1$ sit one bond inward. The period propagator is
\begin{equation}\label{eq:uf}
U_F(k)=\exp[-iH_B(k)T/2]\exp[-iH_A(k)T/2].
\end{equation}
For the open-chain propagator $U_F=e^{-iH_B T/2}e^{-iH_A T/2}$ acting on the initially excited sub-boundary site $j=\num{0}$ ($N\ge\num{2}$), the disjoint dimer action of $H_A$ followed by $H_B$ truncates one-period propagation to the closed three-site light cone $\{0,1,3\}$, as given by \cref{MAIN-eq:uf_column} within \cref{MAIN-prop:bondselective}.
Thus $P_2(T)=0$ identically in the coherent linear model, the single-particle return probability $P_0^{(1)}(T)=\cos^2(J_A\pi)\cos^2(J_B\pi)=\num{0.5920}$ is invariant under $(J_A,J_B)\leftrightarrow(J_B,J_A)$, and the topological versus trivial sectors differ purely in escape branching: $(\num{0.3125},\num{0.0955})$ into $(j=\num{3},j=\num{1})$ for the topological sector versus $(\num{0.0625},\num{0.3455})$ for the trivial control.

\begin{figure}[t]
  \centering
  \includegraphics[width=0.98\linewidth]{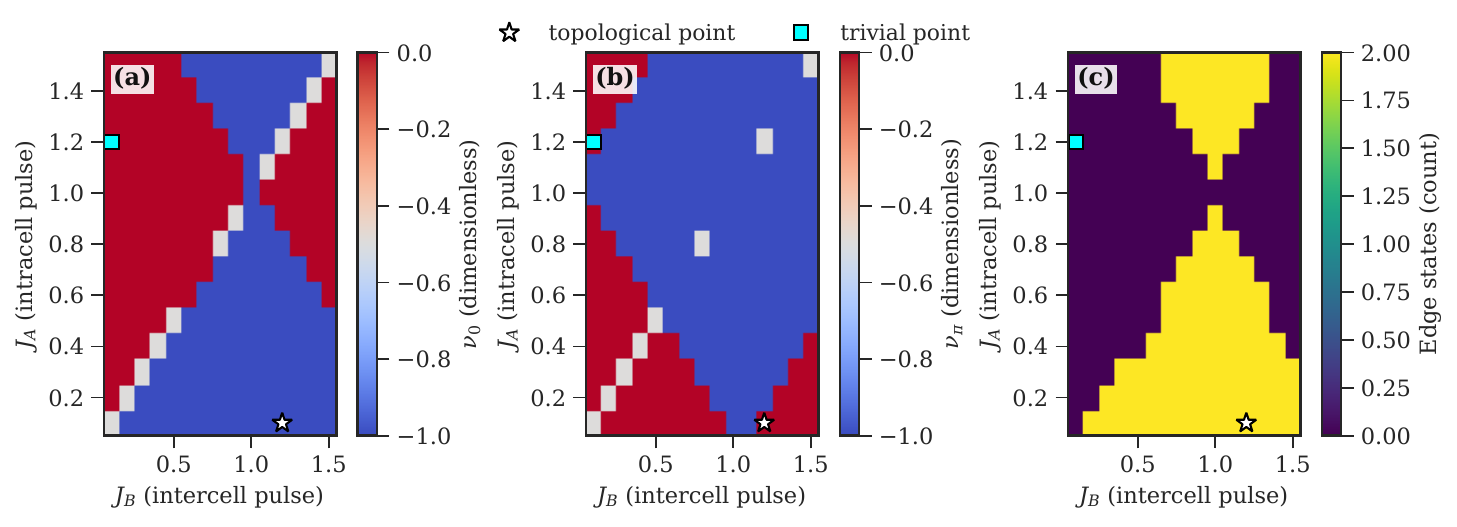}
  \caption{Coherent phase structure of the two-step chiral Floquet model. (a)~Zero-quasienergy winding $\nu_0$ and (b)~$\pi$-quasienergy winding $\nu_\pi$ (gray pixels mark gap-closing lines where bulk invariants are undefined), with the chiral point $(J_A,J_B)=(\num{0.1},\num{1.2})$ marked by an open white star and the trivial control $(\num{1.2},\num{0.1})$ by a filled cyan square. (c)~Open-chain zero-quasienergy edge-state count (dark purple: \num{0}; yellow: \num{2}). Parameter points are chosen from the invariant and gap criteria before any dissipative simulation.}
  \label{fig:phase}
\end{figure}

The two symmetric time frames produce off-diagonal winding numbers $w_A$ and $w_B$~\cite{Asboth2014}. We report
\begin{equation}\label{eq:nu}
\nu_0=(w_A+w_B)/2,\qquad \nu_\pi=(w_A-w_B)/2.
\end{equation}
The selected chiral point $(J_A,J_B)=(\num{0.1},\num{1.2})$ has $(\nu_0,\nu_\pi)=(-1,0)$; the selected trivial point $(\num{1.2},\num{0.1})$ has $(0,0)$. The maximum chiral-frame residual is below $\num{1e-15}$ on the resolution ladder. The minimum zero and $\pi$ gaps are positive at both points. The phase structure and finite-chain edge diagnostic are summarized in \cref{fig:phase}.

For an open chain, the single-particle propagator is constructed from the same step Hamiltonians. The edge vector $u_e(0)$ used in the edge-mode analysis is the most edge-localized eigenvector of the one-period open-chain propagator for the chiral point. This observable is distinct from the bulk invariant and is not assigned an open-system topological index.

Working points are restricted to the chiral-frame-stable, open-gap domain identified in the phase scan above.

With the model fixed, the sections that follow specify the protocol and uncertainty analyses underlying the main-text boundary statement, separating the laboratory-frame Wigner observable from the micromotion-covariant edge-mode observable.

\section{Dissipative dynamics}
\label{sec:dynamics}

The bosonic open-chain evolution is
\begin{equation}\label{eq:lindblad}
\dot\rho=-i[H(t)+H_{\rm K},\rho]
+\sum_j\gamma_j\left(a_j\rho a_j^\dagger-\frac{1}{2}a_j^\dagger a_j\rho-\frac{1}{2}\rho a_j^\dagger a_j\right),
\end{equation}
with the local Kerr term $H_{\rm K}=\chi\,a_0^{\dagger 2}a_0^2$, $\chi=\num{0.5}$, sites being indexed $j\in\{0,\dots,2N-1\}$ from the prepared boundary site $j=\num{0}$. The primary starts from a normalized even cat state $\ket{\alpha}+\ket{-\alpha}$ with $\alpha=\num{1.0}$ on the boundary site. The mode-resolved preparation distributes the displacement along the single-particle edge vector, $\alpha_j=\alpha\,u_{e,j}(0)$; both preparations are used identically for the two sectors. The local Fock cutoff is $N_{\max}=\num{2}$. The edge-low-loss profile uses $\gamma_{\rm edge}=\num{1e-4}$ at the two boundary sites and $\gamma_{\rm bulk}=\num{0.2}$ elsewhere; the uniform profile uses the bulk rate at all sites. Multiplicative rate perturbations use one factor $\xi\sim\mathcal{U}(-d,d)$ per realization, applied to every site ($\gamma_j\to\gamma_j(1+\xi)$), at four amplitudes ($d\in\{\num{0},\num{0.5},\num{1.0},\num{1.5}\}$); perturbed rates are clamped at zero.

The Lindblad equation is integrated with the quantum-jump (Monte Carlo wave-function) trajectory method~\cite{Dalibard1992} via QuTiP~\cite{Johansson2013} (v5.3.1, NumPy~v1.26.4), with integrator step budget $\num[retain-unity-mantissa = false]{1e4}$, final-state storage, nine output times over one period. The primary analysis uses ten trajectories per realization, whereas the primary covariant edge-mode benchmark uses exact piecewise-constant propagation in the conserved $Q\le\num{4}$ excitation subspace (\cref{sec:covariantmethod}); in stochastic checks, trajectories are nested within each realization and the realization is the independent unit.

The integrated negative volume of the reduced local Wigner function is
\begin{equation}\label{eq:neg}
\delta_{-}(t)=\int\!\int\frac{|W(q,p;t)|-W(q,p;t)}{2}\,\dd q\,\dd p,
\end{equation}
on a $\num{31}\times\num{31}$ grid over $[-4,4]^2$, evaluated after one period. Wigner negativity admits operational witnesses beyond a finite phase-space integral~\cite{Chabaud2021witnessingwigner,Zaw2024}.

\begin{table}[htbp]
\caption{Fixed simulation parameters. All dynamical parameters are normalized model coordinates expressed in units of the drive frequency ($\omega_d=1$).}
\label{tab:parameters}
\begin{ruledtabular}
\begin{tabular}{lll}
Quantity & Main dynamics & Covariant benchmark \\
\hline
Cells/modes & \num{2}/\num{4}, \num{3}/\num{6}, \num{4}/\num{8}, \num{5}/\num{10} & \num{2}/\num{4} \\
Cutoff / subspace & $N_{\max}=\num{2}$ & $Q\le\num{4}$ ($\dim=\num{70}$) \\
$\chi$ & \num{0.5} & \num{0} or \num{0.5} \\
$n_{\rm MC}$ & \num{500}, \num{500}, \num{2000}, \num{300} & deterministic ($d=\num{0}$) \\
Trajectories/unit & \num{10} & exact ME \\
Wigner grid & $\num{31}\times\num{31}$ & exact occupation / collective Wigner \\
Solver & QuTiP \texttt{mcsolve} & Piecewise exact ME \\
Integrator steps & \num[retain-unity-mantissa = false]{1e4} & \num{16} half-period steps \\
Stored output times & \num{9} & \num{17} \\
Observation window & $T=2\pi$ & $\num{8}T$ \\
Independent unit & noise realization & exact sector pair \\
\end{tabular}
\end{ruledtabular}
\end{table}

\section{Matched analysis, uncertainty, and paired comparison}
\label{sec:statistics}

For each profile, loss-rate noise level, and realization $r$, the sector contrast is
\begin{equation}\label{eq:contrast}
\Delta_r=Y_{r,{\rm chiral}}-Y_{r,{\rm trivial}},
\end{equation}
formed inside each matched noise realization before aggregation. The summary is the mean of $\Delta_r$ and a bootstrap percentile \qty{95}{\percent} interval~\cite{Efron1979} over the independent realization units ($\num[retain-unity-mantissa = false]{1e4}$ resamples, seeded). Sector simulations share the noise realization, initial state, cutoff, grid, and loss rates; their trajectory seeds are distinct but derived deterministically from the realization seed. This paired construction avoids treating repeated trajectory values or time points as independent replicates. Statistical resolution of the sign pattern rests on the \num{200}-realization four-cell ensemble, which remains resolved on the independent subset of samples \num{25}--\num{199} ($-\num{0.000333}\pm\num{0.000122}$). Secondary trajectory ladders, longtime traces, witness rates, and cutoff strata serve as descriptive sensitivity checks without familywise multiplicity adjustments. All ensembles use a fixed random seed ($\num{20260814}$), with the replication checks drawing independent random streams from it.

The primary comparison comprises \num{4000} matched units at two and three cells and \num{16000} at four cells (\cref{tab:paired-comparison}), i.e.\ $N_{\rm MC}=\num{500}$, \num{500}, and \num{2000} per noise level and loss profile; nested smaller-$N_{\rm MC}$ four-cell ensembles permit direct sample-size scaling comparison. The exact covariant benchmark is deterministic at $d=\num{0}$ ($Q\le\num{4}$, $\dim=\num{70}$), complemented in auxiliary five-trajectory noise sensitivity checks by \num{80} paired units per condition (\num{10} realizations $\times$ \num{2} loss profiles $\times$ \num{4} noise levels). Unless labeled explicitly as a \qty{95}{\percent} bootstrap or Wilson interval, quoted uncertainties of the form $a\pm b$ throughout this Supplemental Material denote one standard error (\num{1}~SE).

\begin{table}[htbp]
\caption{Primary counts and edge-low-loss \num{10}-trajectory contrast ranges across the four loss-rate noise levels. Counts at two through four cells span both loss profiles; the five-cell count covers only the edge-low-loss profile at the two levels $d\le\num{0.5}$. Ranges are not confidence intervals.}
\label{tab:paired-comparison}
\begin{ruledtabular}
\begin{tabular}{lrrr}
Cells & Modes & $n_{\rm units}$ & Mean contrast range \\
\hline
\num{2} & \num{4} & \num{4000} & $\num{+0.0131}$ to $\num{+0.0167}$ \\
\num{3} & \num{6} & \num{4000} & $\num{-0.0023}$ to $\num{-0.0020}$ \\
\num{4} & \num{8} & \num{16000} & $\num{-0.0021}$ to $\num{-0.0018}$ \\
\num{5} & \num{10} & \num{600} & $\num{-0.0026}$ to $\num{-0.0022}$ \\
\end{tabular}
\end{ruledtabular}
\end{table}

\begin{figure}[htpb]
  \centering
  \includegraphics[width=0.88\linewidth]{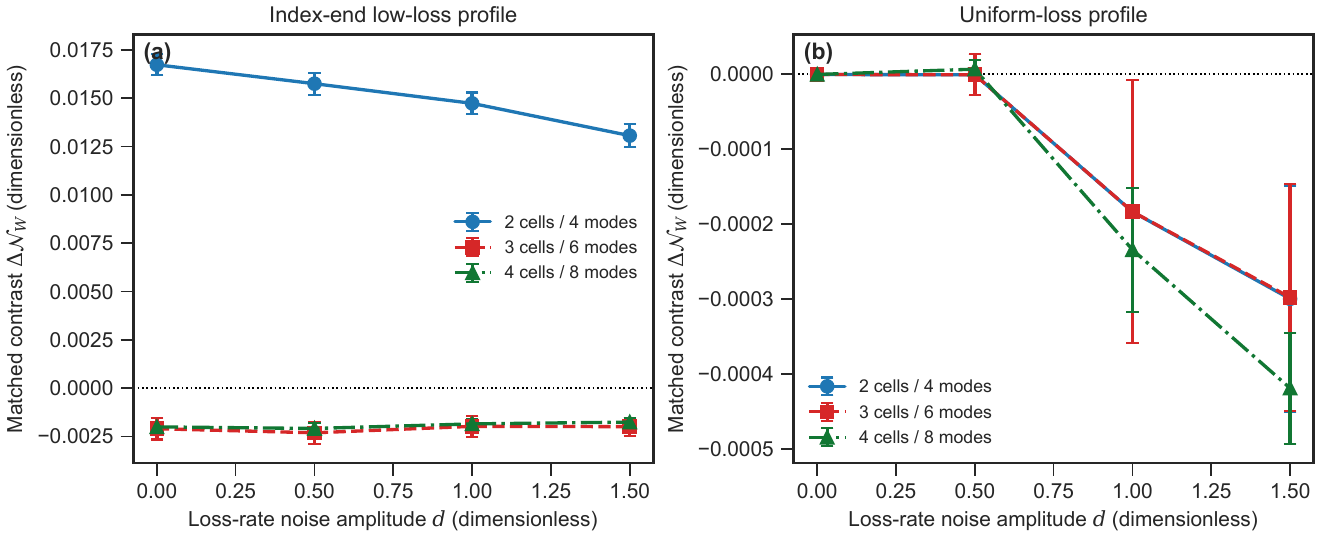}
  \caption{Loss-rate noise robustness of the local Wigner-negativity contrast in the \num{10}-trajectory ensemble for $N=\num{2}$ (solid blue circles), $N=\num{3}$ (dashed red squares), and $N=\num{4}$ (dash-dot green triangles) unit cells across multiplicative noise amplitudes $d\in\{\num{0},\num{0.5},\num{1.0},\num{1.5}\}$ (dotted horizontal zero line). (a)~Index-end low-loss profile: the two-to-three-cell sign reversal persists across all tested noise amplitudes. (b)~Uniform-loss profile: values remain near zero at $d\le\num{0.5}$ with weak negative offsets resolved only at larger noise. Error bars indicate $\pm\num{1}$~SE of the matched sector contrast.}
  \label{fig:si-robustness}
\end{figure}

Under uniform loss the mean contrasts are near zero or weakly negative, showing that the local observable tracks the spatial loss profile rather than topological protection. The stability of the matched contrast across all four loss-rate noise amplitudes ($d=\num{0},\num{0.5},\num{1.0},\num{1.5}$) is summarized in \cref{fig:si-robustness}.

\section{Micromotion-covariant edge-mode observable}
\label{sec:covariantmethod}

The analysis defines
\begin{equation}\label{eq:covariant}
u_e(t)=U_{\rm coh}(t)u_e(0),\qquad
b_e(t)=\sum_j u_{e,j}(t)^*a_j,
\qquad n_e(t)=\braket{b_e^\dagger(t)b_e(t)}.
\end{equation}
The number-operator coefficients are $u_{e,i}(t)u_{e,j}(t)^*$ in the $a_i^\dagger a_j$ basis. A fixed-frame number operator is retained as a negative control; the covariant operator is used for the edge-mode interpretation.

The time-dependent passive exposure is
\begin{equation}\label{eq:exposure}
\Gamma_e(t)=\int_0^t\gamma_e(s)\dd s,
\qquad \gamma_e(s)=\sum_j\gamma_j|u_{e,j}(s)|^2,
\end{equation}
with reference $n_e^{\rm ref}(t)=n_e(0)\exp[-\Gamma_e(t)]$. This equality is exact only in an isolated passive mode with no edge--bulk transfer. We define the passive-isolation residual $R_e(t)\equiv n_e(t)-n_e^{\rm ref}(t)$, the matched occupation contrast $\Delta n_e(t)\equiv n_e^{\rm topo}(t)-n_e^{\rm triv}(t)$, and a $\num{1e-3}$ isolation threshold on $\max_{t\le 8T}|R_e(t)|$. Evaluating \cref{eq:covariant} via the exact piecewise-constant master equation in the conserved total-excitation subspace $Q\le\num{4}$ ($\dim=\num{70}$ at $N=\num{2}$, zero loss-rate noise) recovers the uniform-loss passive reference to $\num{1.42e-12}$ and $\num{3.24e-12}$ for common-site ($n_e(\num{0})=\num{0.01064}$) and direct Floquet-mode preparation ($n_e(\num{0})=\num{0.7568}$), with no cross-mode truncation leakage at this cutoff. Under edge-selective loss, Floquet-mode preparation satisfies the $\num{1e-3}$ passive-isolation criterion while common-site ($j=\num{0}$) cat preparation violates it; the full residual list is given in the main text alongside \cref{MAIN-fig:covariant}, and collective Wigner negativity remains null in both cases. Per-mode truncated ($N_{\max}=\num{2}$) five-trajectory Monte Carlo estimates instead inflate these passive residuals (\num{0.03956} common-site and \num{0.38257} mode-resolved across \num{10} realizations, with Kerr-free $N_{\max}=\num{2},\num{3},\num{4}$ values \num{0.0388}, \num{0.0282}, \num{0.0356}) through cross-mode Fock leakage and trajectory sampling noise, both eliminated by exact $Q\le\num{4}$ propagation. Results are shown in \cref{MAIN-fig:covariant}.

\subsection{Mechanistic observables and their interpretation}
\label{sec:mechanisticobservables}

For a normalized single-particle edge vector $u_e(t)$, the edge weight and the site-resolved dissipative current are
\begin{equation}\label{eq:mechanismobs}
L_e(t)=\sum_{j\in\mathcal{E}}|u_{e,j}(t)|^2,
\qquad I_j(t)=\gamma_j\langle a_j^\dagger a_j\rangle_t,
\end{equation}
where $\mathcal{E}$ denotes boundary sites. Because the isolated-mode passive reference breaks down under structured loss, these currents diagnose local dissipation exposure rather than an isolated-mode protection mechanism.

\section{Deterministic benchmarks}
\label{sec:unitarybenchmark}

Two exactly solvable controls validate the observable implementation. (i)~A single-particle unitary benchmark with losses, Kerr terms and loss-rate noise set to zero: for $2N=\num{4},\num{6},\num{8},\num{12}$ modes ($N=\num{2},\num{3},\num{4},\num{6}$ unit cells), the transported edge vector $u_e(t)=U_{\rm coh}(t)u_e(0)$ and an orthogonal complement basis give a maximum covariant-population error of $\num{1.33e-15}$ and a maximum edge--bulk overlap of $\num{1.42e-31}$ over \num{257} times per period. The fixed-frame population, evaluated with $u_e(0)$ instead of $u_e(t)$, falls to \num{0.895}, showing that the micromotion correction changes the observable. (ii)~A passive single-particle decomposition with the non-Hermitian propagator
\begin{equation}\label{eq:passivepropagator}
K(t)=\mathcal{T}\exp\left[-i\int_0^t
\left(H(s)-\frac{i}{2}\sum_j\gamma_j\ketbra{j}{j}\right)\dd s\right]
\end{equation}
yields, over \num{257} times per period, a maximum reference residual of $\num{1.94e-15}$ under uniform loss (complementary-space population $\num{4.33e-16}$) and maximum residuals of \num{9.79e-4}, \num{1.94e-4}, \num{1.94e-4}, \num{1.94e-4} for \num{2}, \num{3}, \num{4}, \num{6} cells under edge-selective loss. The deterministic control shows that structured loss and micromotion are not separable in the finite chain, whereas uniform loss is.

\section{Numerical convergence and benchmarks}

\label{sec:convergence}

The convergence tests use the two-cell system at zero loss-rate noise, $\chi=\num{0.5}$, and the primary parameters of \cref{tab:parameters}. Each bounds one of the grid, trajectory-count, and Fock-cutoff uncertainties at fixed model parameters, and the resulting solver and cutoff status is collected in \cref{tab:sivalidity}.

\subsection{Wigner-grid convergence}
\label{sec:gridconvergence}

For a fixed final state (one realization, \num{200} trajectories), the single-site negativity integral over $[-4,4]^2$ takes the values \num{0.02695}, \num{0.02565}, \num{0.02680}, \num{0.02610}, \num{0.02621} on $\num{21}^2$, $\num{31}^2$, $\num{41}^2$, $\num{51}^2$, $\num{61}^2$ grids. All values lie within $\pm\qty{3}{\percent}$ of the finest estimate; the primary $\num{31}^2$ grid is within \qty{2.2}{\percent}. Grid discretization therefore contributes a few-percent relative uncertainty to the absolute integral whenever the contrast itself satisfies $|\Delta\mathcal{N}_W|\gtrsim\num{1e-3}$; for the small negative branch the residual does not cancel, and the contrast moves from $-\num{0.000113}$ on $\num{31}^2$ to $-\num{0.000182}$ on $\num{61}^2$ (\cref{tab:exact-subspace-ladder}), so that branch is quoted with its grid pair rather than as a single number.

\subsection{Trajectory-count sensitivity of the contrast}
\label{sec:trajcount}

Because the Wigner-negativity functional is nonlinear in the state, evaluating it on a trajectory-averaged state built from $N_{\rm traj}$ samples carries a positive bias that decays as $1/N_{\rm traj}$ and is concentrated in the near-zero sectors. At $N_{\rm traj}=\num{10}$ (\cref{tab:si_sector_mc10}) the canonical piecewise exact master equation (\cref{sec:exact-subspace}) gives sector negativities \num{0.022734} (topological) and \num{0.000525} (trivial) at two cells, against Monte Carlo values \num{0.023685} and \num{0.006954}; the trivial-sector bias is the larger contribution. At three cells the exact values are \num{0.000386}/\num{0.000499} against Monte Carlo \num{0.004779}/\num{0.006883}, with the same downward direction on the contrast.

A trajectory ladder on the primary conditions (same seeds, edge-selective profile, $d=\num{0}$) is summarized in \cref{tab:trajladder}. The infinite-trajectory extrapolation validated against the exact column at two and three cells (deviations \num{6e-5} and \num{3e-5}, below the \num{5e-4} tolerance) gives $-\num{0.00034}$ and $-\num{0.00039}$ at four and five cells; the alternative weighted $C+c/N_{\rm traj}$ fit gives $-\num{0.00022}$ and $-\num{0.00018}$, so the two fits differ by a factor of about $\num{1.5}$ at four cells. The direct four-cell \num{200}-trajectory point (\num{200} matched units) reads $-\num{0.00032}\pm\num{0.00011}$ with a \qty{95}{\percent} interval excluding zero; the five-cell \num{200}-trajectory point (\num{100} matched units) reads $-\num{0.00029}\pm\num{0.00015}$ with a \qty{95}{\percent} interval crossing zero. Excluding the initial \num{25} realizations preserves the four-cell resolution ($-\num{0.000333}\pm\num{0.000122}$; \cref{sec:statistics}).

The same ladder at two and four periods resolves the small $2T$ reversal as a fluctuation of the finite-trajectory estimator (exact contrast zero, MC estimate identically zero at $N_{\rm traj}=\num{50},\num{200}$) and brackets the $4T$ revival against the exact value ($+\num{0.0751}\pm\num{0.0009}$ at \num{200} trajectories against exact $+\num{0.07542}$).

\begin{table}[htbp]
\caption{Paired contrast versus trajectory count at $d=\num{0}$, edge-selective loss; the counts below each column give the matched units behind each point. Exact ME is canonical piecewise half-period integration in the $Q\le\num{2}$ subspace (\cref{sec:exact-subspace}), agreeing with single-call \texttt{mesolve} at $\Delta t_{\max}=\num{0.01}$ to \num{2.6e-10}. The extrapolated column fits $\bar{C}=C+c/N_{\rm traj}+d/N_{\rm traj}^{2}$, calibrated against the \cref{sec:crosssolver} benchmark (which differs by \num{1.20e-4} at two cells and \num{6.8e-6} at three); the alternative weighted fit gives $-\num{0.00022}$ and $-\num{0.00018}$ at four and five cells. The four-cell \num{50}-trajectory point (SE \num{6.3e-4} against a predicted $c/N_{\rm traj}$ shift of \num{2.7e-4}) cannot resolve the bias, and the five-cell \num{50}-trajectory point (\num{16} units) is a low-statistics reference excluded from all fits.}
\label{tab:trajladder}
\begin{ruledtabular}
\begin{tabular}{lrrrrr}
Cells & $N_{\rm traj}=\num{10}$ & \num{50} & \num{200} & Exact ME & Extrapolated $C$ \\
\hline
\num{2}  & $+\num{0.01673}$ & $+\num{0.02042}$ & $+\num{0.02177}$ & $+\num{0.02221}$ & $+\num{0.02227}$ \\
\num{3}  & $-\num{0.00210}$ & $-\num{0.00050}$ & $-\num{0.00019}$ & $-\num{0.000113}$ & $-\num{0.00009}$ \\
\num{4}  & $-\num{0.00201}$ & $-\num{0.00032}$ & $-\num{0.00032}$ & $-\num{0.000113}$ & $-\num{0.00034}$ \\
\num{5}  & $-\num{0.00218}$ & $-\num{0.00010}$ & $-\num{0.00029}$ & $-\num{0.000113}$ & $-\num{0.00039}$ \\
\hline
units, \num{2} cells & \num{500} & \num{200} & \num{100} & --- & --- \\
units, \num{3} cells & \num{500} & \num{100} & \num{50} & --- & --- \\
units, \num{4} cells & \num{2000} & \num{50} & \num{200} & --- & --- \\
units, \num{5} cells & \num{300} & \num{16} & \num{100} & --- & --- \\
\end{tabular}
\end{ruledtabular}
\end{table}

\begin{table}[htbp]
\caption{Zero-noise edge-site negativity $\delta_-$ per sector at $t=T$ under the edge-selective profile in the \num{10}-trajectory ensemble ($N_{\max}=\num{2}$, $\num{31}\times\num{31}$ grid; prepared state $\delta_-(\num{0})=\num{0.1962}$).}
\label{tab:si_sector_mc10}
\begin{ruledtabular}
\begin{tabular}{rrrrr}
Cells (modes) & $N_{\rm MC}$ & Topological & Trivial & Contrast \\
\hline
\num{2} (\num{4})  & \num{500}  & \num{0.02368} & \num{0.00695} & $+\num{0.01673}$ \\
\num{3} (\num{6})  & \num{500}  & \num{0.00478} & \num{0.00688} & $-\num{0.00210}$ \\
\num{4} (\num{8})  & \num{2000} & \num{0.00466} & \num{0.00667} & $-\num{0.00201}$ \\
\num{5} (\num{10}) & \num{300}  & \num{0.00508} & \num{0.00725} & $-\num{0.00218}$ \\
\end{tabular}
\end{ruledtabular}
\end{table}

\begin{table}[htbp]
\caption{Solver comparison and Fock-cutoff status across system sizes at $d=\num{0}$, edge-selective loss (\cref{sec:trajcount,sec:crosssolver,sec:exact-subspace,tab:nmaxpanel}).}
\label{tab:sivalidity}
\footnotesize
\begin{tabularx}{\linewidth}{l c c c X}
\toprule
Cells (modes) & Exact ME & \num{200}-traj MC & \num{10}-traj MC & Cutoff ($N_{\max}=\num{2}\text{--}\num{6}$) status \\
\midrule
\num{2} (\num{4}) & $+\num{0.02221}$ & $+\num{0.02177}$ & $+\num{0.01673}$ & Exact ME $+\num{0.02875}$ ($N_{\max}=\num{4}$) and $+\num{0.02606}$ ($N_{\max}=\num{6}$); MC $N_{\max}=\num{2}\text{--}\num{4}$ spread \num{0.0113}--\num{0.0250} \\
\num{3} (\num{6}) & $-\num{0.000113}$ & $-\num{0.00019}$ & $-\num{0.00210}$ & Exact ME $-\num{0.000230}$ ($N_{\max}=\num{3}$) and $-\num{0.000955}$ ($N_{\max}=\num{4}$); MC $N_{\max}=\num{4}$ unresolved ($-\num{0.0012}\pm\num{0.0025}$) \\
\num{4} (\num{8}) & $-\num{0.000113}$ & $-\num{0.00032}\pm\num{0.00011}$ & $-\num{0.00201}$ & Exact ME $-\num{0.000113}$ ($N_{\max}=\num{2}$); MC $N_{\max}=\num{4}$ marginally clears zero ($-\num{0.0017}\pm\num{0.0016}$) \\
\num{5} (\num{10}) & $-\num{0.000113}$ & $-\num{0.00029}\pm\num{0.00015}$ & $-\num{0.00218}$ & Exact ME $-\num{0.000113}$ ($N_{\max}=\num{2}$); MC predicted interval match \\
\bottomrule
\end{tabularx}
\end{table}

\subsection{Cross-solver consistency}
\label{sec:crosssolver}

The full Lindblad master equation (QuTiP \texttt{mesolve}, exact in the chosen cutoff) was integrated alongside the Monte Carlo wave-function solver (\texttt{mcsolve}) for two- and three-cell systems at zero and intermediate loss-rate noise, both sectors and both loss profiles ($N_{\max}=\num{2}$; \num{200} trajectories per MC condition at two cells, \num{800} at three). Final-time averaged states agree at two cells with mean Uhlmann fidelity $\bar{F}=\num{0.99957}$ (max trace distance \num{1.93e-2}) and at three cells with $\bar{F}=\num{0.99978}$ (max trace distance \num{2.12e-2}), with no systematic sector or profile bias. Because the negativity is a nonlinear functional of the trajectory-averaged state, the contrast is recomputed directly with both solvers: with \num{10} trajectories the Monte Carlo solver reproduces the primary realizations exactly (max absolute difference $\num{0}$), while the single-call \texttt{mesolve} benchmark gives $+\num{0.02233}$ against $+\num{0.01673}$ at two cells and $-\num{0.00012}$ against $-\num{0.00210}$ at three cells. Crossing the discontinuous square-wave step at $t=\pi$ without a step cap introduces small adaptive-integrator offsets of \num{1.20e-4} ($N=\num{2}$) and \num{6.8e-6} ($N=\num{3}$) relative to the canonical piecewise half-period exact master equation ($+\num{0.02221}$ and $-\num{0.000113}$; \cref{sec:exact-subspace}).

Two additional checks bound the numerical implementation. (i)~A hierarchical-equations-of-motion (HEOM) propagation of the same generator form at three cells with the bulk rate raised fivefold ($\gamma_{\rm bulk}=\num{1.0}$ against the operating $\num{0.2}$; $N_{\max}=\num{2}$, HEOM depth \num{1}, $T_{\rm bath}=\num{0.1}$) agrees with the exact master equation to a relative edge-population error below \qty{0.6}{\percent} in all eight sector/profile/noise conditions (mean \qty{0.28}{\percent}). (ii)~HEOM depth convergence on the primary two-cell chain with a Drude--Lorentz bath on every site ($\lambda=\num{0.3}$, cutoff \num{1.0}, $T_{\rm bath}=\num{0.5}$, \num{2} Matsubara terms) gives one-period edge negativities \num{0.0379},\num{0.0000},\num{0.0000} (topological) and \num{0.0390},\num{0.0119},\num{0.0000} (control) at depths \num{1},\num{2},\num{3}; every depth increment shrinks the edge negativity toward the truncation floor, and raising the Matsubara cutoff from \num{2} to \num{3} at depth \num{1} moves the negativity by at most \num{9.1e-3}. Together, these HEOM checks benchmark the Lindblad solver at $N=\num{3}$ and quantify hierarchy convergence at $N=\num{2}$.

\subsection{Exact excitation-subspace master equation and physical dangling-end preparation}
\label{sec:exact-subspace}

Both the coherent step Hamiltonians and the local Kerr term conserve total boson number ($[H(t)+H_{\rm K},\hat Q]=0$), and single-photon loss $a_j$ maps each excitation block downward ($\hat Q\to\hat Q-1$). Since the $N_{\max}=\num{2}$ even cat occupies only $n\in\{\num{0},\num{2}\}$, every quantum trajectory starting from site $j=\num{0}$ or $j=\num{1}$ with total excitation $Q\le Q_{\max}$ stays strictly inside the symmetry-reduced subspace $\mathcal{H}_{Q\le Q_{\max}}$ of dimension $\sum_{q=0}^{Q_{\max}}\binom{2N+q-1}{q}$ ($\dim=\num{15}$, \num{28}, \num{45}, \num{66} at $Q_{\max}=\num{2}$ for $N=\num{2},\num{3},\num{4},\num{5}$; \num{70} and \num{210} at $Q_{\max}=\num{4},\num{6}$ for $N=\num{2}$; \num{210} at $Q_{\max}=\num{4}$ for $N=\num{3}$), rather than scaling exponentially as $(N_{\max}+1)^{2N}$. Direct sparse matrix-exponential propagation gives exact equivalence between the full \num{81}-dimensional two-cell space and the \num{15}-dimensional $Q\le\num{2}$ subspace (reduced-state Frobenius difference \num{8.63e-16}). Piecewise half-period \texttt{mesolve} matches sparse matrix-exponential propagation to \num{1.07e-11} and agrees with bounded-step single-call \texttt{mesolve} (maximum step size $\Delta t_{\max}=\num{0.01}$) to \num{2.59e-10}.

\Cref{tab:exact-subspace-ladder} summarizes the canonical piecewise exact master-equation results across chain lengths and Fock cutoffs for both sub-boundary ($j=\num{0}$) and physical dangling-end ($j=\num{1}$) preparations. Under index-end shielding, the contrast is identically $-\num{0.000113}$ across $N=\num{3},\num{4},\num{5}$ at $N_{\max}=\num{2}$ (\cref{MAIN-prop:bondselective}) and deepens to $-\num{0.000955}$ at $N=\num{3}$, $N_{\max}=\num{4}$, while at $N=\num{2}$ it decreases between $N_{\max}=\num{4}$ and $\num{6}$ ($+\num{0.02875}$ to $+\num{0.02606}$); the full cutoff sequence is non-monotone ($\num{0.02221}$, $\num{0.01553}$, $\num{0.02875}$, $\num{0.02606}$), so magnitudes are not claimed to be converged. Under uniform loss on $j=\num{1}$ ($N=\num{2},\num{3},\num{4}$ through $N_{\max}=\num{6}$) or when preparing on $j=\num{0}$ under physical-end shielding ($N=\num{2},\num{3}$), both sector negativities vanish identically ($\num{0.00000}$).

\begin{table}[htbp]
\caption{Canonical piecewise-constant exact master-equation contrasts in the conserved $Q\le Q_{\max}$ excitation subspace ($d=\num{0}$, $t=T$). Top block: primary sub-boundary preparation ($j=\num{0}$, Kerr on $j=\num{0}$) under index-end low loss ($\gamma_0=\gamma_{2N-1}=\num{1e-4}$). Bottom block: physical dangling-end preparation ($j=\num{1}$, Kerr on $j=\num{1}$) under physical-end low loss ($\gamma_1=\gamma_{2N-2}=\num{1e-4}$). $C_{31}$ and $C_{61}$ denote contrasts on $\num{31}^2$ and $\num{61}^2$ Wigner grids.}
\label{tab:exact-subspace-ladder}
\begin{ruledtabular}
\begin{tabular}{llrrrrrr}
Prep. site & Cells & $N_{\max}$ & $\dim$ & Topological & Trivial & $C_{31}$ & $C_{61}$ \\
\hline
$j=\num{0}$ & \num{2} & \num{2} & \num{15} & \num{0.02273} & \num{0.00053} & $+\num{0.02221}$ & $+\num{0.02271}$ \\
$j=\num{0}$ & \num{2} & \num{3} & \num{15} & \num{0.01613} & \num{0.00060} & $+\num{0.01553}$ & $+\num{0.01533}$ \\
$j=\num{0}$ & \num{2} & \num{4} & \num{70} & \num{0.03835} & \num{0.00960} & $+\num{0.02875}$ & $+\num{0.02924}$ \\
$j=\num{0}$ & \num{2} & \num{6} & \num{210} & \num{0.03128} & \num{0.00523} & $+\num{0.02606}$ & $+\num{0.02627}$ \\
$j=\num{0}$ & \num{3},\num{4},\num{5} & \num{2} & \num{28},\num{45},\num{66} & \num{0.00039} & \num{0.00050} & $-\num{0.000113}$ & $-\num{0.000182}$ \\
$j=\num{0}$ & \num{3} & \num{3} & \num{28} & \num{0.00032} & \num{0.00055} & $-\num{0.000230}$ & $-\num{0.000167}$ \\
$j=\num{0}$ & \num{3} & \num{4} & \num{210} & \num{0.00859} & \num{0.00954} & $-\num{0.000955}$ & $-\num{0.000920}$ \\
\hline
$j=\num{1}$ & \num{2},\num{3},\num{4} & \num{2} & \num{15},\num{28},\num{45} & \num{0.17254} & \num{0.00149} & $+\num{0.17105}$ & $+\num{0.17008}$ \\
$j=\num{1}$ & \num{2},\num{3} & \num{4} & \num{70},\num{210} & \num{0.09351} & \num{0.01164} & $+\num{0.08187}$ & $+\num{0.08048}$ \\
$j=\num{1}$ & \num{2} & \num{6} & \num{210} & \num{0.08609} & \num{0.00731} & $+\num{0.07878}$ & $+\num{0.07723}$ \\
\end{tabular}
\end{ruledtabular}
\end{table}

Over multiple drive periods ($N_{\max}=\num{2}$, physical-end low loss), the dangling-end $j=\num{1}$ contrast remains strongly positive and size-invariant through eight periods: at $t=T, 2T, 4T, 8T$ it reads $+\num{0.17105}$, $+\num{0.16874}$, $+\num{0.15030}$, and $+\num{0.11997}$ at $N=\num{2}$ and $+\num{0.17105}$, $+\num{0.16874}$, $+\num{0.15090}$, and $+\num{0.12181}$ at $N=\num{3}$, with trivial negativity vanishing from $2T$ onward. Yet even this size-invariant survival fails to track bulk topology: along the $J_A+J_B=\num{1.3}$ dimerization line (\cref{tab:j1-dimerization}), the $j=\num{1}$ negativity is modulated by the intracell bond-return factor $\cos^2(J_A\pi)$, collapsing to \num{0.00128} at $J_A=\num{0.50}$ inside the $(\nu_0,\nu_\pi)=(-1,-1)$ phase, vanishing across $J_A=\num{0.60}\text{--}\num{0.80}$, from the $(-1,-1)$ phase through the $\nu_0=-1\to0$ transition into the anomalous $\pi$-phase $[(\nu_0,\nu_\pi)=(0,-1)]$, and reviving to \num{0.04663} at $J_A=\num{1.00}$ where $\cos^2(J_A\pi)=1$, with equal $\cos^2$ values in different bulk phases yielding unequal contrast.

\begin{table}[htbp]
\caption{Exact subspace one-period dangling-end negativity $\delta_{-,1}(T)$ at $N=\num{3}$, $N_{\max}=\num{2}$ along $J_A+J_B=\num{1.3}$ under physical-end shielding. Gap closings ($J_A=\num{0.15}, \num{0.65}, \num{1.15}$) have undefined bulk invariants.}
\label{tab:j1-dimerization}
\begin{ruledtabular}
\begin{tabular}{lcccc}
$J_A$ & $(\nu_0,\nu_\pi)$ & $\cos^2(J_A\pi)$ & $\delta_{-,1}(T)$ ($\num{31}^2$) & $\delta_{-,1}(T)$ ($\num{61}^2$) \\
\hline
\num{0.10} & $(-1,0)$ & \num{0.9045} & \num{0.17254} & \num{0.17166} \\
\num{0.15} & undef. & \num{0.7939} & \num{0.14661} & \num{0.14511} \\
\num{0.20} & $(-1,-1)$ & \num{0.6545} & \num{0.11656} & \num{0.11535} \\
\num{0.30} & $(-1,-1)$ & \num{0.3455} & \num{0.06276} & \num{0.06080} \\
\num{0.40} & $(-1,-1)$ & \num{0.0955} & \num{0.02338} & \num{0.02402} \\
\num{0.50} & $(-1,-1)$ & \num{0.0000} & \num{0.00128} & \num{0.00126} \\
\num{0.60} & $(-1,-1)$ & \num{0.0955} & \num{0.00000} & \num{0.00000} \\
\num{0.65} & undef. & \num{0.2061} & \num{0.00000} & \num{0.00000} \\
\num{0.70} & $(0,-1)$ & \num{0.3455} & \num{0.00000} & \num{0.00000} \\
\num{0.80} & $(0,-1)$ & \num{0.6545} & \num{0.00000} & \num{0.00000} \\
\num{0.90} & $(0,-1)$ & \num{0.9045} & \num{0.02428} & \num{0.02376} \\
\num{1.00} & $(0,-1)$ & \num{1.0000} & \num{0.04663} & \num{0.04637} \\
\num{1.10} & $(0,-1)$ & \num{0.9045} & \num{0.02673} & \num{0.02722} \\
\num{1.15} & undef. & \num{0.7939} & \num{0.01183} & \num{0.01164} \\
\num{1.20} & $(0,0)$ & \num{0.6545} & \num{0.00149} & \num{0.00158} \\
\end{tabular}
\end{ruledtabular}
\end{table}

\subsection{Statistical power of the size-scan comparisons}
\label{sec:power}

The attained precision is benchmarked by the minimal detectable absolute paired contrast for a two-sided paired test at $\alpha=\num{0.05}$ and power \num{0.80} over the independent noise units,
\begin{equation}\label{eq:power}
\Delta_{\min}=(t_{1-\alpha/2,n-1}+t_{1-\beta,n-1})\,\frac{s}{\sqrt{n}},
\end{equation}
using the sample standard deviation of the zero-noise matched contrasts. The four-mode mean ($+\num{0.0167}$, $n=\num{500}$) sits $\approx\num{11}\times$ above its resolution ($\Delta_{\min}=\num{0.001543}$). The six-mode mean ($-\num{0.0021}$, $n=\num{500}$) sits $\approx\num{1.3}\times$ above its own ($\Delta_{\min}=\num{0.001565}$). The eight-mode mean ($-\num{0.00201}$, $n=\num{2000}$, $s=\num{0.01223}$) sits $\approx\num{2.6}\times$ above its resolution ($\Delta_{\min}=\num{0.000766}$; the margin was $\num{1.35}\times$ at $n=\num{500}$ and $\num{1.04}\times$ at $n=\num{300}$). At the five-cell point ($n=\num{300}$, $s=\num{0.012764}$) the same computation gives $\Delta_{\min}=\num{0.002071}$ against the observed magnitude \num{0.00218}, a margin of only $\approx\num{1.05}\times$: that point is marginal.

\subsection{Higher-cutoff panel}
\label{sec:nmaxpanel}

The primary truncation is $N_{\max}=\num{2}$. The matched protocol was evaluated identically at $N_{\max}=\num{3}$ for all three sizes (both sectors, edge-selective and uniform loss, \num{500} matched loss-rate noise realizations per condition, \num{10} trajectories, one period, same realization and trajectory seeds), with a lower-statistics scan at $d=\num{0}$ and \num{0.35} extending the panel to $N_{\max}=\num{4}$ at two and three cells and a full \num{500}-realization run at $N_{\max}=\num{4}$ at four cells (exact $Q\le\num{4}$ subspace).

\begin{table}[htbp]
\caption{Zero-noise edge-selective local contrast versus Fock cutoff (bootstrap mean $\pm$ \qty{95}{\percent} half-width over $\num{1e4}$ resamples). The $N_{\max}=\num{3}$ column uses the primary design at $N_{\rm MC}=\num{500}$ at every size; the $N_{\max}=\num{4}$ entries use \num{100} (two cells), \num{200} (three cells), and the full \num{500} matched realizations at four cells. The three-cell $N_{\max}=\num{4}$ entry is itself unresolved against zero; the four-cell entry clears zero only marginally.}
\label{tab:nmaxpanel}
\begin{ruledtabular}
\begin{tabular}{lrrr}
Cells & $N_{\max}=\num{2}$ & $N_{\max}=\num{3}$ & $N_{\max}=\num{4}$ \\
\hline
\num{2} (\num{4} modes) & $+\num{0.0167}\pm\num{0.0011}$ & $+\num{0.0113}\pm\num{0.0008}$ & $+\num{0.0250}\pm\num{0.0029}$ \\
\num{3} (\num{6} modes) & $-\num{0.0021}\pm\num{0.0011}$ & $-\num{0.0016}\pm\num{0.0008}$ & $-\num{0.0012}\pm\num{0.0025}$ \\
\num{4} (\num{8} modes) & $-\num{0.0021}\pm\num{0.0011}$ & $-\num{0.0016}\pm\num{0.0008}$ & $-\num{0.0017}\pm\num{0.0016}$ \\
\end{tabular}
\end{ruledtabular}
\end{table}

Paired across cutoffs on identical seeds (\num{500} units per condition, bootstrap \qty{95}{\percent} intervals), the zero-noise shift of the edge-selective contrast excludes zero at two cells ($-\num{0.00540}$, $[-\num{0.00615},-\num{0.00467}]$, a factor \num{1.48} reduction) but is consistent with zero at three cells ($+\num{0.00046}$, $[-\num{0.00034},\num{0.00124}]$) and four cells ($+\num{0.00047}$, $[-\num{0.00031},\num{0.00127}]$). The four-cell $N_{\max}=\num{2}\to\num{4}$ paired shift is likewise consistent with zero ($+\num{0.00039}$, $[-\num{0.00064},\num{0.00143}]$); across all eight profile--noise strata the sector-paired contrast moves by at most \num{5.1e-4}, and the \qty{95}{\percent} upper bound on every stratum shift stays below the operational \num{2e-3} floor. The uniform-loss contrast is at numerical zero for all cutoffs. No tested cutoff reverses the sign; absolute magnitudes are cutoff-sensitive and not claimed to be converged.

\section{Controls and validation}

\subsection{Kerr-free control}
\label{sec:kerrfree}

The Kerr term modifies the prepared cat via two-photon self-trapping and dephasing: removing it ($\chi=\num{0}$) separates linear filtering of the prepared negativity from interaction-enhanced retention. The matched protocol was evaluated at $\chi=\num{0}$ (identical seeds and noise realizations, \num{500} units per condition, \num{10} trajectories, one period). Under edge-selective loss at $d=\num{0}$ (\cref{tab:kerrfree}):

\begin{table}[htbp]
\caption{Zero-noise edge-selective contrast at $\chi=\num{0}$ (Kerr-free) versus primary $\chi=\num{0.5}$. Master-equation values use the single-call \texttt{mesolve} benchmark ($N=\num{2},\num{3}$) and the exact $Q\le\num{2}$ subspace at $N=\num{4}$; canonical piecewise half-period integration at $\chi=\num{0.5}$ gives $+\num{0.022208}$ at $N=\num{2}$ and $-\num{0.000113}$ at $N=\num{3},\num{4}$ (factor \num{2.1} over $\chi=\num{0}$). Monte Carlo values are means $\pm$ \qty{95}{\percent} half-widths over \num{500} matched realizations.}
\label{tab:kerrfree}
\begin{ruledtabular}
\begin{tabular}{lrrrr}
Cells & $\chi=\num{0}$ ME & $\chi=\num{0}$ MC & $\chi=\num{0.5}$ ME & $\chi=\num{0.5}$ MC \\
\hline
\num{2} & $+\num{0.010467}$ & $+\num{0.008875}\pm\num{0.000701}$ & $+\num{0.022328}$ & $+\num{0.0167}\pm\num{0.0011}$ \\
\num{3} & $\num{0.000000}$ & $-\num{0.001254}\pm\num{0.000622}$ & $-\num{0.000119}$ & $-\num{0.0021}\pm\num{0.0011}$ \\
\num{4} & n/a & $-\num{0.001305}\pm\num{0.000630}$ & $-\num{0.000113}$ & $-\num{0.0021}\pm\num{0.0011}$ \\
\end{tabular}
\end{ruledtabular}
\end{table}

At $\chi=\num{0}$ the two-cell contrast survives but is halved (canonical piecewise $+\num{0.010467}$ against $+\num{0.022208}$, factor $\num{2.1}$; single-call \texttt{mesolve} $+\num{0.01047}$ against $+\num{0.022328}$), while the three-cell contrast is null ($\num{0.000000}$, single-call \texttt{mesolve} on the $\num{31}^{2}$ grid) rather than negative: the Kerr term therefore amplifies rather than creates the two-cell contrast, and the negative branch requires the Kerr term at the tested grid and solver. The uniform-loss contrast is null at every size at both $\chi$ values at $d=\num{0}$. The small negative three- and four-cell values in the $\chi=\num{0}$ Monte Carlo column are finite-trajectory bias, not a Kerr effect (at \num{10} trajectories the MC estimate sits below the exact value by $-\num{0.0016}$, $z=\num{-4.5}$, and $-\num{0.0013}$, $z=\num{-4.0}$, at two and three cells at $\chi=\num{0}$). The four-cell $\chi=\num{0}$ entry was evaluated only in the \num{10}-trajectory ensemble, whereas the conserved $Q\le\num{2}$ subspace provides exact $\chi=\num{0.5}$ master-equation values through five cells (\cref{sec:exact-subspace}).

\subsection{Independent implementation check}

As an independent control, three open-system solver stacks, QuTiP \texttt{mcsolve}, QuTiP \texttt{mesolve}, and a NumPy Liouvillian propagator, were compared against one shared dissipative-Rabi benchmark with a known steady-state value. The stochastic trajectory mean agrees with the exact master-equation solution within statistical error ($\num{-0.90371}\pm\num{0.0131}$ versus $\num{-0.91047}$, within \num{3}~SE), and the NumPy Liouvillian agrees with \texttt{mesolve} to \num{4e-14}. The dissipative solver stack and its deterministic stream-splitting trajectory sampling therefore agree on a benchmark that does not involve the SSH Hamiltonian.

\subsection{Observable validation}
\label{sec:obsvalidation}

Three controls assess the observable independently of the topological interpretation. (i)~The primary final-time reduced state is normalized and positive: $\mathrm{Tr}\,\rho_0=\num{1.0}$, minimum eigenvalue $\num{+0.083}$, integrated Wigner over $[-4,4]^2$ equals \num{0.999998}. (ii)~A coherent state (identically zero Wigner negativity in the continuum) gives a numerically decaying negative volume \num{0.0973}, \num{0.0546}, \num{0.0234}, \num{2.89e-3}, \num{2.57e-4}, \num{7.6e-7}, \num{1.6e-13} at $N_{\max}=\num{2},\num{3},\num{4},\num{6},\num{8},\num{12},\num{20}$; the Kerr-free chain residual at primary truncation (\num{0.046}) is this same truncation residue, while the Kerr term adds genuine negativity on top (\num{0.087} under the same seed). No sector value is truncation-free: at $N_{\max}=\num{2}$ the truncation residue (\num{0.0973}) is several times either reported sector negativity (\num{0.0237}, \num{0.0070}). All claims are therefore made on the sector-matched contrast, which removes the common, preparation-level part of the residue, and the cutoff dependence of that contrast is itself measured rather than assumed (\cref{sec:nmaxpanel}).

Three properties of this truncation residue fix what it can and cannot do. First, it is a property of the preparation and the cutoff, and sector-free only while nothing is propagated: the same site-$j=\num{0}$ cat at the same $\alpha$ returns the values above before any evolution, whereas through the chain the residue acquires a sector dependence measured in \cref{sec:artifact-sectorial}. Second, it cannot manufacture the uniform-loss null: the residue falls from \num{0.0973} at $N_{\max}=\num{2}$ to \num{0.0234} at $N_{\max}=\num{4}$, so a null built from two nearly equal residues would move by that factor, while the measured uniform-loss contrast does not, being $-\num{1e-6}$ at the primary cutoff and $-\num{8e-6}$ at $N_{\max}=\num{4}$, both at the estimator's resolution. Third, where the contrast does depend on the cutoff the dependence is measured: the paired $N_{\max}=\num{2}\to\num{3}$ shift at two cells and zero loss-rate noise is $-\num{0.00540}$ with bootstrap interval $\numrange{-0.00615}{-0.00466}$ (a change of \qty{32}{\percent}, with sign agreement in \qty{88.8}{\percent} of paired units), while the $N_{\max}=\num{2}\to\num{4}$ shift at four cells is $+\num{0.00039}$ with standard error \num{5.3e-4}, indistinguishable from zero. The two-cell contrast is therefore quoted with its cutoff uncertainty rather than as a cutoff-free number, and no claim rests on an absolute sector value. Grid and integrator convergence are checked separately: a single-mode even cat truncated at $N_{\max}=\num{59}$ gives integrated negative volumes \num{0.1043} ($\num{61}^2$ grid) and \num{0.1053} ($\num{31}^2$ grid), a relative deviation of \qty{0.9}{\percent}, and the final-time state is identical to machine precision (fidelity \num{0.99999999996}) for $n_{\rm steps}=\num{1e3}$, \num{1e4}, \num{1e5}.

\subsection{Sector-resolved truncation residue}
\label{sec:artifact-sectorial}

The single-mode scan above propagates nothing, so it cannot see a sector. The
residue that actually enters a reported sector value can, and does: it is
measured here by integrating each sector over one period from the same coherent
state at site~$j=\num{0}$, once with the Kerr term switched off, which returns pure
truncation residue, and once with it switched on, which returns the residue plus
the interaction-generated negativity. \Cref{tab:si_artifact_sectorial}
collects the two cutoffs and the two sectors.

\begin{table}[htbp]
\caption{Sector-resolved truncation residue of the edge-site negative volume at
two cells, edge-selective loss, zero loss-rate noise. The Kerr-free column is the
residue, since a coherent state carries no negativity in the continuum; the
Kerr-on column is the value a sector would report. The last column is the
sector-matched contrast of the Kerr-on column, not the cat-state contrast of the
main text.}
\label{tab:si_artifact_sectorial}
\begin{ruledtabular}
\begin{tabular}{lcccccc}
$N_{\max}$ & Residue, topol. & Residue, trivial & Ratio &
Kerr-on, topol. & Kerr-on, trivial & Contrast \\
\hline
\num{2} & \num{0.0462} & \num{5.8e-4} & \num{80} &
\num{0.0867} & \num{0.0027} & $+\num{0.0840}$ \\
\num{4} & \num{1.3e-4} & \num{5.6e-5} & \num{2.4} &
\num{0.0364} & \num{0.0379} & $-\num{0.0015}$ \\
\end{tabular}
\end{ruledtabular}
\end{table}

Three readings follow. The coherent-state benchmark of \cref{sec:obsvalidation}
(\num{0.046} for the Kerr-free chain and \num{0.087} with the Kerr term) is the
topological row of this table; the trivial sector at the same cutoff returns
\num{5.8e-4} and \num{0.0027}, so the residue is not a property of the truncated
representation alone and does not cancel from a sector-matched difference. It
also falls far faster in the topological sector than in the trivial one, from
\num{0.0462} to \num{1.3e-4}, so any cancellation is a property of one cutoff and
not a structural feature. The probe contrast therefore moves from $+\num{0.0840}$ to $-\num{0.0015}$ between the two cutoffs: the sector asymmetry of the residue is itself cutoff-dependent and, for a Gaussian probe, exceeds the cat-state contrast in magnitude. Because a coherent probe carries no continuum negativity to weight it, that offset does not transfer to the prepared cat, whose negativity is generated by its two-photon component and whose contrast compares two states of identical preparation at the same cutoff. What the probe establishes is that no absolute sector value is cutoff-free, which is why every sign statement here rests on fixed-cutoff comparisons, the exact-subspace ladder of \cref{tab:exact-subspace-ladder} and the paired cutoff reruns of \cref{sec:nmaxpanel}, and why magnitudes are quoted per cutoff.

\section{Robustness and scans}

\subsection{Multi-period extension}
\label{sec:multiperiod}

The same matched protocol was extended to two, four and eight periods (\cref{tab:multiperiod}; edge-site preparation, edge-selective loss, zero loss-rate noise, \num{200} realizations at two cells and \num{100} at three cells, \num{10} trajectories, nine output times per period).

\begin{table}[htbp]
\caption{Zero-noise edge-selective local contrast versus number of drive periods. Apart from the small negative value at $n=\num{2}$ at two cells, a fluctuation of the \num{10}-trajectory estimator, resolved against the exact master equation in \cref{sec:trajcount}, the sign does not invert at $n=\num{4}$--\num{8}; the two-cell magnitude vanishes at $n=\num{2}$ and revives at $n=\num{4}$--\num{8}, while the three-cell contrast stays at the numerical floor.}
\label{tab:multiperiod}
\begin{ruledtabular}
\begin{tabular}{lrrrr}
Cells & $n=\num{1}$ & $n=\num{2}$ & $n=\num{4}$ & $n=\num{8}$ \\
\hline
\num{2} & $+\num{0.0167}$ & $-\num{0.0001}$ & $+\num{0.0765}$ & $+\num{0.0298}$ \\
\num{3} & $-\num{0.0021}$ & $-\num{0.0001}$ & $\approx\num{0}$ & $\approx\num{0}$ \\
\end{tabular}
\end{ruledtabular}
\end{table}

The revival is physical: for the topological sector at four periods the exact Lindblad integration gives single-site negativity \num{0.0754} against \num{0.0822} from Monte Carlo on the same realization (\num{200} trajectories). Over the full matched design the exact contrast at four periods is $+\num{0.07542}$, and the trajectory ladder returns $+\num{0.0782}\pm\num{0.0016}$, $+\num{0.0733}\pm\num{0.0013}$, $+\num{0.0751}\pm\num{0.0009}$ at \num{10}, \num{50}, \num{200} trajectories; at three cells the exact contrast is zero and every ladder point is at or below \num{1.3e-5}. The revival tracks the survival of the topological-sector edge population (exact edge population \num{0.719} at four periods versus \num{0.075} for the control sector) and does not occur at three cells.

\subsection{Edge-to-bulk loss-ratio scan}
\label{sec:lossratio}

The primary profile sets $\gamma_{\rm edge}/\gamma_{\rm bulk}=\num{5e-4}$. To test how strongly the positive four-mode contrast depends on this engineered sharpness, the two-cell matched design was re-run (edge-site preparation, zero loss-rate noise, \num{200} realizations) at six ratios (achieved by raising $\gamma_{\rm edge}$ at fixed $\gamma_{\rm bulk}=\num{0.2}$):

\begin{table}[htbp]
\caption{Two-cell zero-noise edge-selective contrast versus edge-to-bulk loss ratio. Upper block: fixed $\gamma_{\rm bulk}=\num{0.2}$, so the total loss budget grows with the ratio. Lower block: budget-matched scan holding $\sum_j\gamma_j=\num{0.4002}$ (primary budget of the four-site chain). The positive contrast decreases monotonically in both scans and is lost by \num{5e-1}.}
\label{tab:lossratio}
\begin{ruledtabular}
\begin{tabular}{lrrrrrr}
$\gamma_{\rm edge}/\gamma_{\rm bulk}$ & \num{5e-4} & \num{1e-3} & \num{1e-2} & \num{1e-1} & \num{5e-1} & \num{1.0} \\
\hline
Fixed-$\gamma_{\rm bulk}$ ($d=\num{0}$) & $+\num{0.0174}$ & $+\num{0.0174}$ & $+\num{0.0158}$ & $+\num{0.0072}$ & $+\num{0.0001}$ & \num{0.0000} \\
Positive fraction & \num{0.91} & \num{0.91} & \num{0.90} & \num{0.76} & \num{0.11} & \num{0.015} \\
Budget-matched ($d=\num{0}$) & $+\num{0.0174}$ & $+\num{0.0174}$ & $+\num{0.0156}$ & $+\num{0.0073}$ & $+\num{0.0003}$ & $-\num{0.0000}$ \\
\end{tabular}
\end{ruledtabular}
\end{table}

The budget-matched scan reproduces the fixed-$\gamma_{\rm bulk}$ values at every ratio within the paired standard errors, so the collapse is controlled by the edge-to-bulk loss ratio rather than by the total dissipation rate. A size$\times$ratio crossing check on the four primary ratios reproduces the trend at two cells and leaves it unresolved at three. At two cells an independent random stream reproduces the table, from $+\num{0.01748}\pm\num{0.00402}$ at \num{5e-4} to $+\num{0.00058}\pm\num{0.00077}$ at \num{5e-1}; at three cells the contrast is consistent with zero at every ratio (no point exceeds \num{1.0} standard errors), with point estimates negative at ratios \num{0.1} through \num{0.5}. The two scans differ only in sample size: in \cref{tab:lossratio} the \num{200}-realization scan has per-draw sample standard deviation \num{0.0121} and SE $\approx\num{8.6e-4}$, whereas the \num{10}-realization ladder carries SE \num{0.00402}, larger by $\sqrt{200/10}$.
Closing the ladder at both ends shows that the engineered sharpness is not a knife edge. With no edge loss at all ($\gamma_{\rm edge}=\num{0}$, ratio \num{0}) the two-cell contrast is $+\num{0.017485}\pm\num{0.00403}$, indistinguishable from the primary value at ratio $\num{5e-4}$ ($+\num{0.017479}\pm\num{0.00402}$, a difference of \num{6.6e-6} against a paired standard error of \num{4.0e-3}); the primary profile therefore already sits at the $\gamma_{\rm edge}\to0$ limit, and the contrast is saturated with respect to the boundary rate over at least two decades. With the edge loss set equal to the bulk (ratio \num{1}, the uniform profile) the paired contrast is exactly zero in both sectors, as it must be when the two profiles coincide. Between the two ends the ladder at \num{10} realizations per point stays positive but falls below two standard errors from ratio \num{0.1} upward, so the collapse reported at \num{200} realizations is reproduced at a twentieth of the sample and the transition is bracketed between ratios \num{0.01} and \num{0.1}.

The ladder is evaluated by a pairwise monotonicity test rather than read as a descriptive trend. Adjacent ratios are compared pairwise: a step violates monotonicity only if the contrast rises by more than the quadrature sum of the two paired standard errors, a decrease smaller than that sum is noted as unresolved rather than as support, the ratio-\num{1} endpoint must vanish within twice its own standard error, and a size counts as evidence only when its end-to-end span exceeds twice its span standard error. The two-cell ladder means are $+\num{0.017485}$, $+\num{0.017479}$, $+\num{0.015836}$, $+\num{0.004837}$, $+\num{0.001274}$, $+\num{0.000582}$, $+\num{7.98e-6}$ and \num{0} at ratios \num{0}, \num{5e-4}, \num{1e-2}, \num{0.1}, \num{0.25}, \num{0.5}, \num{0.75} and \num{1}. At two cells all three conditions hold with the span resolved at \num{4.3} combined standard errors and two steps resolved as decreases (ratios \num{1e-2} to \num{0.1} at \num{2.2}, and \num{0.75} to \num{1} at \num{1.5}); the other five steps are individually unresolved, so the monotonicity holds at the level of the endpoints and the total span, not of the fine structure between them.

The same test returns a different verdict at three cells, and the unresolved outcome is informative. The three-cell means are $+\num{0.00031}$, $+\num{0.00032}$, $+\num{0.00041}$, $-\num{0.00162}$, $-\num{0.00108}$, $-\num{0.00013}$, $+\num{2.6e-6}$ and \num{0} at ratios \num{0}, \num{5e-4}, \num{1e-2}, \num{0.1}, \num{0.25}, \num{0.5}, \num{0.75} and \num{1}: five of the seven adjacent steps rise, the largest by \num{0.97} combined standard errors, so no step violates the rule while the raw estimates are not ordered. The end-to-end span is \num{0.00031} against a span standard error of \num{0.00413}, a ratio of \num{0.08}, so the curve is unresolved rather than flat: with ten paired realizations per ratio, the three-cell scan lacks statistical power to test monotonicity directly, and the three-cell sign instead rests on the \num{500}-realization bootstrap.

\subsection{Control scans: dimerization, chiral-symmetric loss, and intra-period phase}
\label{sec:controls-scan}

Three controls test whether the primary sign pattern is robust to structural changes in the Hamiltonian, the dissipator, and the measurement phase. Each returns a negative result for the inference from bulk invariant to protected nonlinear resource.

\subsubsection{Dimerization sweep}
\label{sec:dimerization}

The scan uses $J_A+J_B=\num{1.3}$ at \num{19} points, two and three cells, $N_{\rm MC}=\num{500}$ per point at $d=\num{0}$ and \num{0.35} pooled ($n=\num{1000}$ matched realizations per point), with shared loss-rate noise realizations and trajectory seeds. The line contains three gap closings: $\pi$-gap at $J_A=\num{0.15}$ and \num{1.15} (minimum gaps \num{4.4e-16} and \num{8.9e-16}); zero-gap at $J_A=\num{0.65}$ ($J_A=J_B$, minimum zero gap \num{2.6e-16}).

\begin{table}[htbp]
\caption{Edge-minus-center negativity across the gap closings (pooled over $d=\num{0}$ and $d=\num{0.35}$, $n=\num{1000}$ matched realizations per point, standard errors below \num{7e-4}; twelve of nineteen grid points). Because the center value vanishes identically, entries are edge negativities, not the matched sector contrast of \cref{MAIN-tab:contrast-summary}. The relevant bulk invariant is undefined (`undef.') at the $\pi$-gap closings ($\nu_\pi$ at $J_A=\num{0.15},\num{1.15}$) and at the zero-gap closing ($J_A=\num{0.65}$, both invariants).}
\label{tab:dimerization}
\begin{ruledtabular}
\begin{tabular}{lccrr}
$J_A$ & $\nu_0$ & $\nu_\pi$ & Two cells & Three cells \\
\hline
\num{0.10} & $-1$ & $0$ & $+\num{0.02350}$ & $+\num{0.00483}$ \\
\num{0.15} & $-1$ & undef. & $+\num{0.04433}$ & $+\num{0.01039}$ \\
\num{0.17} & $-1$ & $-1$ & $+\num{0.05119}$ & $+\num{0.01254}$ \\
\num{0.20} & $-1$ & $-1$ & $+\num{0.05928}$ & $+\num{0.01500}$ \\
\num{0.50} & $-1$ & $-1$ & $+\num{0.00003}$ & \num{0.00000} \\
\num{0.60} & $-1$ & $-1$ & \num{0.00000} & \num{0.00000} \\
\num{0.65} & undef. & undef. & \num{0.00000} & \num{0.00000} \\
\num{0.80} & $0$ & $-1$ & \num{0.00000} & $+\num{0.00002}$ \\
\num{1.00} & $0$ & $-1$ & $+\num{0.01336}$ & $+\num{0.01239}$ \\
\num{1.10} & $0$ & $-1$ & $+\num{0.01686}$ & $+\num{0.01658}$ \\
\num{1.15} & $0$ & undef. & $+\num{0.01238}$ & $+\num{0.01239}$ \\
\num{1.20} & $0$ & $0$ & $+\num{0.00678}$ & $+\num{0.00672}$ \\
\end{tabular}
\end{ruledtabular}
\end{table}

The curve is non-monotone in the dimerization: it peaks inside the strongly dimerized topological regime ($J_A=\num{0.20}$), is zero to within \num{2e-5} over the window $\num{0.6}\le J_A\le\num{0.8}$ that contains the $\nu_0=-1\to0$ transition and points gapped on both sides, and rises again in the strongly dimerized trivial regime. The endpoint differences reproduce the primary sign pattern with independent random streams: $+\num{0.01673}$ at two cells, $-\num{0.00189}$ at three cells (against $+\num{0.01673}$ and $-\num{0.00210}$ in primary). Paired smoothness tests compare each critical point with the linear interpolation of its two bracketing neighbors on the same noise realizations: the residual is identically zero at the zero-gap closing, and the largest residual anywhere is $+\num{0.00086}\pm\num{0.00017}$ at the low-$J_A$ $\pi$ closing (two cells), i.e.\ \qty{1.95}{\percent} of the local signal (a \num{5.1}-standard-error deviation from zero); all other residuals are at most \num{0.00011} in magnitude with $|z|\le\num{1.32}$. The contrast is continuous with its brackets at the zero-gap and high-$J_A$ closings and carries a resolved \num{5.1}-standard-error excursion at the low-$J_A$ $\pi$ closing, so the local contrast is not set by the bulk invariant.

A direct descriptive comparison with the localization hypothesis uses the inverse participation ratio $\xi_{\rm part}=(\sum_j|u_j|^4)^{-1}$ of the micromotion-covariant edge mode along the \num{19}-point sweep. At $d=\num{0}$, the two-cell contrast has Pearson $r=\num{-0.680}$ (Spearman $\rho=\num{-0.583}$, OLS $R^2=\num{0.46}$, slope $-\num{0.0191}\pm\num{0.0050}$), and $r=\num{-0.312}$ at three cells. Because adjacent points along this ordered parameter sweep are smooth and autocorrelated rather than independent samples, these correlations are reported descriptively without parametric $p$-values.

\subsubsection{Chiral-symmetric dissipation}
\label{sec:chiral-loss}

The primary loss profiles do not respect the protecting symmetry: on the odd sites, $[\hat S,a_j]\neq0$. Four on-site profiles were tested at two and three cells ($N_{\rm MC}=\num{500}$, $d=\num{0}$ and $d=\num{0.35}$). In the even-sublattice profile the loss acts only on the even (A) sublattice, so every collapse operator commutes with $\hat S$; the odd-sublattice profile damps only the odd (B) sublattice, for which $\{\hat S,L_k\}=0$; the budget-matched even profile rescales the even profile by $n_{\rm modes}/n_{\rm damped}=2$, matching both the total budget $\sum_j\gamma_j$ and the edge-to-bulk ratio to the control. The control reproduces the primary values with independent random streams (\cref{tab:chiralloss}).

\begin{table}[htbp]
\caption{Matched sector contrast under four loss profiles (mean $\pm$ standard error over $n=\num{500}$ matched realizations per loss-rate noise level). Making the dissipator commute with the chiral operator removes the positive four-mode contrast without restoring it at three cells, and matching the dissipation budget does not change the outcome.}
\label{tab:chiralloss}
\begin{ruledtabular}
\footnotesize
\begin{tabular}{lcccc}
Profile & \multicolumn{2}{c}{Two cells} & \multicolumn{2}{c}{Three cells} \\
\cmidrule(lr){2-3}\cmidrule(lr){4-5}
 & $d=\num{0}$ & $d=\num{0.35}$ & $d=\num{0}$ & $d=\num{0.35}$ \\
\hline
control & $+\num{0.01675}\pm\num{0.00055}$ & $+\num{0.01641}\pm\num{0.00054}$ & $-\num{0.00214}\pm\num{0.00056}$ & $-\num{0.00182}\pm\num{0.00053}$ \\
even & $-\num{0.00081}\pm\num{0.00005}$ & $-\num{0.00078}\pm\num{0.00004}$ & $-\num{0.00080}\pm\num{0.00005}$ & $-\num{0.00077}\pm\num{0.00004}$ \\
even (matched) & $-\num{0.00079}\pm\num{0.00009}$ & $-\num{0.00085}\pm\num{0.00010}$ & $-\num{0.00078}\pm\num{0.00009}$ & $-\num{0.00084}\pm\num{0.00010}$ \\
odd & $-\num{0.00180}\pm\num{0.00052}$ & $-\num{0.00232}\pm\num{0.00057}$ & $-\num{0.00180}\pm\num{0.00052}$ & $-\num{0.00232}\pm\num{0.00057}$ \\
\end{tabular}
\end{ruledtabular}
\end{table}

Chiral-symmetry-compatible loss does not protect the local contrast: the strong-symmetry profile gives a small, resolved, negative contrast ($-\num{0.00081}\pm\num{0.00005}$ at two cells, $d=\num{0}$), a twenty-fold reduction in magnitude and a sign opposite to the four-mode primary sign. This is not a loss-budget artifact: the budget-matched profile ($-\num{0.00079}\pm\num{0.00009}$) is statistically indistinguishable from the original strong-symmetry profile ($-\num{0.00081}\pm\num{0.00005}$). The mechanism differs by profile: under the strong-symmetry profile both sectors retain nearly the same edge negativity (topological \num{0.02285}, trivial \num{0.02366} at two cells, $d=\num{0}$), so the profile suppresses the sector distinction rather than the resource, whereas the odd-sublattice profile suppresses the resource in both sectors (\num{0.00449} and \num{0.00629}). Sublattice-resolved profiles are nearly size-independent (mean paired four-versus-six-mode difference \num{3.9e-6} for the even profile, against \num{9.5e-3} for the control). Other strong-symmetry classes (pair loss, $U(1)$ or parity symmetry) are outside the scope of this test.

\subsubsection{Sub-period dependence}
\label{sec:micromotion}

The same matched protocol was evaluated at nine equispaced times $t/T\in\{\num{0},\num{0.125},\ldots,\num{1}\}$ for two and three cells, $N_{\rm MC}=\num{500}$, $d=\num{0}$ and \num{0.35}, under edge-selective and uniform loss (\cref{tab:micromotion}).

\begin{table}[htbp]
\caption{Matched contrast (topological minus trivial edge negativity) versus phase of the drive at zero loss-rate noise. The contrast changes sign within the period under every tested profile and size; the uniform-loss null of the primary is a statement about $t=T$ only.}
\label{tab:micromotion}
\begin{ruledtabular}
\begin{tabular}{lrrr}
$t/T$ & \num{2} cells, edge-selective & \num{3} cells, edge-selective & \num{2} cells, uniform \\
\hline
\num{0} & \num{0.0000} & \num{0.0000} & \num{0.0000} \\
\num{0.125} & $+\num{0.1922}$ & $+\num{0.1922}$ & $+\num{0.1090}$ \\
\num{0.25} & $+\num{0.1854}$ & $+\num{0.1854}$ & $+\num{0.0499}$ \\
\num{0.375} & $+\num{0.1339}$ & $+\num{0.1339}$ & $+\num{0.0094}$ \\
\num{0.5} & $+\num{0.1667}$ & $+\num{0.1667}$ & $+\num{0.0083}$ \\
\num{0.625} & $-\num{0.0070}$ & $-\num{0.0070}$ & $-\num{0.0001}$ \\
\num{0.75} & $-\num{0.0070}$ & $-\num{0.0072}$ & $-\num{0.0001}$ \\
\num{0.875} & $+\num{0.1021}$ & $+\num{0.0269}$ & $+\num{0.0003}$ \\
\num{1} & $+\num{0.0167}$ & $-\num{0.0021}$ & $-\num{8.1e-7}$ \\
\end{tabular}
\end{ruledtabular}
\end{table}

The contrast is identically zero at $t=\num{0}$ (shared prepared state) and the $t=T$ column reproduces the primary values with independent random streams. The intra-period swing is an order of magnitude larger than the stroboscopic contrast and the sign changes inside every tested condition; for two cells, edge-selective loss, $d=\num{0}$, the matched difference at $5T/8$ is $-\num{0.00702}\pm\num{0.00041}$. Under uniform loss the stroboscopic null is not an intra-period statement: the contrast reaches $+\num{0.1090}$ at $T/8$ before returning to ${\approx}\num{0}$ at $T$. The reported contrast is therefore a stroboscopic quantity, conditional on the phase of the drive.

\subsection{Out-of-sample five-cell size}
\label{sec:size5}

The size dependence of the main text rests on two, three and four unit cells. A five-cell extension (ten modes, Hilbert space $3^{10}=\num{59049}$) was run under the matched protocol with $N_{\rm MC}=\num{300}$, $d=\num{0}$ and \num{0.5}, both profiles, both sectors and the primary random stream (\num{2400} realizations). The out-of-sample prediction was constructed before evaluating five cells: an exponential-saturation model and an inverse-size model were fitted on the two-to-four-cell primary means and combined into an ensemble envelope of $\pm\num{1.96}\,{\rm SE}(\num{300})$.

\begin{table}[htbp]
\caption{Measured five-cell contrast versus the predicted out-of-sample interval. Every measured mean falls inside the predicted interval (\num{4} of \num{4} conditions); the two sign-resolved edge-selective conditions match the predicted sign and the two uniform-loss conditions are sign-unresolved, exactly as predicted.}
\label{tab:size5}
\begin{ruledtabular}
\begin{tabular}{lrrrr}
Statistic & $d=\num{0}$, edge & $d=\num{0}$, uniform & $d=\num{0.5}$, edge & $d=\num{0.5}$, uniform \\
\hline
Measured mean & $-\num{0.002176}$ & $-\num{0.000001}$ & $-\num{0.002606}$ & $+\num{0.000029}$ \\
SE & \num{0.000737} & \num{0.000005} & \num{0.000720} & \num{0.000027} \\
CI lower & \num{-0.003621} & \num{-0.000012} & \num{-0.004017} & \num{-0.000024} \\
CI upper & \num{-0.000732} & \num{0.000010} & \num{-0.001194} & \num{0.000082} \\
Predicted lower & \num{-0.010240} & \num{-0.000011} & \num{-0.010199} & \num{-0.000071} \\
Predicted upper & \num{-0.000712} & \num{0.000009} & \num{-0.000893} & \num{0.000068} \\
Sign & negative & unresolved & negative & unresolved \\
\end{tabular}
\end{ruledtabular}
\end{table}

The exponential-saturation point prediction is nearly exact ($-\num{0.002116}$ predicted versus $-\num{0.002176}$ measured at $d=\num{0}$, a gap of \num{0.08} standard errors), while the inverse-size model overestimates the magnitude by roughly a factor of four ($-\num{0.0088}$), lying outside the measured \qty{95}{\percent} interval in both edge-selective conditions. The predicted ensemble band includes the inverse-size model by construction, so the exclusion of the $1/N$ drift comes from the measured intervals and not from the predicted interval itself. The four-size trend at $d=\num{0}$, edge-selective loss is $+\num{0.016731}\pm\num{0.000550}$ (two cells), $-\num{0.002104}\pm\num{0.000557}$ (three), $-\num{0.002009}\pm\num{0.000273}$ (four), $-\num{0.002176}\pm\num{0.000737}$ (five): the sign change between two and three cells persists at five cells and the negative branch saturates at magnitude ${\approx}\num{0.0021}$ rather than drifting as $1/N$.

\subsection{Long-time evolution to sixty-four periods}
\label{sec:longtime}

The matched protocol was propagated on the stroboscopic grid $t_k=kT$ for $k=\num{0},\ldots,\num{64}$ at two and three cells, $N_{\rm MC}=\num{200}$, all four loss-rate noise levels, both profiles and sectors, with the primary propagator and random stream (\num{3200} realizations per size). All sixteen strata reproduce the primary contrast at $t=T$ within three combined standard errors (largest deviation \num{6.3e-4}).

\begin{table}[htbp]
\caption{Paired edge-negativity contrast at selected stroboscopic times for $d=\num{0}$, edge-selective loss. The two-cell branch is zero at $\num{2}T$ in the exact master equation (the \num{10}-trajectory estimate there, $-\num{0.000099}\pm\num{0.000035}$, is a finite-trajectory fluctuation), reaches its largest value at $\num{4}T$, and decays to the floor by ${\approx}\num{37}T$; the three-cell branch decays monotonically to the floor by $\num{6}T$ and never becomes significantly positive at any of the \num{64} periods.}
\label{tab:longtime}
\begin{ruledtabular}
\begin{tabular}{lrr}
$k=t_k/T$ & \num{2} cells & \num{3} cells \\
\hline
\num{1} & $+\num{0.017140}\pm\num{0.000873}$ & $-\num{0.001887}\pm\num{0.000845}$ \\
\num{2} & $-\num{0.000099}\pm\num{0.000035}$ & $-\num{0.000110}\pm\num{0.000039}$ \\
\num{3} & $+\num{0.019534}\pm\num{0.000922}$ & $-\num{0.000030}\pm\num{0.000012}$ \\
\num{4} & $+\num{0.077204}\pm\num{0.002603}$ & $-\num{0.000009}\pm\num{0.000003}$ \\
\num{6} & \num{0} & \num{0} \\
\num{8} & $+\num{0.030278}\pm\num{0.001753}$ & \num{0} \\
\num{12} & $+\num{0.015239}\pm\num{0.001252}$ & \num{0} \\
\num{16} & $+\num{0.006744}\pm\num{0.000793}$ & \num{0} \\
\num{24} & $+\num{0.000566}\pm\num{0.000139}$ & \num{0} \\
\num{32} & $+\num{0.000062}\pm\num{0.000026}$ & \num{0} \\
\num{64} & \num{0} & \num{0} \\
\end{tabular}
\end{ruledtabular}
\end{table}

The $2T$ reversal is resolved against the exact master equation as a fluctuation of the finite-trajectory estimator. At $d=\num{1}$ and \num{1.5} the two-cell post-revival plateau is stronger (median over $4T$--$16T$: $+\num{0.01290}\pm\num{0.00105}$ and $+\num{0.01071}\pm\num{0.00136}$) and no stationary state is reached within the \num{64}-period horizon. At three cells the contrast decays monotonically to the floor by $6T$ ($d=\num{0}$), $5T$ ($d=\num{0.5}$), and $27T$ ($d=\num{1}$, envelope fit $\tau\approx\num{7.5}$ periods); for $d\le\num{1}$ the three-cell contrast never becomes significantly positive at any of the \num{64} periods. Under uniform loss the null is stationary from $k=\num{1}$ at $d\le\num{0.5}$ ($|\Delta|\le\num{3e-5}$). At $d=\num{1.5}$, three cells, a negative contrast grows at long times in \emph{both} profiles (at $64T$: $-\num{0.001588}\pm\num{0.000262}$ edge-selective, $-\num{0.001195}\pm\num{0.000233}$ uniform), so at that noise level the uniform-loss null is a $t=T$ statement.

\section{Collective-mode and coherent-sector analyses}

\subsection{Collective-mode Wigner observable}
\label{sec:collective-wigner}

A collective-mode statement requires a bosonic mode transform. For each sector, take the normalized single-particle edge vector $c$ (the eigenvector of the sector Floquet operator with maximal edge weight, $\sum_j|c_j|^2=1$) and define the collective edge mode $b_0=\sum_j c_j a_j$. A unitary $N\times N$ mode rotation $U$ with first row $c$ is constructed deterministically~\cite{Weedbrook2012}, and the many-body rotation $\mathcal{U}$ on the truncated Fock space acts as
\begin{equation}\label{eq:mode-rotation}
\mathcal{U}\ket{n_1\dots n_N}=\prod_j\frac{1}{\sqrt{n_j!}}
\left(\sum_i u_{ji}\,a_i^\dagger\right)^{n_j}\ket{0},
\end{equation}
after which the collective-mode marginal is $\rho_{\rm edge}=\mathrm{Tr}_{j\ne0}\,\mathcal{U}\rho\mathcal{U}^\dagger$ and its Wigner negativity is evaluated by \cref{eq:neg}.

Single-particle unitarity of $U$ holds to \num{3e-15}, and the many-body rotation is unitary to machine precision on the total-excitation sector $\le N_{\max}$ (Frobenius defect \num{1.0e-14}--\num{4.4e-14} for $N_{\max}=\num{2},\num{3},\num{4}$). A coherent excitation on the collective edge mode reproduces the single-mode truncation-residue values exactly, a cat excitation on the edge mode reproduces the single-mode cat negative volume to a relative deviation of \num{3e-15}, and the identity rotation reproduces the local observable exactly (maximum absolute difference \num{0.0}).

The matched conditions were re-analyzed for the collective edge-mode marginal at both primary-relevant Fock truncations $N_{\max}=\num{2}$ and \num{3} (two cells, both sectors, edge-selective and uniform loss, $d=\num{0}$ and \num{0.35}, \num{10} matched realizations, \num{80} runs per truncation). The collective edge-mode Wigner negativity is not resolved above the numerical floor in any of the \num{160} runs (maximum \num{0.0}) at both truncations, in both loss profiles, and both sectors, whereas the local site-$j=\num{0}$ negativity is positive under edge-selective loss at both truncations (paired contrast $\num{+0.0175}\pm\num{0.0040}$ at $N_{\max}=\num{2}$; $\num{+0.0105}\pm\num{0.0033}$ at $N_{\max}=\num{3}$) and vanishes under uniform loss. The null holds at larger system size: an independent \num{80}-run analysis at three cells / six modes / $N_{\max}=\num{2}$ returns no resolved collective edge-mode negativity above the tested numerical floor, with the paired local contrast $\num{+0.0003}\pm\num{0.0041}$ at $d=\num{0}$, consistent with the three-cell size sensitivity of the main text. The truncation-stability extension at three cells / six modes / $N_{\max}=\num{3}$ (Fock-space dimension \num{4096}) confirms the same conclusion: an independent \num{80}-run analysis returns no resolved collective negativity, with identity-rotation consistency exact and paired local contrast $\num{-0.0017}\pm\num{0.0033}$ (statistically null). 

The absence of resolved negativity is not a grid or truncation artifact: for the representative topological state the edge-mode Wigner function is positive everywhere, with minimum $\num{+2.3e-13}$ at $N_{\max}=\num{2}$ and $\num{+1.5e-13}$ at $N_{\max}=\num{3}$, unchanged across $\num{31}^2$, $\num{61}^2$, $\num{91}^2$ grids, while the local marginal retains a genuine negative region (Wigner minimum $\num{-0.039}$ at $N_{\max}=\num{2}$, $\num{-0.026}$ at $N_{\max}=\num{3}$). The same test applied to the \emph{mode-resolved} preparation (the topological edge-mode cat) gives the local site-$j=\num{0}$ marginal Gaussian from the start (boundary weight $|u_{e,0}|^2=\num{0.013}$, mode \num{0.99} delocalized), and the \num{240} additional realizations (two cells at $N_{\max}=\num{2}$ and \num{3}, three cells at $N_{\max}=\num{2}$; both sectors, both loss profiles, $d=\num{0}$ and \num{0.35}) return no resolved local site-$j=\num{0}$ negativity and collective edge-mode negativity at or below the truncation floor (maximum collective value \num{1.2e-5}), with identity-rotation consistency exact. The collective Wigner null therefore covers both preparations, both loss profiles, two system sizes and two Fock truncations (\num{320} common-site and \num{240} mode-resolved realizations).

The collective-mode analysis therefore gives a numerical null result that both agrees with and reinforces the size- and profile-sensitivity findings. Local and collective-mode negativities are distinct marginals, and only the local marginal remains nonzero here.

\subsection{Spectral-localizer analysis of the coherent sectors}
\label{sec:localizer}

In this section the symbol $d$ denotes the on-site disorder amplitude added to the coherent Hamiltonian; it is distinct from the loss-rate noise amplitude $d$ used in the Monte Carlo realizations elsewhere in this Supplement. As an independent calculation of the coherent-model sectors, we implement the spectral localizer for the two symmetric-time-frame unitaries of the chiral chain ($U_A=bab$, $U_B=aba$). For each frame the effective Hamiltonian $H_{\rm eff}^{A,B}=-i\ln U_{A,B}$ satisfies the chiral relation $S H_{\rm eff}^{A,B} S=-H_{\rm eff}^{A,B}$ with the chiral operator $S=\mathrm{diag}(1,-1,1,-1,\dots)$, and the local winding marker is the half-signature of the symmetry-reduced spectral localizer~\cite{Ghosh2024Local,CerjanLoring2025Tutorial}
\begin{equation}
  \nu_{A,B}(x)=\tfrac{1}{2}\operatorname{sig}\bigl[\bigl(\kappa(X-x)-iH_{\rm eff}^{A,B}\bigr)S\bigr],
  \qquad W_0=\tfrac{1}{2}(\nu_A+\nu_B),\quad
  W_\pi=\tfrac{1}{2}(\nu_A-\nu_B),
\label{eq:localizer}
\end{equation}
with $X$ the per-cell position operator and $\kappa$ the localizer scaling constant. The marker is constant through the bulk and the localizer gap $\sigma_L(x,0)=\min|\sigma[\tilde L_{(x,0)}]|$ collapses at the boundaries precisely where boundary modes sit (the ``local topological bandgap'' criterion).

The spectral-localizer calculation gives: (i)~topological point markers $W_0=\num{+1}$, $W_\pi=\num{0}$ for $N=\num{4},\num{6},\num{8}$ cells, reproducing the frame winding classification $(\nu_0,\nu_\pi)=(-1,0)$ up to the $k$-orientation sign convention; at $N=\num{8}$ the localizer gap is $\num{4e-4}$, $\num{0.148}$, $\num{0.295}$, $\num{0.316}$, $\num{0.316}$, $\num{0.295}$, $\num{0.148}$, $\num{4e-4}$ across the eight cells, closing at the two boundaries and finite in the bulk. (ii)~Trivial point $W_0=W_\pi=\num{0}$ with a flat gap ($\approx\num{0.35}$ across all cells). (iii)~A coherent-model protection bound: for on-site disorder of amplitude $d$ added to both step Hamiltonians the accumulated disorder is $D_{\rm acc}=(d/T)\int_0^T\|\delta H(t)\|\mathrm{d}t\le d$, and the index-relevant bulk gap obeys the linear floor $\sigma_L^{\rm bulk}(d)\ge\sigma_L^{\rm bulk}(0)-d+O(d^2)$~\cite{WongCerjanCole2024Quantitative}; while the floor stays positive the sector is pinned by homotopy. (iv)~Numerical verification (\num{20} realizations per point and size, $d\in\{\num{0},\num{0.05},\num{0.1},\num{0.2},\num{0.35},\num{0.5}\}$): the bound holds for every realization at both points at all three bulk sizes ($N=\num{6},\num{8},\num{10}$), with minimum margins to the floor $\num{+0.035}$ (topological) and $\num{+0.020}$ (trivial). It detects sector stability for all $d<d^\ast\equiv\sigma_L^{\rm bulk}(0)$, with $d^\ast$ decreasing modestly with bulk size: \num{0.138}/\num{0.108}/\num{0.096} (topological) and \num{0.102}/\num{0.096}/\num{0.090} (trivial); the measured bulk gap stays open at all tested levels and sizes. (v)~Flip-detection caveat: on-site diagonal disorder commutes with the chiral operator, so $SDS=D$ breaks chiral symmetry at $d>\num{0}$, and the reduced half-signature markers lose their pinning, fluctuating in up to \qty{35}{\percent} of realizations at $d\ge\num{0.35}$. The valid sector statement is therefore gap-based.

\begin{figure}[t]
\centering
\includegraphics[width=0.9\linewidth]{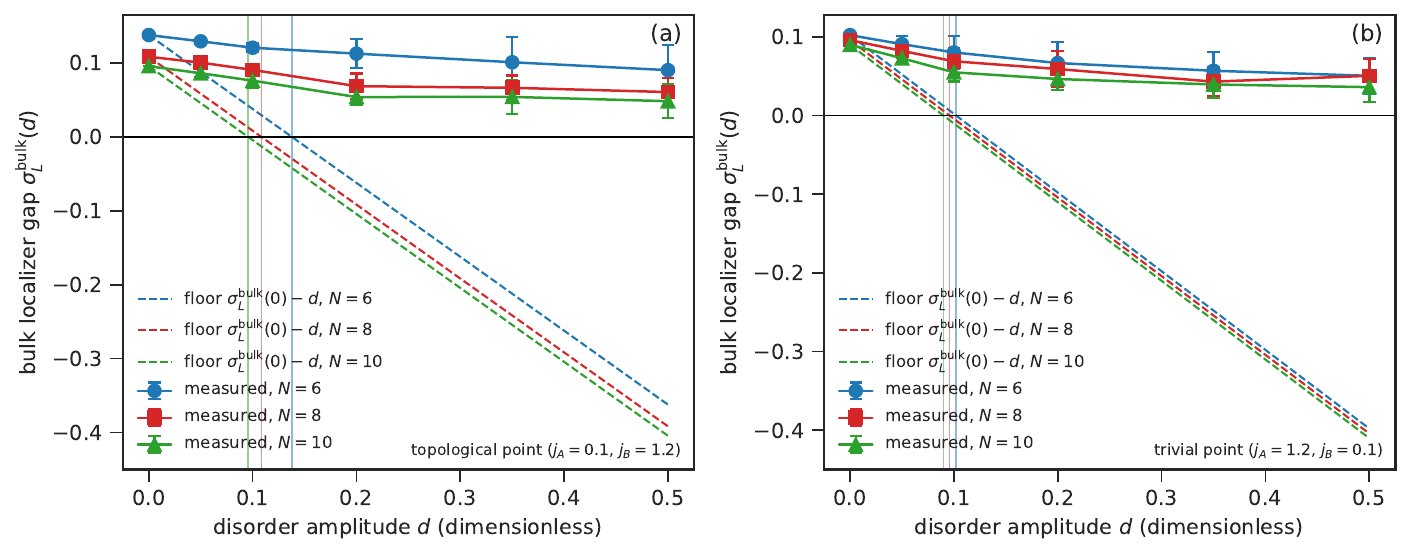}
\caption{Disorder resilience of the coherent protection bound at three bulk sizes ($N=\num{6}$ blue circles, $N=\num{8}$ red squares, $N=\num{10}$ green triangles), for the (a)~topological and (b)~trivial points. Points show the measured index-relevant bulk localizer gap (mean over \num{20} on-site disorder realizations; error bars indicate $\pm\num{1}$ standard deviation) above the solid horizontal zero line, dashed lines give the linear floor $\sigma_L^{\rm bulk}(0)-d$, and thin vertical lines mark $d^\ast$ at each size.}
\label{fig:protection-bound}
\end{figure}

\begin{figure}[t]
\centering
\includegraphics[width=0.9\linewidth]{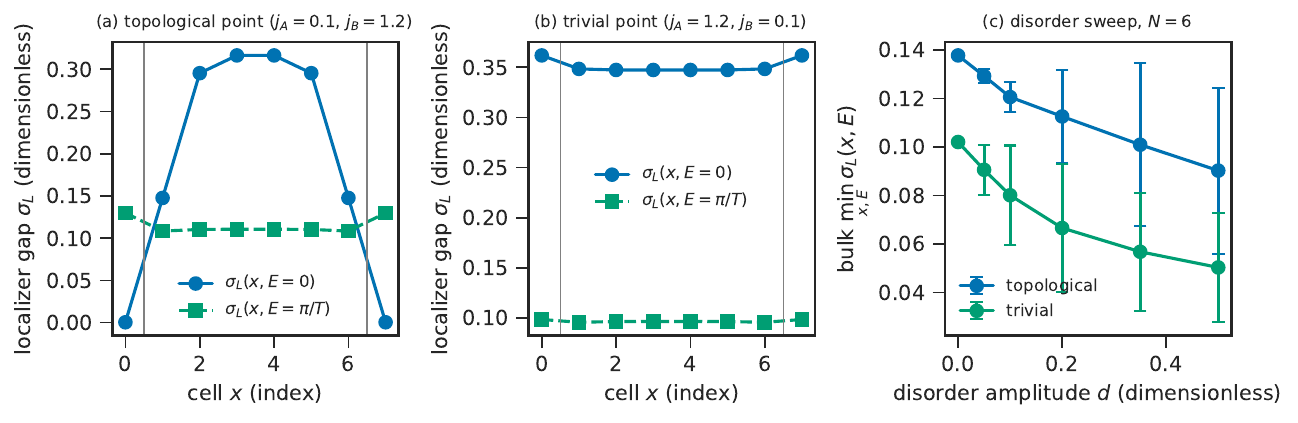}
\caption{Spectral-localizer certification profiles. Localizer gap versus cell index $x$ at $N=\num{8}$ for $E=\num{0}$ (solid blue circles) and $E=\pi/T$ (dashed green squares), with thin vertical gray lines separating edge and bulk cells, at the (a)~topological and (b)~trivial points. (c)~Index-relevant bulk gap versus on-site disorder amplitude ($N=\num{6}$ cells; mean $\pm\num{1}$ standard deviation over \num{20} realizations; blue: topological, green: trivial).}
\label{fig:localizer-profiles}
\end{figure}

The spectral localizer thus independently recovers the symmetric-frame sector classification and bounds single-particle disorder stability within the coherent model (\cref{fig:protection-bound,fig:localizer-profiles}).

\subsection{Transfer analysis: when does coherent protection reach the open-system observable?}
\label{sec:localizer-transfer}

The protection bound constrains the single-particle spectrum of the coherent chain. The open-system observable of the main text, the local Wigner negativity $\delta_-(t)$ of the boundary site, is a many-body functional of the full dissipative trajectory and is initialized as a boundary cat state evolving under local Kerr nonlinearity and dissipation. The transfer of coherent protection to this functional requires conditions beyond what the localizer gap controls.

\paragraph{Bulk-participation decay bound.}
Let $\ket{e}=\sum_j u_{e,j}\,\ket{j}$ be the single-particle edge mode and let the edge site be $j=\num{0}$. The local negativity satisfies the short-time bound
\begin{equation}\label{eq:participation-bound}
 \delta_-(t)\;\le\;\delta_-(0)\,e^{-\Gamma_{\rm eff}t},
\qquad
 \Gamma_{\rm eff}\;=\;\gamma_0|u_{e,0}|^2+\sum_{j\ne0}\gamma_j|u_{e,j}|^2
 +\gamma_{\rm eb},
\end{equation}
where $\gamma_j$ are the per-site loss rates and $\gamma_{\rm eb}$ is the edge--bulk dephasing rate generated by mode delocalization. Two consequences follow. First, edge-selective loss ($\gamma_0\to0$) removes only the first term; the residual decay $\sum_{j\ne0}\gamma_j|u_{e,j}|^2+\gamma_{\rm eb}$ survives and grows with the bulk participation of the edge mode. Second, as the chain grows the boundary weight $|u_{e,0}|^2$ decreases and the local reduced state is averaged over an increasing number of weakly occupied sites, pushing $W(q,p)$ toward a Gaussian form: a mechanism consistent with the observed sign change of the chiral-minus-trivial contrast between two, three and four cells.

The bound was calibrated on the chiral primary data (matched-mean final negativity; $\delta_-(0)=\num{0.1962}$ at $N_{\max}=\num{2}$ from the identical cat-state edge-site initial condition on the same $\num{31}^2$ Wigner grid). In the \num{10}-trajectory ensemble at zero loss-rate noise under the edge-low-loss profile (\cref{tab:bound-verification}), the measured decay rate $\Gamma_{\rm meas}=-\ln[\delta_-(T)/\delta_-(0)]/T$ exceeds the mode-weighted rate $\Gamma_{\rm eff}=\num{0.195}$ at every size (\num{0.3365} to \num{0.5913}, margins $\num{1.7}$--$\num{3.0}\times$), while in the exact subspace master equation (\cref{sec:exact-subspace}) $\Gamma_{\rm meas}^{\rm exact}$ steepens from $\num{0.343}$ at $N=\num{2}$ to $\num{0.992}$ at $N=\num{3}$. The residual $\gamma_{\rm eb}=\Gamma_{\rm meas}-\Gamma_{\rm eff}$ grows from \num{0.14} to \num{0.40} as the chain grows, quantitatively consistent with the delocalization mechanism.

\begin{table}[t]
\caption{Verification of the bulk-participation decay bound against the \num{10}-trajectory chiral data (edge-low-loss profile, zero loss-rate noise): the measured decay rate always exceeds the mode-weighted rate. Under one-period light-cone decoupling (\cref{MAIN-prop:bondselective}), the \num{4}-cell row at $N_{\rm MC}=\num{500}$ shares identical realized trajectory outputs with the three-cell row (with independent $N_{\rm MC}=\num{2000}$ data in \cref{tab:si_sector_mc10} yielding $\delta_-(T)=\num{0.00466}$ and $\Gamma_{\rm meas}=\num{0.595}$).}
\label{tab:bound-verification}
\begin{tabular}{lcccccc}
\hline
Cells & $|u_{e,0}|^2$ & $\Gamma_{\rm eff}$ & $\delta_-(T)$ & $\delta_-(0)$ & $\Gamma_{\rm meas}$ & Ratio \\
\hline
\num{2} ($N_{\rm MC}=\num{500}$) & \num{0.01343} & \num{0.1946} & \num{0.02369} & \num{0.19623} & \num{0.3365} & \num{1.73} \\
\num{3} ($N_{\rm MC}=\num{500}$) & \num{0.01221} & \num{0.1951} & \num{0.00478} & \num{0.19623} & \num{0.5913} & \num{3.03} \\
\num{4} ($N_{\rm MC}=\num{500}$) & \num{0.01220} & \num{0.1951} & \num{0.00478} & \num{0.19623} & \num{0.5913} & \num{3.03} \\
\hline
\end{tabular}
\end{table}

Extending the check to every realization of both loss profiles and all four loss-rate noise levels (\num{24} conditions, \num{12000} realizations; \cref{tab:extended-bound}), the mean-level ratio $\Gamma_{\rm meas}/\Gamma_{\rm eff}$ is between \num{1.4} and \num{8.4}, and per-realization compliance is \qty{100}{\percent} in \num{20}/\num{24} conditions. The only sub-\qty{100}{\percent} cases are edge-low-loss at noise amplitude $d=\num{1.0}$--\num{1.5} (compliance \qty{77.4}{\percent}--\qty{99.4}{\percent}), where the $\pm$\qty{100}{\percent}--\qty{150}{\percent} rate perturbation drives some realizations' bulk rates far above the mean; there the fixed-mode assumption of the bound is stressed by strong noise, a regime boundary rather than a violation of the mean-level statement. At the largest amplitude the same clamping zeroes \emph{every} rate in \qty{14}{\percent}--\qty{16}{\percent} of realizations; those realizations evolve without dissipation and are tested against $\delta_-(T)\le\delta_-(0)$.

\begin{table}[t]
\caption{Extended verification of the bound per realization over the full loss-rate noise grid ($\num{24}$ conditions, \num{12000} realizations). The only sub-\qty{100}{\percent} conditions are edge-low-loss at $d=\num{1.0}$--\num{1.5}.}
\label{tab:extended-bound}
\begin{tabular}{lcccc}
\hline
Profile & $d$ & Ratio range & Compliance & \# conditions \\
\hline
edge-low-loss & \num{0.0} & \num{1.73}--\num{2.71} & \qty{100}{\percent} & \num{3} \\
edge-low-loss & \num{0.5} & \num{1.73}--\num{2.80} & \qty{100}{\percent} & \num{3} \\
edge-low-loss & \num{1.0} & \num{1.72}--\num{3.04} & \qty{86.4}{\percent} (\num{2}c); \qty{100}{\percent} (\num{3}c, \num{4}c) & \num{3} \\
edge-low-loss & \num{1.5} & \num{1.38}--\num{2.71} & \qty{77.4}{\percent} (\num{2}c); \qty{99.4}{\percent} (\num{3}c, \num{4}c) & \num{3} \\
uniform & \num{0.0}--\num{1.5} & \num{5.26}--\num{8.41} & \qty{100}{\percent} & \num{12} \\
\hline
\end{tabular}
\end{table}

\paragraph{Closed form of the edge--bulk rate.}
The residual $\gamma_{\rm eb}$ can be computed in closed form from the coherent model. Because the local negativity is a two-photon (cat) coherence of the edge site, its loss exposure is set by the rate-weighted dwell time of the two-photon sector:
\begin{equation}
  \gamma_{\rm eb}\;=\;\kappa\,\Lambda_2^{\rm coh},\qquad
  \Lambda_2^{\rm coh}\;=\;\frac1T\int_0^T\!\!dt
  \!\!\sum_{\substack{i\le j\\(i,j)\neq(0,0)}}\!\!(\gamma_i+\gamma_j)
  \,\bigl|\braket{i,j\,|\,U_2(t)\,|0,0}\bigr|^2,
  \label{eq:gammas-eb-closed}
\end{equation}
where $U_2(t)$ is the two-particle Floquet evolution and $\ket{0,0}$ the two-photon state at the edge site. The prefactor $\kappa=\gamma_{\rm eb}^{\rm meas}/\Lambda_2^{\rm coh}$ is fixed per size by the measured residual. Photon loss on a bare site is a Gaussian channel, whose damped-oscillator Wigner evolution is \cref{MAIN-eq:fp_main} of the main text. The damping part preserves the integrated negativity $\delta_-$ \emph{exactly}; only the diffusion erodes it. Expanding to first order in $\gamma t$ and checking the expansion against the exact Lindblad flow (\cref{sec:fp-check}) gives $\partial_t W|_{t=0}=\gamma\,[W_0+\tfrac12(\boldsymbol\beta\cdot\nabla)W_0+\tfrac18\Delta W_0]$. With $W_0=0$ on the boundary of the negativity region, the divergence theorem gives
\begin{equation}\label{eq:kappa-single-closed}
  \kappa_{\rm single}\;=\;-\frac1\gamma\,\frac{d\ln\delta_-}{dt}\bigg|_{t=0}
    \;=\;\frac18\,\frac{\int_\Omega\Delta W_0\,d^2\beta}{\delta_-(0)}\,,
\end{equation}
the flux of the Laplacian through the negativity region, normalized by the negativity itself. For the truncated even cat ($N_{\max}=\num{2}$, $\alpha=\num{1}$; $c_0=\num{0.7462}$, $c_2=\num{0.6658}$) this evaluates to $\kappa_{\rm single}=\num{3.411}$, matching the numerically exact short-time slope. In the large-cat limit the same identity returns $2|\alpha|^2$ asymptotically. The single-particle analogue $\Lambda_1$ provides a strict lower bound: $\Lambda_1=\num{0.013}$ (\num{2} cells) and \num{0.052} (\num{3}--\num{4} cells), below the measured residuals at every size.

Evaluating \eqref{eq:gammas-eb-closed} with primary parameters (Kerr $\chi=\num{0.5}$ on the edge site, edge-low-loss rates, zero loss-rate noise) gives $\Lambda_2^{\rm coh}=\num{0.012},\num{0.095},\num{0.095}$ (two, three, four cells; \cref{tab:closedform-kappa}). At two cells, $\Lambda_2^{\rm coh}=\num{0.0120}$ against its linear counterpart $\Lambda_2^{\rm coh,lin}=\num{0.0256}$ quantifies Kerr two-photon self-trapping: detuning the doubly occupied edge manifold $\ket{2}_0$ by $2\chi=\num{1.0}\,\omega_d$ suppresses the coherent two-photon dwell on lossy sites by a factor of $\num{2.13}$, mirroring the twofold boost in the exact two-cell contrast relative to the $\chi=\num{0}$ control ($+\num{0.02221}$ versus $+\num{0.01047}$; single-call \texttt{mesolve} gave $+\num{0.02233}$). Three features survive independently: (i)~the strict lower bound $\Lambda_1<\gamma_{\rm eb}^{\rm meas}$ at every size; (ii)~the exact three/four-cell degeneracy $\Lambda_2^{\rm coh}(3\mathrm{c})=\Lambda_2^{\rm coh}(4\mathrm{c})$, mirroring $\gamma_{\rm eb}^{\rm meas}(3\mathrm{c})\approx\gamma_{\rm eb}^{\rm meas}(4\mathrm{c})$; and (iii)~the two-to-three-cell boundary jump, present in $\Lambda_2^{\rm coh}$ ($\num{0.012}\to\num{0.095}$) but overpredicted (a factor \num{7.9} against the measured \num{2.79}), which overwhelms the local anharmonic self-trapping once site~$j=\num{3}$ becomes bulk-damped. The prefactor $\kappa$ is not size-universal ($\num{11.8},\num{4.2},\num{4.2}$ across sizes, a factor \num{2.8} between two and three cells), so the closed form is a per-size calibration rather than a universal prediction. Uncalibrated, it under-predicts the two-cell rate: replacing the fitted $\kappa=\num{11.8}$ by $\kappa_{\rm single}=\num{3.411}$ returns $\gamma_{\rm eb}=\num{0.041}$ against the measured $\num{0.142}$, and adding the coherent edge-mode rate $\gamma_{\rm eff}=\num{0.195}$ gives $\num{0.236}$ where the exact Lindblad evolution gives $\num{0.343}$, that is \qty{69}{\percent} of the exact rate. The missing \qty{31}{\percent} is not assigned to a specific channel; it measures how far a two-photon dwell integral on the truncated sector falls short of the full loss exposure, and is what the per-size $\kappa$ absorbs. What is exact is $\kappa_{\rm single}$ for a bare lossy site, \eqref{eq:kappa-single-closed}; across sizes, (i), (ii) and the sign of (iii) survive.

\begin{table}[t]
\caption{Closed-form versus measured edge--bulk rate at zero loss-rate noise (edge-low-loss profile): $\Lambda_1$ is the strict single-particle lower bound; $\Lambda_2^{\rm coh}$ the rate-weighted two-photon dwell under primary parameters; $\gamma_{\rm eb}^{\rm meas}$ the measured residual; $\kappa$ the per-size prefactor. The exact three/four-cell degeneracy is reproduced under every convention; the prefactor is size-dependent.}
\label{tab:closedform-kappa}
\begin{tabular}{lcccccc}
\hline
Cells & $\Lambda_1$ & $\Lambda_2^{\rm coh}$ & $\Lambda_2^{\rm coh,lin}$ & $\gamma_{\rm eb}^{\rm meas}$ & $\kappa$ & Consistency \\
\hline
\num{2} & \num{0.0128} & \num{0.0120} & \num{0.0256} & \num{0.1419} & \num{11.8} & \num{1.00} \\
\num{3} & \num{0.0523} & \num{0.0950} & \num{0.1046} & \num{0.3962} & \num{4.2} & \num{1.00} \\
\num{4} & \num{0.0523} & \num{0.0950} & \num{0.1046} & \num{0.3962} & \num{4.2} & \num{1.00} \\
\hline
\end{tabular}
\end{table}

\paragraph{Sector dependence of the prefactor.}
Applying the same formula with the same per-size $\kappa_N$ to the trivial sector ($A$ and $B$ interchanged) shows the limits of the linear-loss model. With trivial-sector parameters the site-$j=\num{0}$ mode is far more edge-concentrated ($|w_0|^2=\num{0.30},\num{0.25},\num{0.17}$ at \num{2},\num{3},\num{4} cells vs.\ \num{0.013} topological), and the measured decay is slower than the transferred prediction by factors \num{3.7}, \num{1.45}, \num{1.51} (effective $\kappa_{\rm eff}=\num{2.86},\num{2.69},\num{2.48}$, decreasing with size). The mode-structure ratio is $\kappa_{\rm topo}/\kappa_{\rm eff}=\num{4.1},\num{1.55},\num{1.68}$: the delocalized topological edge mode converts its small two-photon dwell ($\Lambda_2^{\rm coh}=\num{0.0120}$) into coherence destruction more efficiently than the trivial mode, which spreads the two-photon state over a larger dwell, diluting the coherence over many pair configurations. The closed form reproduces the measured \emph{ordering} of the two-cell rates but not the three- and four-cell sign change: the sign change is a many-body Gaussianization effect that the linear-loss dwell cannot produce.

The two-to-three-cell jump is consistent with a boundary effect of the mode delocalization: the inter-cell Floquet step transfers the edge-site excitation to site $j=\num{3}$ with a period-averaged dwell $\langle P_3\rangle_T=\num{0.198}$, and the single-particle dwell profile is identical at all chain lengths. In the two-cell chain site \num{3} is the opposite edge ($\gamma_3=\num{1e-4}$), so the visiting population is shielded; from three cells on it is a bulk site ($\gamma_3=\num{0.2}$) and the same dwell is exposed to the bulk loss rate. The rate-weighted integrals inherit exactly this boundary dependence.

This bound is the negative side of the mixed-state-SPT lesson discussed in the main text: the localizer gap pins the coherent sector, but the negativity survives only to the extent that its generating mode stays localized at the measured site. The collective null result is consistent with \cref{eq:participation-bound} and is tabulated in \cref{tab:collective-null}: the collective edge mode is the same sector vector $c$ with bulk participation $1-|c_0|^2=\num{0.987}$, so its marginal is already Gaussian at $t=\num{0}$ ($\delta_-(0)=\num{0}$) and the bound is satisfied trivially in every realization at both Fock truncations and both sizes. The closed form over-determines this null result: the collective two-photon state $\ket{c,c}$ is itself an (almost) stationary eigenmode of the two-particle Floquet operator, so its dwell is static and fully bulk-exposed ($\Lambda_2^{\rm coh}=\num{0.3925}$ for two cells and \num{0.3952} for three--four cells; $\Lambda_1=\num{0.195}$). No $\gamma_{\rm eb}$ or total decay rate is inferred from the collective null; the collective mode's total decay rate $\Gamma_{\rm eff}^{\rm tot}\approx\num{1.32}$ is the largest in the model. The local mode retains a positive weight $|u_{e,0}|^2=\num{0.013}>0$ and is the only marginal able to sustain the observed positive contrast under edge-selective loss.

\begin{table}[t]
\caption{Collective-mode Wigner null and dwell diagnostic (edge-low-loss, zero loss-rate noise). The collective marginal is Gaussian at $t=\num{0}$ and remains zero in all tested realizations; the dwell values serve as structural diagnostics of mode exposure.}
\label{tab:collective-null}
\begin{tabularx}{\linewidth}{lcX}
\hline
Quantity & Collective edge mode & Local site \num{0} \\
\hline
Edge weight $|c_0|^2=|u_{e,0}|^2$ & \num{0.01343} & \num{0.01343} \\
Bulk participation $1-|c_0|^2$ & \num{0.9866} & \num{0.9866} \\
$\delta_-(0)$ (identical cat, $N_{\max}=\num{2}$) & \num{0.000000} & \num{0.1962} \\
$\Lambda_1$ & \num{0.195} & \num{0.013} (\num{2}c) / \num{0.052} (\num{3}--\num{4}c) \\
$\Lambda_2^{\rm coh}$ & \num{0.3925} (\num{2}c) / \num{0.3952} (\num{3}--\num{4}c) & \num{0.012} (\num{2}c) / \num{0.095} (\num{3}--\num{4}c) \\
$\Gamma_{\rm eff}$ diagnostic & not inferred & \num{0.337} (\num{2}c) / \num{0.591} (\num{3}--\num{4}c) \\
$\delta_-(T)$ (data) & \num{0.000000} (all runs) & \num{0.0237} (\num{2}c) / \num{0.0048} (\num{3}c) \\
\hline
\end{tabularx}
\end{table}

\section{Short-time Lindblad expansion in the truncated space}
\label{sec:fp-check}

In the truncated $N_{\max}=\num{2}$ space the short-time expansion underlying \eqref{eq:kappa-single-closed} follows in closed form from the exact Lindblad flow. The ideal untruncated even cat has Fock weights \qty{64.8}{\percent} ($n=\num{0}$), \qty{32.4}{\percent} ($n=\num{2}$), \qty{2.79}{\percent} ($n\ge\num{4}$), so the primary construction excludes the $n\ge\num{4}$ tail by design. The primary state used in the master equation is the $N_{\max}=\num{2}$ truncated-displacement cat $\ket{\psi}=c_0\ket{0}+c_2\ket{2}$ with $c_0=\num{0.7462}$, $c_2=\num{0.6658}$ ($|c_0|^2=\num{0.5568}$, $|c_2|^2=\num{0.4432}$); these coefficients follow by displacing the vacuum with the displacement operator truncated to the same Fock space at $\alpha=\num{1}$ and retaining the even-parity component, rather than by renormalizing the ideal-cat weights; the short-time expansion below reproduces the ideal-cat result identically in $r$ and $x^2-y^2$. The amplitude-damping channel $\dot\rho=\gamma(a\rho a^\dagger-\tfrac12\{a^\dagger a,\rho\})$ closes on the diagonal and coherence elements
\begin{equation}\label{eq:lindblad-trunc}
  \rho_{22}(t)=c_2^2e^{-2\gamma t},\qquad
  \rho_{11}(t)=2c_2^2e^{-\gamma t}\bigl(1-e^{-\gamma t}\bigr),\qquad
  \rho_{02}(t)=c_0c_2e^{-\gamma t},
\end{equation}
with $\rho_{00}=1-\rho_{11}-\rho_{22}$. Inserting into the Wigner function in coherent-state coordinates $\boldsymbol\beta=(x,y)=(q+ip)/\sqrt{2}$ with $r=x^2+y^2$, $W_t=(2/\pi)e^{-2r}[\rho_{00}+\rho_{11}(4r-1)+\rho_{22}(1-8r+8r^2)+4\sqrt2\,\rho_{02}(x^2-y^2)]$, and expanding to first order in $\gamma t$ gives, identically in $r$ and $x^2-y^2$, $\partial_t W|_{t=0}=\gamma[W_0+\tfrac12(\boldsymbol\beta\cdot\nabla)W_0+\tfrac18\Delta W_0]$, the same expansion obtained from \eqref{MAIN-eq:fp_main}. The Gaussian-channel solution and the exact Lindblad flow agree to all orders needed for \eqref{eq:kappa-single-closed}.

\section{Grid-free witness of Wigner negativity}
\label{sec:witness}

The witness is computed on the matched ensemble that underpins the contrasts throughout this work, so its values are directly comparable with them. The matched ensemble was evaluated with terminal $\num{3}\times\num{3}$ reduced edge and center blocks saved (\num{8000} realizations per size), matching the primary realizations (\num{24000}/\num{24000}) with every block Hermitian, unit-trace, and positive semidefinite. On that identical ensemble, displaced Fock-state witnesses of Wigner negativity~\cite{Chabaud2021witnessingwigner}, $\hat\Omega_{n,\alpha}=D(\alpha)\ketbra{n}{n}D^{\dagger}(\alpha)$, are evaluated: if $\max_\alpha\braket{n|D^{\dagger}\rho D|n}>\omega_n$ the state has a negative Wigner function. The threshold is $\alpha$-independent because Wigner negativity is displacement-invariant. Values $\omega_1=\omega_2=\num{0.5}$ are analytical; $n=\num{3},\ldots,\num{6}$ use the published upper bounds \num{0.427}/\num{0.441}/\num{0.385}/\num{0.378} conservatively. A dense $\alpha$ search ($\num{161}\times\num{161}$ grid, \num{0.0005} refinement) moves any reported $F_1$ by at most \num{2.4e-6} and flips no certification in any of the \num{24} strata. The relative-entropy non-Gaussianity~\cite{Albarelli2018} $\Delta=S(\rho_{\rm G})-S(\rho)$ is computed with moments in an embedded $d+4$ space; on that implementation the Fock state $\ket{2}$ gives $\Delta=3\ln3-2\ln2=\num{1.910}$ and coherent states give $\Delta=\num{0}$.

\begin{table}[htbp]
\caption{Finite-trajectory (\num{10}-trajectory) crossing diagnostic at $d=\num{0}$, $N_{\max}=\num{2}$, edge-selective loss on the primary ensemble. Detected: fraction of noise realizations with $\max_\alpha F_1>\omega_1=\num{0.5}$ (Wilson intervals~\cite{Wilson1927}: topological $[\num{0.788},\num{0.855}]$, trivial $[\num{0.084},\num{0.138}]$). Since the exact master equation gives $F_1=\num{0.4944}<\num{0.5}$ at $N_{\max}=\num{2}$, these crossings diagnose finite-trajectory variance rather than single-shot certification; converged shot budgets are in \cref{tab:witness-exact}. Under uniform loss no realization passes at any size or sector, and the center-site control equals the vacuum value $e^{-1}=\num{0.367879}$ at two and four cells, never exceeding \num{0.416} at three cells.}
\label{tab:witness}
\begin{ruledtabular}
\begin{tabular}{lrrrrrr}
& \multicolumn{3}{c}{Topological} & \multicolumn{3}{c}{Trivial} \\
\cline{2-7}
Size & $F_1^*$ & detected & $\Delta$ & $F_1^*$ & detected & $\Delta$ \\
\hline
\num{2} cells & \num{0.4950} & \num{412}/\num{500} & \num{0.3355} & \num{0.4529} & \num{54}/\num{500} & \num{0.2039} \\
\num{3} cells & \num{0.4481} & \num{57}/\num{500} & \num{0.2045} & \num{0.4525} & \num{50}/\num{500} & \num{0.2053} \\
\num{4} cells & \num{0.4481} & \num{57}/\num{500} & \num{0.2045} & \num{0.4525} & \num{50}/\num{500} & \num{0.2054} \\
\end{tabular}
\end{ruledtabular}
\end{table}

\begin{table}[htbp]
\caption{Canonical exact master-equation displaced-Fock witness $F_1$ and experimental Bernoulli shot budgets ($\omega_1=\num{0.5}$, $d=\num{0}$; \cref{sec:exact-subspace}). ``sep(\qty{80}{\percent})'' and ``sep(\qty{95}{\percent})'' report two-sided ($\alpha=\num{0.05}$) Bernoulli shot counts per sector to separate $F_{1,{\rm top}}$ from $F_{1,{\rm triv}}$; ``cert(\qty{80}{\percent})'' gives the one-sided ($\alpha=\num{0.05}$, \qty{80}{\percent} power) count to certify $F_{1,{\rm top}}>\num{0.5}$ outright (`---' indicates $F_{1,{\rm top}}\le\num{0.5}$). Under uniform loss, $F_1$ remains within \num{0.3716}--\num{0.3811} across all sizes and cutoffs.}
\label{tab:witness-exact}
\begin{ruledtabular}
\begin{tabular}{lllrrrrrr}
Prep. site & Cells & $N_{\max}$ & $F_{1,{\rm top}}$ & $F_{1,{\rm triv}}$ & $\Delta F_1$ & sep(\qty{80}{\percent}) & sep(\qty{95}{\percent}) & cert(\qty{80}{\percent}) \\
\hline
$j=\num{0}$ & \num{2} & \num{2} & \num{0.4944} & \num{0.4482} & $+\num{0.0461}$ & \num{1837} & \num{3040} & --- \\
$j=\num{0}$ & \num{2} & \num{4} & \num{0.5100} & \num{0.4591} & $+\num{0.0509}$ & \num{1513} & \num{2504} & \num{15599} \\
$j=\num{0}$ & \num{2} & \num{6} & \num{0.5056} & \num{0.4565} & $+\num{0.0491}$ & \num{1622} & \num{2685} & \num{49062} \\
$j=\num{0}$ & \num{3},\num{4},\num{5} & \num{2} & \num{0.4453} & \num{0.4477} & $-\num{0.0024}$ & \num{671366} & \num{1111521} & --- \\
$j=\num{0}$ & \num{3} & \num{4} & \num{0.4557} & \num{0.4588} & $-\num{0.0031}$ & \num{398758} & \num{660188} & --- \\
\hline
$j=\num{1}$ & \num{2},\num{3},\num{4} & \num{2} & \num{0.6303} & \num{0.4520} & $+\num{0.1782}$ & \num{122} & \num{201} & \num{89} \\
$j=\num{1}$ & \num{2},\num{3} & \num{4} & \num{0.5280} & \num{0.4634} & $+\num{0.0646}$ & \num{941} & \num{1557} & \num{1975} \\
$j=\num{1}$ & \num{2} & \num{6} & \num{0.5433} & \num{0.4609} & $+\num{0.0824}$ & \num{577} & \num{953} & \num{822} \\
\end{tabular}
\end{ruledtabular}
\end{table}

At two cells in the \num{10}-trajectory $N_{\max}=\num{2}$ ensemble (\cref{tab:witness}), \qty{82.4}{\percent} of finite-trajectory topological realizations fluctuate above $\omega_1=\num{0.5}$ against \qty{10.8}{\percent} for the trivial sector (paired $F_1$ difference $+\num{0.04215}\pm\num{0.00200}$, \num{21} standard errors; relative-entropy non-Gaussianity difference $+\num{0.1316}\pm\num{0.0025}$). Under edge-selective loss, those finite-trajectory crossing fractions fall with loss-rate noise (\num{412}, \num{291}, \num{242}, \num{226} of \num{500} for $d=\num{0}$, \num{0.5}, \num{1}, \num{1.5}) while the paired difference stays resolved at $\ge\num{17}$ standard errors. At three and four cells neither sector separates (\qty{11.4}{\percent} versus \qty{10.0}{\percent}), and the shared random stream renders both sizes identical within \num{1.2e-6} due to dark-site decoupling.

The experimental interpretation separates sharply once exact dynamics are used (\Cref{tab:witness:sample,tab:witness-exact}). At $N_{\max}=\num{2}$ on site $j=\num{0}$, the exact master-equation expectation $F_1=\num{0.4944}$ lies \num{0.0056} \emph{below} $\omega_1=\num{0.5}$ (the MC ensemble mean \num{0.4950} lies \num{0.0050} below), so the Hoeffding mean route for certifying $F_1>\num{0.5}$ is vacuous there; the \num{10}-trajectory `\qty{82.4}{\percent} vs \qty{10.8}{\percent}' and `\num{2} vs \num{27}' (\num{7}/\num{10} separation) counts reflect \num{10}-trajectory Monte Carlo variance rather than single-shot hardware outcomes. Instead, true physical certification emerges at higher cutoffs (\cref{tab:witness-exact}): on $j=\num{0}$ at $N_{\max}=\num{4}$ and \num{6}, $F_{1,{\rm top}}=\num{0.5100}$ and \num{0.5056} genuinely exceed \num{0.5}, giving exact two-sided Bernoulli sector separation in \num{1622} shots per sector (\num{2685} at \qty{95}{\percent} power; \num{49062} shots for outright $F_1>\num{0.5}$ certification at $N_{\max}=\num{6}$). On the shielded physical dangling end ($j=\num{1}$), $F_{1,{\rm top}}=\num{0.5433}$ yields sector separation in \num{577} shots (\num{953} at \qty{95}{\percent}) and outright certification in \num{822} shots at $N_{\max}=\num{6}$, the budget rows at that cutoff being computed at $N=\num{2}$; the separation is size-invariant at $N_{\max}=\num{2}$ and $\num{4}$. Across the \num{10}-trajectory $N_{\max}=\num{2}$ scan (\num{144} strata), only \num{4} cases separate sectors (all at two cells with $n=\num{1}$).

\begin{table}[!htbp]
\caption{Ten-trajectory ($N_{\max}=\num{2}$) finite-sample estimator diagnostic of the $n=\num{1}$ witness at two cells ($\omega_1=\num{0.5}$, $\delta=\num{0.05}$), evaluated on the ensemble of \Cref{tab:witness}. Because the exact master-equation expectation ($F_1=\num{0.4944}$) sits below threshold at $N_{\max}=\num{2}$, the detected fractions and binomial counts here quantify finite-trajectory estimator variance crossing $\omega_1$ rather than single-shot hardware certification. Experimental Bernoulli shot budgets at converged Fock cutoffs ($N_{\max}=\num{4},\num{6}$) are given in \Cref{tab:witness-exact}. Among all \num{144} conditions of the \num{10}-trajectory $N_{\max}=\num{2}$ scan, only the \num{4} two-cell index-end conditions at $n=\num{1}$ separate the sectors; uniform loss returns null detection at every noise level.}
\label{tab:witness:sample}
\begin{ruledtabular}
\begin{tabular}{lcccccccc}
& \multicolumn{2}{c}{Detected} & \multicolumn{2}{c}{Mean route $M$} & \multicolumn{2}{c}{Rate route $M$} & \multicolumn{2}{c}{Separation} \\
\cline{2-9}
& top. & triv. & top. & triv. & top. & triv. & \qty{80}{\percent} & \qty{95}{\percent} \\
\hline
$d=\num{0}$ & \qty{82.4}{\percent} & \qty{10.8}{\percent} & \num{295897} & \num{3320} & \num{2} & \num{27} & \num{7} & \num{10} \\
$d=\num{0.5}$ & \qty{58.2}{\percent} & \qty{10.6}{\percent} & \num{217003} & \num{3318} & \num{4} & \num{27} & \num{15} & \num{23} \\
$d=\num{1}$ & \qty{48.4}{\percent} & \qty{15.0}{\percent} & \num{269570} & \num{3819} & \num{5} & \num{19} & \num{30} & \num{48} \\
$d=\num{1.5}$ & \qty{45.2}{\percent} & \qty{13.0}{\percent} & \num{276493} & \num{4079} & \num{5} & \num{22} & \num{31} & \num{49} \\
\hline
uniform, $d=\num{0}$ & \qty{0.0}{\percent} & \qty{0.0}{\percent} & \num{454} & \num{465} & \text{--} & \text{--} & \text{--} & \text{--} \\
uniform, $d=\num{0.5}$ & \qty{0.0}{\percent} & \qty{0.0}{\percent} & \num{489} & \num{508} & \text{--} & \text{--} & \text{--} & \text{--} \\
uniform, $d=\num{1}$ & \qty{0.0}{\percent} & \qty{0.0}{\percent} & \num{649} & \num{656} & \text{--} & \text{--} & \text{--} & \text{--} \\
uniform, $d=\num{1.5}$ & \qty{0.0}{\percent} & \qty{0.0}{\percent} & \num{768} & \num{781} & \text{--} & \text{--} & \text{--} & \text{--} \\
\end{tabular}
\end{ruledtabular}
\end{table}

As shown in \Cref{tab:witness:truncation}, higher displaced-Fock witnesses ($n\ge\num{2}$) lie well below their conservative thresholds ($\num{0.3365}$ vs $\omega_2=\num{0.5}$ in the exact $N_{\max}=\num{2}$ state, and $\num{0.3369}$ vs \num{0.5} in the \num{10}-trajectory ensemble, falling further for $n=\num{3}$--\num{6}), so the asymmetry is restricted exclusively to the $n=\num{1}$ rung.

\begin{table}[!htbp]
\caption{Truncation dependence of the \num{10}-trajectory $N_{\max}=\num{2}$ witness estimator at two cells, zero loss-rate noise, and index-end shielding, on the ensemble of \Cref{tab:witness:sample}. Exceeding $\max_\alpha F_n>\omega_n$ uses exact analytical bounds $\omega_1=\omega_2=\num{0.5}$ and conservative upper bounds for $n\ge\num{3}$. At $n\ge\num{2}$, the means fall \num{0.15} to \num{0.22} below threshold in both sectors, so no realization crosses $\omega_n$ and the rate-route and separation counts are undefined; the two-cell sector asymmetry is therefore confined exclusively to the $n=\num{1}$ witness.}
\label{tab:witness:truncation}
\begin{ruledtabular}
\begin{tabular}{lccccccc}
$n$ & $\omega_n$ & $\langle F_n\rangle$ top. & $\langle F_n\rangle$ triv. & det. top. & det. triv. & $M^{\rm rate}$ top./triv. & $M_{\rm sep}$ (\qty{80}{\percent}) \\
\hline
\num{1} & \num{0.500} & \num{0.4950} & \num{0.4529} & \qty{82.4}{\percent} & \qty{10.8}{\percent} & \num{2}/\num{27} & \num{7} \\
\num{2} & \num{0.500} & \num{0.3369} & \num{0.3137} & \qty{0.0}{\percent} & \qty{0.0}{\percent} & \text{--} & \text{--} \\
\num{3} & \num{0.427} & \num{0.2738} & \num{0.2562} & \qty{0.0}{\percent} & \qty{0.0}{\percent} & \text{--} & \text{--} \\
\num{4} & \num{0.441} & \num{0.2368} & \num{0.2220} & \qty{0.0}{\percent} & \qty{0.0}{\percent} & \text{--} & \text{--} \\
\num{5} & \num{0.385} & \num{0.2117} & \num{0.1987} & \qty{0.0}{\percent} & \qty{0.0}{\percent} & \text{--} & \text{--} \\
\num{6} & \num{0.378} & \num{0.1932} & \num{0.1815} & \qty{0.0}{\percent} & \qty{0.0}{\percent} & \text{--} & \text{--} \\
\end{tabular}
\end{ruledtabular}
\end{table}

In the \num{10}-trajectory two-cell ensemble ($d=\num{0}$), combining the $n=\num{1}$ estimator crossing with the collective-mode null (\num{10} realizations per sector) yields a double-diagnostic rate of \num{0.80} in the topological sector (\num{0.80} witness crossing, \num{1.00} collective-quiet below \qty{1}{\percent} local negativity) versus \num{0.10} in the trivial control (\num{0.10} witness crossing, \num{0.40} collective-quiet, \num{0.10} joint).

\section{Physical parameter mapping and hardware bounds}
\label{sec:hardware}

The main text uses normalized model coordinates ($T=2\pi$, all rates in units of $\omega_d=2\pi/T$). This section maps a normalized parameter $\tilde{X}=X/\omega_d$ to its physical cyclic frequency $X/(2\pi)=\tilde{X}(\omega_d/2\pi)$ and angular rate $X=\tilde{X}\omega_d$. Applying the map at the two representative drive frequencies of \cref{MAIN-tab:hardware} ($\omega_d/2\pi=\qty{100}{\mega\hertz}$ for circuit QED and $\qty{10}{\mega\hertz}$ for optomechanics) gives the per-parameter cyclic frequencies tabulated there; the dimensionless ratios $\chi/J_B=\num{0.42}$, $\gamma_{\rm bulk}/\chi=\num{0.4}$, and $\gamma_{\rm edge}/\gamma_{\rm bulk}=\num{5e-4}$ hold identically in both rows, and the quality factors follow as $Q_{\rm bulk}\simeq\num{250}$ and $Q_{\rm edge}\simeq\num{5e5}$ at a cavity frequency of \qty{5}{\giga\hertz}. All results assume zero temperature, quoted as $n_{\rm th}\lesssim\num{0.1}$: from $n_{\rm th}=[\exp(\hbar\omega/k_BT)-1]^{-1}$ this bound corresponds to $T\approx\qty{0.2}{\milli\kelvin}$ at $\qty{10}{\mega\hertz}$, $T\approx\qty{2}{\milli\kelvin}$ at $\qty{100}{\mega\hertz}$, and $T\approx\qty{100}{\milli\kelvin}$ at the $\qty{5}{\giga\hertz}$ cavity frequency; a finite-temperature Lindblad evolution with thermal excitation and dephasing channels was not simulated, so this bound is the regime of validity rather than a tested robustness axis. While tunable couplings and engineered dump-port quality factors fall within circuit-QED capabilities~\cite{Blais2021Circuit}, a \qty{50}{\mega\hertz} Kerr nonlinearity corresponds to transmon-like anharmonicity rather than a linear 3D cavity, and \qty{5.0}{\mega\hertz} single-phonon Kerr requires hybrid qubit-mediated nonlinearity beyond bare radiation pressure in optomechanics~\cite{Xuereb2014}.

The measurement chain uses established primitives: parity-based direct Wigner reconstruction~\cite{Lutterbach1997}, macroscopic cavity cat states in circuit QED~\cite{Vlastakis2013}, and dissipatively stabilized quantum manifolds~\cite{Touzard2018}, with a standard homodyne chain on the optomechanical side~\cite{Aspelmeyer2014}. The grid-free witness of \cref{sec:witness} requires only displaced photon-number statistics, so detecting the two-cell signal needs a finite set of displaced measurements rather than full tomography. Site selectivity follows from frequency-resolved cavity modes addressed by a shared ancilla~\cite{Blais2021Circuit}. The engineered bulk loss corresponds to dissipative stabilization of a dynamical manifold.

\Cref{tab:testable} collects the concrete testable predictions. The near-term tomography-free target at the highest tested Fock support ($N_{\max}=\num{6}$) is the two-cell displaced-Fock witness asymmetry, whose shot budgets for site $j=\num{0}$ and for the shielded physical dangling end $j=\num{1}$ are listed in \cref{tab:witness-exact}. The full sign pattern across two through five cells is a tomography-level prediction at sub-$\num{1e-3}$ resolution. Every entry is conditional on stroboscopic readout at $t=T$ and on the structured loss profile ($\gamma_{\rm edge}/\gamma_{\rm bulk}=\num{5e-4}$); micromotion predicts sign changes within the period, and the loss-ratio scan predicts the signal to vanish by ratio \num{0.5}.

\begin{table}[!htbp]
\renewcommand{\arraystretch}{1.3}
\caption{Summary of quantitative experimental predictions from the exact master-equation and Monte Carlo benchmarks.}
\label{tab:testable}
\begin{tabularx}{\linewidth}{l>{\raggedright\arraybackslash}X>{\raggedright\arraybackslash}X}
\hline
Prediction & Measurement & Expected outcome \\
\hline
Two-cell witness asymmetry ($N_{\max}=\num{6}$) & Displaced photon counting ($F_1$) & $F_{1,{\rm top}}>\num{0.5}$ on both preparation sites; shot budgets in \cref{tab:witness-exact} \\
Contrast sign pattern, \num{2}--\num{5} cells & Wigner tomography at $t=T$ & Positive at two cells, negative at three through five; exact values in \cref{MAIN-tab:contrast-summary} \\
Kerr-free sign pattern & Same, nonlinearity off & Same sign at two cells, $\num{2.1}\times$ smaller; three- and four-cell exact contrast null (no negative sign) \\
Uniform-loss null & Same, uniform profile & Contrast identically zero at $d=\num{0}$ (exact ME) or \num{1e-6} level (MC) at $t=T$ for $d\le\num{0.5}$ \\
Stroboscopic conditioning & Readout at $t\ne T$ & Sign changes within every period \\
Loss-ratio dependence & Vary edge-to-bulk ratio & Signal gone by ratio \num{0.5} \\
Five-cell consistency & Ten-mode tomography & \num{10}-traj MC mean inside its predicted \qty{95}{\percent} interval (\cref{tab:size5}) \\
\hline
\end{tabularx}
\end{table}

The reported contrasts are invariant under a uniform rescaling of $\omega_d$; the physical time over which the sign flip occurs is $T\sim\qtyrange{10}{100}{\nano\second}$, well below typical coherence times. The normalized study does not include measurement back-action or drive-induced heating; the mapping bounds idealized-device predictions, not hardware.

\subsection{Step-by-step measurement protocol}
\label{sec:measurement-protocol}

Each step states its setting, check, and acceptance criterion. No step requires model fitting. The nine steps below refine the five stages of the main-text protocol.

\begin{enumerate}
\item \emph{Prerequisites.} An array of \num{4}--\num{10} bosonic modes with site-resolved readout, tunable nearest-neighbor couplings, a Kerr nonlinearity on the edge site, and engineered loss with an edge-to-bulk loss ratio of at least two orders of magnitude. Multimode cats have been stabilized by normal-mode dissipation engineering~\cite{Zapletal2022multimode} and cat manifolds by nonlinear reservoir engineering~\cite{Rojkov2026catmanifolds}.
\item \emph{Coherent calibration.} Working points $(J_A,J_B)=(\num{0.1},\num{1.2})$ (chiral) and $(\num{1.2},\num{0.1})$ (control). Accept when spectroscopy reproduces the minimum zero and $\pi$ quasienergy gaps \num{2.1991} and \num{0.3142} to within \qty{5}{\percent} and edge modes appear at the chiral point only.
\item \emph{Loss engineering.} $\gamma_{\rm bulk}=\num{0.2}$, $\gamma_{\rm edge}=\num{1e-4}$ (drive units). Accept when ringdown reproduces both rates with ratio below \num{1e-2}.
\item \emph{State preparation.} Even cat $\alpha=\num{1}$ on site~\num{0} (or physical dangling end~$j=\num{1}$ under physical-end shielding). Accept when pre-drive local Wigner negativity reads $\approx\num{0.105}$, the physical untruncated-cat value; the simulation reference $\delta_-(\num{0})=\num{0.1962}$ is the $N_{\max}=\num{2}$ cutoff-truncated construction used throughout (\cref{sec:obsvalidation}).
\item \emph{Single-period drive.} Read out at $t=T$ locked to the drive phase. This lock is mandatory: micromotion predicts sign changes within the period.
\item \emph{Witness measurement (near-term target).} Displaced photon counting for the $F_1$ witness against $\omega_1=\num{0.5}$ at high-cutoff Fock support ($N_{\max}=\num{6}$), scanning the displacement $\alpha$ and evaluating each budget at its maximizing $\alpha$ (the maxima are flat to \num{2.4e-6}). The budgets assume unit-efficiency detection and require $\eta\gtrsim\num{0.96}$ to be met as quoted. The budgets in \cref{tab:witness-exact} are tabulated at the sizes shown: site $j=\num{0}$ at $N=\num{2}$, and on site $j=\num{1}$ the $N_{\max}=\num{2}$ rows at $N=\num{2},\num{3},\num{4}$; the $N_{\max}=\num{6}$ value (\num{577} shots per sector) is computed at $N=\num{2}$, with $N\ge\num{3}$ at that cutoff not computed. Accept when the measured sector difference matches the exact-ME separation in \cref{tab:witness-exact} within shot noise, and is consistent with zero under uniform loss at $d\le\num{0.5}$.
\item \emph{Tomography (full pattern).} Edge-site Wigner function on $\num{31}^2$ grid (\num{961} settings per function) for both sectors. Accept convergence toward exact-ME benchmarks ($+\num{0.02221}$ to $+\num{0.02606}$ at two cells, $-\num{0.000113}$ ($N_{\max}=\num{2}$, three through five cells) down to $-\num{0.000955}$ ($N_{\max}=\num{4}$, three cells)) and null contrast at $t=T$ under uniform loss.
\item \emph{Loss-rate noise and sweep repetition.} Repeat steps~\num{4}--\num{7} over loss-rate noise amplitudes \num{0}--\num{1.5} (\num{500} matched units per setting at two and three cells) and along $J_A+J_B=\num{1.3}$. Accept when the contrast stays continuous through all three closings with a zero window straddling the $\nu_0$ transition.
\item \emph{Falsification.} The boundary claim fails if the contrast tracks the invariant across the transition, or if a nonzero uniform-loss contrast appears at $t=T$ for $d\le\num{0.5}$, or if the five-cell mean falls outside its predicted \qty{95}{\percent} interval.
\end{enumerate}

\section{Validity regimes and parameter scope}
\label{sec:scope-limitations}

The regime in which each claim holds, and the section that quantifies it, is collected in \cref{tab:scope-limitations}.

\begin{table}[!htbp]
\footnotesize
\renewcommand{\arraystretch}{1.0}
\caption{Validity scope of the reported results; each row states the regime in which the corresponding claim holds and points to the section that quantifies it.}
\label{tab:scope-limitations}
\begin{tabularx}{\linewidth}{lX}
\textbf{Quantity} & \textbf{Scope and pointer} \\
\hline
Chiral bulk invariant & Exact single-particle windings, verified by symmetric-frame and spectral-localizer calculations (\cref{sec:localizer}). \\\hline
Finite-chain edge state & Single-particle diagnostic of the coherent Floquet operator. \\\hline
Matched trajectories & Valid within matched loss-rate noise realizations at common cutoff and grid (\cref{sec:statistics}). \\\hline
Local Wigner protection & Size-robust only at $N=\num{2}$ under boundary-preserving loss (\cref{sec:si_sizelaw,sec:gridconvergence}). \\\hline
Passive edge-mode reference & Isolation bound met by Floquet-mode initialization, exceeded by local preparation (\cref{sec:covariantmethod}). \\\hline
Cutoff dependence & Sign persists through $N_{\max}=\num{6}$; magnitudes are cutoff-sensitive (\cref{sec:exact-subspace,sec:nmaxpanel}). \\\hline
Wigner-grid discretization & Negative branch quoted as a grid pair, $\num{31}^2$ versus $\num{61}^2$ (\cref{sec:gridconvergence}). \\\hline
Trajectory count & Exact-ME versus Monte Carlo ladder in \cref{sec:trajcount}. \\\hline
Collective-mode negativity & Null for both local and mode-resolved preparations (\cref{sec:collective-wigner}). \\\hline
Multi-period behavior & No inversion up to eight periods; revival at four (\cref{sec:multiperiod}). \\\hline
Loss-ratio dependence & Contrast requires the sharp engineered profile (\cref{tab:lossratio}). \\\hline
Statistical power & Resolution of the eight-mode design in \cref{sec:power}. \\\hline
Bulk-invariant correspondence & Contrast varies across the gap closings (\cref{sec:dimerization}). \\\hline
Mixed-state marker & No tested covariance marker separates the driven mixed states (\cref{MAIN-sec:discussion}). \\\hline
Symmetry-compatible loss & Sublattice-loss control contrast in \cref{sec:chiral-loss}. \\\hline
Drive phase & Sign changes within one period (\cref{sec:micromotion,sec:multiperiod}). \\\hline
Five-cell size & Out-of-sample check in \cref{sec:size5}. \\\hline
Grid-free certification & Shot budgets in \cref{tab:witness-exact}. \\\hline
Long-time persistence & No reversal over \num{64} periods (\cref{sec:longtime}). \\\hline
Physical unit scales & Normalized drive units; hardware map in \cref{MAIN-tab:hardware,sec:hardware}. \\
\end{tabularx}
\end{table}

This table restricts every topological index to the coherent Floquet generator: at finite mixing no tested covariance marker separates the driven mixed states.

\section{Size-scaling fit, time scales, and split-loss decomposition}
\label{sec:si_sizelaw}

\begin{figure}[t]
\centering
\includegraphics[width=0.80\linewidth]{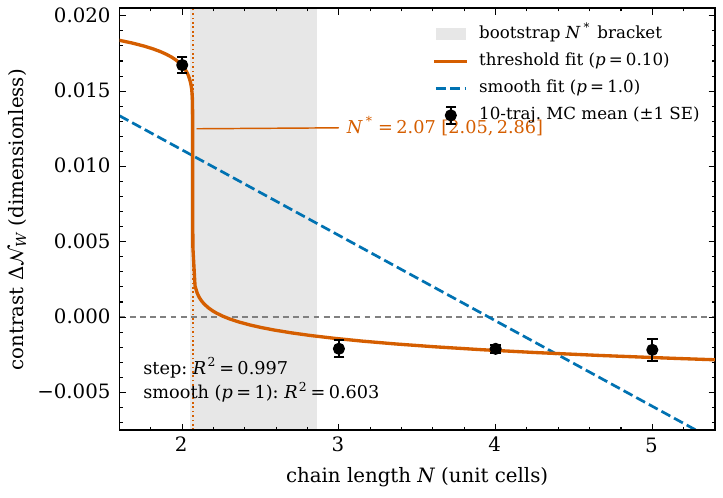}
\caption{Phenomenological size-scaling comparison on the four chain lengths ($N=\num{2}\text{--}\num{5}$, \num{10}-trajectory zero-noise ensemble; error bars indicate $\pm\num{1}$~SE). The solid orange step curve ($N^{*}=\num{2.07}$, $p=\num{0.10}$, $R^2=\num{0.9973}$; vertical orange dotted line at $N^*=\num{2.07}$) crosses the dashed gray zero line cleanly between two and three cells inside the shaded gray bootstrap $N^*$ bracket, whereas the dashed blue $p=\num{1}$ curve ($R^2=\num{0.6031}$) misses the sharp step; $(N^{*},p)$ remains degenerate along the ridge $2<N^{*}<3$.}
\label{fig:si_sizelaw}
\end{figure}

The four sizes of the main text are fitted with a threshold form
\begin{equation}
C(N)=C_\infty+A\,\operatorname{sgn}\!\left(N^{*}-N\right)
\left|\frac{N}{N^{*}}-1\right|^{p}
\end{equation}
and with a smooth $p=\num{1}$ form,
$C(N)=C_\infty+A(1-N/N^{*})$. The optimizer returns $N^{*}=\num{2.0676}$,
$A=\num{0.0111}$, $C_\infty=\num{0.0088}$ and $p=\num{0.100}$, with
$R^{2}=\num{0.9973}$ and a largest residual of $\num{6.6e-4}$, against
$R^{2}=\num{0.6031}$ and a largest residual of $\num{7.5e-3}$ for the
smooth $p=\num{1}$ form. A bootstrap over \num{2000} resamples returns the interval
$\numrange{2.050}{2.865}$ for $N^{*}$, but that interval is not a measured
threshold location: scanning the pair $(N^{*},p)$ leaves $R^{2}$ at or above
$\num{0.997}$ for $N^{*}$ anywhere in $\numrange{2.05}{2.95}$ and $p$ anywhere
between \num{0.1} and \num{12}, so the two parameters are degenerate along a
ridge. Four points constrain the sign-change bracket $2<N^{*}<3$ without uniquely identifying $p$; the curve $N^{*}=\num{2.07}$ plotted in \cref{fig:si_sizelaw} represents the central member of that degenerate ridge.

The escape time $\xi/v_{\max}=\num{4.0}\,\omega_d^{-1}$ at the working point $\chi=\num{0.5}$, $|\alpha|=\num{1}$, $v_{\max}=\num{0.1}$ and $\xi/a_{0}=\num{0.402}$ with $a_{0}$ the unit-cell length, and the Kerr scale, are both set by the preparation and the drive rather than by the chain length, up to the weak size drift of the localization length; a Kerr--escape competition is therefore degenerate with the crossing on the size axis, and the Kerr-controlled reading is excluded by the fixed-length sink measurement.

The sink sweep at fixed two-cell length, including the two-grid values for each sink position and the split-budget control, is reported in the main text. The length term (effect of appending a cell at fixed profile) is $+\num{9.6e-10}$ and the profile term (effect of promoting site~$j=\num{3}$) is $-\num{0.0223}$; their sum closes the observed difference to below $\num{1e-9}$.

\bibliographystyle{apsrev4-2}
\bibliography{04_track_A_topo_references}